\documentclass[final,5p,times,twocolumn]{elsarticle}

\usepackage{graphicx}

\usepackage{amssymb}
\usepackage{amsthm}

\biboptions{comma,square,numbers,sort&compress}

\usepackage[utf8]{inputenc}
\usepackage[T1]{fontenc}
\usepackage{lmodern}
\usepackage{microtype}
\usepackage[english]{babel}
\usepackage[hyperindex=true,colorlinks=true]{hyperref}
\usepackage{amsmath}
\usepackage{amsfonts}
\usepackage[force]{feynmp-auto}

\usepackage{algorithm}
\usepackage[noend]{algpseudocode}

\usepackage{mathtools, cuted}

\newtheorem*{proposition*}{Proposition}

\newcounter{bla}

\journal{Computer Physics Communications}

 {
   \begin{list}{\bf\arabic{enumi} --}
    {\usecounter{enumi} \setlength{\parsep}{0pt}
     \setlength{\itemsep}{3pt} \settowidth{\labelwidth}{10 --}
     \sloppy
    }}{\end{list}}

 {
   \begin{list}{\bf\Alph{enumii}/}
    {\usecounter{enumii} \setlength{\parsep}{0pt}
     \setlength{\itemsep}{3pt} \settowidth{\labelwidth}{A/}
     \sloppy
    }}{\end{list}}

 {
   \begin{list}{\bf\alph{enumii}/}
    {\usecounter{enumii} \setlength{\parsep}{0pt}
     \setlength{\itemsep}{3pt} \settowidth{\labelwidth}{a/}
     \sloppy
    }}{\end{list}}

 {
   \begin{list}{\bf\roman{enumiii}/}
    {\usecounter{enumiii} \setlength{\parsep}{0pt}
     \setlength{\itemsep}{3pt} \settowidth{\labelwidth}{a/}
     \sloppy
    }}{\end{list}}

 {
   \begin{list}{$\bullet$}
    {\setlength{\parsep}{3pt}
     \setlength{\itemsep}{3pt} \settowidth{\labelwidth}{a/}
     \sloppy
    }}{\end{list}}

\makeatletter     
\def\overbracket#1{\mathop{\vbox{\ialign{##\crcr\noalign{\kern2\p@}
\downbracketfill\crcr\noalign{\kern6\p@\nointerlineskip}
$\hfil\displaystyle{#1}\hfil$\crcr}}}\limits}
\def\downbracketfill{$\m@th
\makesm@sh{\kern4\p@\llap{\vrule\@height.7\p@\@depth4.\p@\@width.7\p@}}%
\leaders\vrule\@height.7\p@\hfill
\makesm@sh{\rlap{\vrule\@height.7\p@\@depth4.\p@\@width.7\p@}\kern4\p@}$}
\def\underbracket#1{\mathop{\vtop{\ialign{##\crcr
$\hfil\displaystyle{#1}\hfil$\crcr\noalign{\kern3\p@\nointerlineskip}
\upbracketfill\crcr\noalign{\kern3\p@}}}}\limits}
\def\overparenthesis#1{\mathop{\vbox{\ialign{##\crcr\noalign{\kern3\p@}
\downparenthfill\crcr\noalign{\kern3\p@\nointerlineskip}
$\hfil\displaystyle{#1}\hfil$\crcr}}}\limits}
\def\underparenthesis#1{\mathop{\vtop{\ialign{##\crcr
$\hfil\displaystyle{#1}\hfil$\crcr\noalign{\kern3\p@\nointerlineskip}
\upparenthfill\crcr\noalign{\kern3\p@}}}}\limits}
\def\downparenthfill{$\m@th\braceld\leaders\vrule\hfill\bracerd$}
\def\upparenthfill{$\m@th\bracelu\leaders\vrule\hfill\braceru$}
\def\upbracketfill{$\m@th\makesm@sh{\llap{\vrule\@height3\p@\@width.7\p@}}%
\leaders\vrule\@height.7\p@\hfill
\makesm@sh{\rlap{\vrule\@height3\p@\@width.7\p@}}$}
\makeatother
\newcommand{\ra}   {\rangle}

\newcommand{\ket}[1]{{|#1\rangle}}

\newcommand{\be}{\begin{equation}}
\newcommand{\ee}{\end{equation}}
\newcommand{\bd}{\begin{displaymath}}
\newcommand{\ed}{\end{displaymath}}
\newcommand{\bqr}{\begin{eqnarray}}
\newcommand{\eqr}{\end{eqnarray}}
\newcommand{\bqrr}{\begin{eqnarray*}}
\newcommand{\eqrr}{\end{eqnarray*}}
\newcommand{\bt}{\begin{tabular}}
\newcommand{\et}{\end{tabular}}
\newcommand{\bc}{\begin{center}}
\newcommand{\ec}{\end{center}}

\begin{document}

\allowdisplaybreaks

\begin{frontmatter}



\title{\texttt{ADG}: Automated generation and evaluation of many-body diagrams \\
II. Particle-number projected Bogoliubov many-body perturbation theory}


\author[a,b,c]{P.~Arthuis}
\ead{p.arthuis@surrey.ac.uk}
\author[d,e,f,g]{A.~Tichai}
\ead{a\_tichai@theorie.ikp.physik.tu-darmstadt.de}
\author[c]{J.~Ripoche}
\ead{julien.ripoche@u-psud.fr}
\author[b,h]{T.~Duguet}
\ead{thomas.duguet@cea.fr}

\address[a]{Department of Physics, University of Surrey, Guildford GU2 7XH, United Kingdom}
\address[b]{IRFU, CEA, Université Paris-Saclay, F-91191 Gif-sur-Yvette, France}
\address[c]{CEA, DAM, DIF, F-91297 Arpajon, France}
\address[d]{Max-Planck-Institut f\"ur Kernphysik, Heidelberg, Germany}
\address[e]{Institut f\"ur Kernphysik, Technische Universit\"at Darmstadt, Darmstadt, Germany}
\address[f]{ExtreMe Matter Institute EMMI, GSI Helmholtzzentrum f\"ur Schwerionenforschung GmbH, Darmstadt, Germany}
\address[g]{ESNT, IRFU, CEA, Université Paris-Saclay, F-91191 Gif-sur-Yvette, France}
\address[h]{KU Leuven, Instituut voor Kern- en Stralingsfysica, 3001 Leuven, Belgium}

\begin{abstract}
We describe the second version (v2.0.0) of the code \textbf{\texttt{ADG}} that automatically (1) generates all valid \emph{off-diagonal} Bogoliubov many-body perturbation theory diagrams at play in particle-number projected Bogoliubov many-body perturbation theory (PNP-BMBPT) and (2) evaluates their algebraic expression to be implemented for numerical applications. This is achieved at any perturbative order $p$ for a Hamiltonian containing both two-body (four-legs) and three-body (six-legs) interactions (vertices). All valid off-diagonal BMBPT diagrams of order $p$ are systematically generated from the set of \emph{diagonal}, i.e.,\@ unprojected, BMBPT diagrams. The production of the latter were described at length in Ref.~\cite{arthuis18a} dealing with the first version of \textbf{\texttt{ADG}}. The automated evaluation of off-diagonal BMBPT diagrams relies both on the application of algebraic Feynman's rules and on the identification of a powerful diagrammatic rule providing the result of the remaining $p$-tuple time integral. The new diagrammatic rule generalizes the one already identified in Ref.~\cite{arthuis18a} to evaluate diagonal BMBPT diagrams independently of their perturbative order and topology. The code \textbf{\texttt{ADG}} is written in \emph{Python3} and uses the graph manipulation package \emph{NetworkX}. The code is kept flexible enough to be further expanded throughout the years to tackle the diagrammatics at play in various many-body formalisms that already exist or are yet to be formulated. 

\end{abstract}


\end{frontmatter}



{\bf NEW VERSION PROGRAM SUMMARY}

\begin{small}
\noindent
{\em Program Title:} \texttt{ADG}                                         \\
{\em Licensing provisions:} GPLv3                                   \\
{\em Programming language:} Python3                                   \\
{\em Supplementary material:} Yes.                                \\
{\em Journal reference of previous version:} P.\@ Arthuis, T.\@ Duguet, A.\@ Tichai, R.-D.\@ Lasseri and J.-P.\@ Ebran, "ADG: Automated generation and evaluation of many-body diagrams I. Bogoliubov many-body perturbation theory", Computer Physics Communications 240 (2019), pp. 202-227.                 \\
{\em Does the new version supersede the previous version?:} Yes.  \\
{\em Reasons for the new version:} Incorporation of a new formalism into the program. \\
{\em Summary of revisions:} Addition of off-diagonal BMBPT to the formalisms for which diagrams can be generated, fix of a wrong symmetry factor, move of the codebase from Python2 to Python3 while maintaining support for Python2, various optimizations to reduce the time and memory necessary to the program. \\
{\em Nature of problem:} As formal and numerical developments in many-body-perturbation-theory-based {\it ab initio} methods make higher orders reachable, manually producing and evaluating all the diagrams becomes rapidly untractable as both their number and complexity grow quickly, making it prone to mistakes and oversights.\\
{\em Solution method:} Diagonal BMBPT diagrams are encoded as square matrices known as oriented adjacency matrices in graph theory, and then turned into graph objects using the \emph{NetworkX} package. Off-diagonal BMBPT diagrams can then be generated from those graphs. Checks on the diagrams and evaluation of their time-integrated expression are eventually done on a purely diagrammatic basis. HF-MBPT diagrams are produced and evaluated as well using the same principle.\\
   \\

\end{small}

\section{Introduction}
\label{s:intro}

In Ref.~\cite{arthuis18a}, the first version of the code \textbf{\texttt{ADG}} was described. The code was designed to automatically (1) generate all valid Bogoliubov many-body perturbation theory (BMBPT) diagrams and (2) evaluate their algebraic expression to be implemented for numerical applications. This was achieved at any perturbative order $p$ for a Hamiltonian containing both two-body (four-legs) and three-body (six-legs) interactions (vertices). This code development took place in the context of the rapidly evolving field of quantum many-body calculations for fermionic systems, i.e.,\@ atomic nuclei, molecules, atoms or solids. In nuclear physics in particular, the past decade has witnessed the development and/or the application of formalisms~\cite{Tsukiyama:2010rj,Soma:2011aj,HeBo13,Hergert:2014iaa,Duguet:2014jja,Bogner:2014baa,Jansen:2014qxa,Duguet:2015yle,Gebrerufael:2016xih,Ti17,Tichai:2018mll,Tichai:2020review} among which is Bogoliubov many-body perturbation theory~\cite{Tichai:2018mll,Demol:2020}. This profusion of diagrammatic methods, along with the rapid progress of computational power allowing for high-order implementations, welcomes the development of a versatile code capable of both generating and evaluating many-body diagrams. This was the goal achieved in Ref.~\cite{arthuis18a} for BMBPT.

The aim of BMBPT is to tackle (near) degenerate Fermi systems, e.g.,\@ open-shell nuclei displaying a superfluid character, by perturbatively expanding the exact solution of the Schrödinger equation around a so-called Bogoliubov reference state. This tremendous benefit is obtained at the price of using a reference state that breaks $U(1)$ global-gauge symmetry associated with the conservation of particle number in the system. Given that the breaking of a symmetry cannot actually be realized in a finite quantum system, BMBPT calculations come with a (hopefully mild) contamination of physical observables. In this context, BMBPT must actually be seen as a first step towards a more general diagrammatic method, the particle-number projected Bogoliubov many-body perturbation theory (PNP-BMBPT)~\cite{Duguet:2015yle}, in which the broken symmetry is exactly restored at any truncation order. 

It is thus the goal of the present paper to extend the formal and code developments performed in Ref.~\cite{arthuis18a} to PNP-BMBPT.  For a reason that will become clear later on, the diagrammatics at play in PNP-BMBPT is coined as the \emph{off-diagonal} BMBPT diagrammatics from which the \emph{diagonal} BMBPT diagrammatics encountered in straight BMBPT is recovered in a particular limit, i.e.,\@ diagonal BMBPT diagrams characterize the subset of off-diagonal BMBPT diagrams that remains non-zero in that limit. In this context, the new version of the code \textbf{\texttt{ADG}} is capable of automatically (1) generating all valid \emph{off-diagonal} BMBPT diagrams and of (2) evaluating their algebraic expression to be implemented for numerical applications. This is achieved at any perturbative order $p$ for a Hamiltonian containing both two-body (four-legs) and three-body (six-legs) interactions (vertices). The numerical tool remains flexible enough to be expanded throughout the years to tackle the diagrammatics at play in yet other many-body formalisms (already existing or yet to be formulated). 

The paper is organized as follows. Section~\ref{s:basicsingredients} introduces the basic ingredients that are needed to elaborate on the PNP-BMBPT formalism in Sec.~\ref{definitions}. The master equations, the associated diagrammatics and the difficulties to overcome in order to achieve an automated generation and evaluation of diagrams of arbitrary orders are detailed. Building on this, Secs.~\ref{secautomaticgeneration} and~\ref{secautomaticevaluation} detail the method developped to reach such an objective. While Sec.~\ref{programuse} details how the \textbf{\texttt{ADG}} code operates, conclusions are given in Sec.~\ref{conclusions}. Three appendices follow to provide details regarding the formalism and the structure of the program.

\section{Basic ingredients}
\label{s:basicsingredients}

Expansion many-body methods consist of expanding exact A-body quantities, e.g.,\@ the ground-state energy, around a reference state. By default, the reference state is naturally chosen to carry the symmetry quantum numbers of the exact target state dictacted by the Hamiltonian. As for $U(1)$ global gauge symmetry associated with particle-number conservation, this typically leads to using a Slater determinant as a reference state. In open-shell nuclei, however, the reference Slater determinant is degenerate with respect to elementary, i.e.,\@ particle-hole, excitations, which eventually makes the many-body expansion singular. This singularity relates to Cooper pair's instability associated with the superfluid character of singly (doubly) open-shell nuclei.

One option to address Cooper pair's instability is to allow the reference state to break $U(1)$ symmetry, i.e.,\@ to be chosen under the form of a more general Bogoliubov product state. Doing so, the degeneracy of the Slater determinant with respect to particle-hole excitations is lifted and commuted into a degeneracy with respect to symmetry transformations of the group. As a consequence, the ill-defined (i.e., singular) expansion of exact quantities is replaced by a well-behaved one. This task is indeed achieved by straight BMBPT~\cite{Tichai:2018mll,arthuis18a} in a perturbative setting.

Eventually, however, the degeneracy with respect to $U(1)$ transformations must also be lifted via the restoration of the symmetry. Indeed, the breaking of a symmetry is fictitious in finite quantum systems and can at best be characterized as being \emph{emergent}~\cite{yannouleas07a,Papenbrock:2013cra}. In this regards, straight BMBPT only restores the symmetry in the limit of an all-orders resummation, and, thus, includes a symmetry contamination at any finite order. Thus, an explicit extension of the formalism is necessary in practice to restore good particle-number at any truncation order. Within a perturbative setting, this task is achieved by PNP-BMBPT~\cite{Duguet:2015yle}. Before detailing this many-body formalism and its associated diagrammatics, the present section introduces the necessary ingredients.

\subsection{$U(1)$ group}

Let us consider the abelian compact Lie group $U(1)\equiv \{S(\varphi), \varphi \in  [0,2\pi]\}$ associated with the global rotation of an A-body fermion system in gauge space. As $U(1)$ is considered to be a symmetry group of the Hamiltonian $H$, commutation relations
\begin{equation}
\left[H,S(\varphi)\right] = \left[A,S(\varphi)\right] = 0 \, , \label{commutation}
\end{equation}
hold for any $\varphi \in  [0,2\pi]$, where $A$ denotes the particle-number operator. We utilize the unitary representation of $U(1)$ on Fock space $\mathcal{F}$ given by  
\begin{equation}
S(\varphi) = e^{iA\varphi} \, .
\end{equation}

One is presently interested in evaluating a ground-state observable $\text{O}^{\text{A}}_0$ whose associated operator $O$ commutes with $H$ and $A$. Generically speaking, this means that the eigenequations of interest in PNP-BMBPT\footnote{While the focus is presently on ground-state quantities, extensions of the formalism to excited states or transitions of non-scalar operators can be envisioned.} are given by
\begin{align}
O | \Psi^{\text{A}}_{0} \rangle &=  \text{O}^{\text{A}}_{0} \, | \Psi^{\text{A}}_{0} \rangle \, , \label{schroed} 
\end{align}
where $O\equiv H$ or $A$ and where Eq.~\eqref{schroed} explicitly stipulates that energy eigenstates $| \Psi^{\text{A}}_{\mu} \rangle$ carry good symmetry quantum number $\text{A}$. 

Given these many-body eigenstates, matrix elements of the irreducible representations (IRREPs) of $U(1)$ are given by 
\begin{equation}
\langle \Psi^{\text{A}}_{\mu} |  S(\varphi)  |\Psi^{\text{A}'}_{\mu'} \rangle \equiv e^{i\text{A}\varphi} \, \delta_{\text{A}\text{A}'} \, \delta_{\mu\mu'} \, . \label{irreps}
\end{equation}
The volume of the group is given by
\begin{equation*}
v_{U(1)} \equiv \int_{0}^{2\pi} \!d\varphi = 2\pi \, ,
\end{equation*}
while the orthogonality of IRREPs reads as
\begin{equation}
\frac{1}{2\pi}\int_{0}^{2\pi} \!d\varphi \,e^{-i\text{A}\varphi}\,e^{+i\text{A}'\varphi}  = \delta_{\text{A}\text{A}'} \, . \label{orthogonality}
\end{equation}

Whenever a many-body state is not an eigenstate of the particle-number operator, it is easy to demonstrate that the eigen-component of $A$ associated with particle-number $\text{A}$ can be extracted from it via the application of the projection operator $P^{\text{A}}$ defined through 
\begin{align}
P^{\text{A}} &\equiv  \frac{1}{2\pi}\int_{0}^{2\pi} \!d\varphi \, e^{-i\text{A}\varphi}  \, S(\varphi)  \, . \label{projAoperator}
\end{align}

\subsection{Bogoliubov vacuum}
\label{subs:bogovacuum}

The set up of the many-body formalism starts with the introduction of the Bogoliubov reference state
\begin{align}
\ket{\Phi} \equiv \mathcal{C} \prod_k \beta_k \ket{0} \, ,
\end{align}
where $\mathcal{C}$ is a complex normalization constant and $\vert 0 \ra$ denotes the physical vacuum. The Bogoliubov state is a vacuum for the set of quasi-particle operators obtained from those associated with a basis of the one-body Hilbert space via a unitary linear transformation of the form~\cite{ring80a}
\begin{subequations}
\begin{align}
\beta_k &\equiv \sum_p U^*_{pk} c_p + V^*_{pk} c^\dagger_p \, , \\
\beta_k^\dagger &\equiv \sum_p U_{pk} c^\dagger_p + V_{pk} c_p \, ,
\end{align}
\end{subequations}
i.e., $\beta_k \ket{\Phi} = 0$ for all $k$. One possiblity to specify the Bogoliubov reference state $\ket{\Phi}$ is to require that it solves the Hartree-Fock-Bogoliubov (HFB) variational problem. This fixes the transformation matrices $(U,V)$~\cite{ring80a} and delivers the set of quasi-particle energies $\{E_k > 0\}$ defining the unperturbed part of the Hamiltonian later on (see Eqs.~\eqref{split2}-\eqref{onebodypiece}). We do not impose this choice here such that the reference state and the associated unperturbed Hamiltonian can be defined more generally.

The Bogoliubov reference state is not an eigenstate of the particle-number operator $A$. Although the present objective is to perturbatively correct $\ket{\Phi}$ while exactly restoring the particle number at any truncation order, the fact that the reference state breaks the symmetry in the first place requires to work with the grand potential $\Omega \equiv H - \lambda A$ rather than with $H$ itself in the set up of the many-body formalism~\cite{Duguet:2015yle}, where $\lambda$ denotes the chemical potential.

\subsection{Normal-ordered operator}
\label{subs:NOop}

The operator $O$ of interest typically contains one-body, two-body and three-body terms\footnote{Higher-body operators can be employed as well. From the formal point of view, it poses no fundamental difficulty but further complicates the diagrammatic formalism. As for the automated generation of diagrams, it poses no fundamental difficulty but requires to handle the memory needed to deal with the increased combinatorial complexity.}. The operator in the Schrödinger representation is expressed in an arbitrary basis of the one-body Hilbert space as
\begin{align}
O &\equiv o^{[2]} + o^{[4]} + o^{[6]} \label{e:ham}  \\
&\equiv  o^{11} +  o^{22} +  o^{33}  \nonumber \\
&\equiv  \frac{1}{(1!)^2} \sum _{p_1p_2} o^{11}_{p_1p_2} c^{\dagger}_{p_1} c_{p_2} \nonumber \\
&\phantom{=}+\frac{1}{(2!)^2} \sum _{p_1p_2p_3p_4} o^{22}_{p_1p_2p_3p_4}  c^{\dagger}_{p_1} c^{\dagger}_{p_2} c_{p_4} c_{p_3} \nonumber \\
&\phantom{=}+\frac{1}{(3!)^2} \sum _{p_1p_2p_3p_4p_5p_6} o^{33}_{p_1p_2p_3p_4p_5p_6}  c^{\dagger}_{p_1} c^{\dagger}_{p_2} c^{\dagger}_{p_3} c_{p_6} c_{p_5} c_{p_4}  \, . \nonumber
\end{align} 
Each term $o^{kk}$ of the particle-number conserving operator $O$ is characterized by the equal number $k$ of particle creation and annihilation operators. The class $o^{[2k]}$ is nothing but the term $o^{kk}$ of $k$-body character. The maximum value $\texttt{deg\_max}\equiv \text{Max} \, 2k$ defines the rank of the operator $O$. Matrix elements are fully antisymmetric, i.e.,
\begin{equation}
o^{kk}_{p_1 \ldots p_{k} p_{k+1} \ldots p_{2k}} = (-1)^{\sigma(P)} o^{kk}_{P(p_1 \ldots p_{k} | p_{k+1} \ldots p_{2k})} \, ,
\end{equation}
where $\sigma(P)$ refers to the signature of the permutation $P$.  The notation $P(\ldots | \ldots)$ denotes a separation into the $k$ particle-creation operators and the $k$ particle-annihilation operators such that permutations are only considered among members of the same group. 

The next step consists of normal ordering $O$ with respect to the Bogoliubov vacuum $\ket{\Phi}$, thus, obtaining
\begin{align}
\label{e:oqpas}
O &\equiv O^{[0]} + O^{[2]} + O^{[4]} + O^{[6]} \\
&\equiv O^{00} + \Big[O^{11} + \{O^{20} + O^{02}\}\Big]  \nonumber\\
&\phantom{=} + \Big[O^{22} + \{O^{31} + O^{13}\} + \{O^{40} + O^{04}\}\Big] \nonumber\\
   & \phantom{=} + \Big[O^{33} + \{O^{42} + O^{24}\} + \{O^{51} + O^{15}\} + \{O^{60} + O^{06}\}\Big] \nonumber \\
&= O^{00} \nonumber\\
& \phantom{=} + \frac{1}{(1!)^2}\sum_{k_1 k_2} O^{11}_{k_1 k_2}\beta^{\dagger}_{k_1} \beta_{k_2} \nonumber\\
& \phantom{=} + \frac{1}{2!}\sum_{k_1 k_2} \Big \{O^{20}_{k_1 k_2} \beta^{\dagger}_{k_1}
 \beta^{\dagger}_{k_2} + O^{02}_{k_1 k_2}   \beta_{k_2} \beta_{k_1} \Big \} \nonumber\\
& \phantom{=} + \frac{1}{(2!)^{2}} \sum_{k_1 k_2 k_3 k_4} \hspace{-5pt}O^{22}_{k_1 k_2 k_3 k_4}
   \beta^{\dagger}_{k_1} \beta^{\dagger}_{k_2} \beta_{k_4}\beta_{k_3} \nonumber\\
   & \phantom{=} + \frac{1}{3!1!}\sum_{k_1 k_2 k_3 k_4}\hspace{-5pt}\Big \{ O^{31}_{k_1 k_2 k_3 k_4}
   \beta^{\dagger}_{k_1}\beta^{\dagger}_{k_2}\beta^{\dagger}_{k_3}\beta_{k_4} \nonumber\\
   & \phantom{= + \frac{1}{1!3!}\sum_{k_1 k_2 k_3 k_4}} + O^{13}_{k_1 k_2 k_3 k_4} \beta^{\dagger}_{k_1} \beta_{k_4} \beta_{k_3} \beta_{k_2}  \Big \} \nonumber\\
  & \phantom{=} +  \frac{1}{4!} \sum_{k_1 k_2 k_3 k_4}\hspace{-5pt}\Big \{ O^{40}_{k_1 k_2 k_3 k_4}
   \beta^{\dagger}_{k_1}\beta^{\dagger}_{k_2}\beta^{\dagger}_{k_3}\beta^{\dagger}_{k_4}  \nonumber\\
   & \phantom{= +  \frac{1}{4!} \sum_{k_1 k_2 k_3 k_4}} + O^{04}_{k_1 k_2 k_3 k_4}  \beta_{k_4} \beta_{k_3} \beta_{k_2} \beta_{k_1}  \Big \}  \nonumber\\
   & \phantom{=} + \frac{1}{(3!)^2} \sum_{k_1 k_2 k_3 k_4 k_5 k_6} \hspace{-5pt}
   O^{33}_{k_1 k_2 k_3 k_4 k_5 k_6}
   \beta^{\dagger}_{k_1}\beta^{\dagger}_{k_2}\beta^{\dagger}_{k_3}\beta_{k_6}\beta_{k_5}\beta_{k_4} \nonumber\\
   & \phantom{=} + \frac{1}{2! \, 4!} \sum_{k_1 k_2 k_3 k_4 k_5 k_6} \hspace{-5pt}\Big \{
   O^{42}_{k_1 k_2 k_3 k_4 k_5 k_6}\beta^{\dagger}_{k_1}\beta^{\dagger}_{k_2}\beta^{\dagger}_{k_3}
   \beta^{\dagger}_{k_4}\beta_{k_6}\beta_{k_5} \nonumber\\
   & \phantom{= + \frac{1}{2! \, 4!} \sum_{k_1 k_2 k_3 k_4 k_5 k_6}} +  O^{24}_{k_1 k_2 k_3 k_4 k_5 k_6}
   \beta^{\dagger}_{k_1}\beta^{\dagger}_{k_2}\beta_{k_6}\beta_{k_5}\beta_{k_4}\beta_{k_3} \Big \} \nonumber\\
   & \phantom{=} + \frac{1}{5!1!} \sum_{k_1 k_2 k_3 k_4 k_5 k_6} \hspace{-5pt}\Big \{
   O^{51}_{k_1 k_2 k_3 k_4 k_5 k_6}\beta^{\dagger}_{k_1}\beta^{\dagger}_{k_2}\beta^{\dagger}_{k_3}
   \beta^{\dagger}_{k_4}\beta^{\dagger}_{k_5}\beta_{k_6} \nonumber\\
   & \phantom{= + \frac{1}{1!5!} \sum_{k_1 k_2 k_3 k_4 k_5 k_6}}  +  O^{15}_{k_1 k_2 k_3 k_4 k_5 k_6}
   \beta^{\dagger}_{k_1}\beta_{k_6}\beta_{k_5}\beta_{k_4}\beta_{k_3}\beta_{k_2} \Big \} \nonumber\\
   & \phantom{=} + \frac{1}{6!} \sum_{k_1 k_2 k_3 k_4 k_5 k_6} \hspace{-5pt} \Big \{
   O^{60}_{k_1 k_2 k_3 k_4 k_5 k_6}\beta^{\dagger}_{k_1}\beta^{\dagger}_{k_2}\beta^{\dagger}_{k_3}
   \beta^{\dagger}_{k_4}\beta^{\dagger}_{k_5}\beta^{\dagger}_{k_6} \nonumber\\
   & \phantom{= + \frac{1}{6!} \sum_{k_1 k_2 k_3 k_4 k_5 k_6}} +  O^{06}_{k_1 k_2 k_3 k_4 k_5 k_6}
   \beta_{k_6}\beta_{k_5}\beta_{k_4}\beta_{k_3}\beta_{k_2}\beta_{k_1} \Big \} \ , \nonumber\\
   \end{align}
where the expressions of the matrix elements of each operator $O^{ij}$ in terms of those of the operators $o^{kk}$ and of the $(U,V)$ matrices can be found in Ref.~\cite{Si15} for terms up to $O^{[4]}$ and in Ref.~\cite{Ripoche:2019nmy} for  $O^{[6]}$. Each term $O^{ij}$ is characterized by its number $i$ ($j$) of quasi-particle creation (annihilation) operators. Because $O$ has been normal-ordered  with respect to $| \Phi \rangle$, all quasi-particle creation operators (if any) are located to the left of all quasi-particle annihilation operators (if any).  The class $O^{[2k]}$ groups all the terms $O^{ij}$ of \emph{effective} $k$-body character, i.e.,\@ with $i+j=2k$. The operator being overall unchanged by the normal-ordering procedure, its rank $\texttt{deg\_max}\equiv \text{Max} \, 2k$ remains itself unchanged. Matrix elements are fully antisymmetric, i.e.,
\begin{equation}
O^{ij}_{k_1 \ldots k_{i} k_{i+1} \ldots k_{i+j}} = (-1)^{\sigma(P)} O^{ij}_{P(k_1 \ldots k_i | k_{i+1} \ldots k_{i+j})}  \, .
\end{equation}
More details and properties can be found in Refs.~\cite{Si15,Duguet:2015yle,Ripoche:2019nmy}.

State-of-the-art many-body calculations are typically performed within the normal-ordered two-body (NO2B) approximation~\cite{Hagen2007,Roth:2011vt,Binder:2012mk,Binder:2013oea}. However, the naive adaptation of the NO2B approximation to many-body formalisms based on a particle-number breaking reference state, which would results in neglecting the residual three-body part $O^{[6]}$, has been shown to be fundamentally inappropriate. As a result, a particle-number conserving normal-ordered two-body (PNO2B) approximation was designed~\cite{Ripoche:2019nmy}. The net effect of the PNO2B approximation is to modify in a specific way the matrix elements at play in $O^{[4]}$ in addition to fully neglecting $O^{[6]}$. In the present work, however, the diagrammatic is anyway worked out in presence of the effective three-body part, i.e.,\@ in presence of six-legs vertices (see below), which significantly increases the number of possible diagrams at a given order and the complexity of their topology. Correspondingly, the code can eventually be run with or without including the effective three-body part of the operators.

\section{Many-body formalism}
\label{definitions}

\subsection{Projective eigenequations}

Taking the Hermitian conjugate of Eq.~\eqref{schroed} and right-multiplying by an arbitrary auxiliary state $| \Theta \rangle$ (such that $\langle \Psi^{\text{A}}_{0}  |  \Theta\rangle\neq 0$), one obtains a projective equation of the form
\begin{align}
\text{O}^{\text{A}}_{0} &= \frac{\langle \Psi^{\text{A}}_{0} | O | \Theta \rangle}{\langle \Psi^{\text{A}}_{0}  |  \Theta\rangle} \, . \label{projective1}
\end{align}
Choosing $| \Theta \rangle \equiv | \Phi \rangle$ and expanding $\langle \Psi^{\text{A}}_{0}  |$ around it leads to straight, i.e.,\@ symmetry-breaking, BMBPT~\cite{arthuis18a}. In the present work, the auxiliary state is taken as $| \Theta \rangle \equiv P^{\text{A}} | \Phi \rangle$ such that the symmetry is explicitly restored by the presence of the projection operator $P^{\text{A}}$ even after expanding and truncating $\langle \Psi^{\text{A}}_{0}  |$ around the Bogoliubov reference state. 

In this context, Eq.~\eqref{projective1} becomes
\begin{align}
\text{O}^{\text{A}}_{0} &= \frac{\langle \Psi^{\text{A}}_{0} | O P^{\text{A}} | \Phi \rangle}{\langle \Psi^{\text{A}}_{0}  |  P^{\text{A}} | \Phi \rangle} \, , \label{projective2}
\end{align}
such that inserting Eq.~\eqref{projAoperator} as well as introducing the so-called off-diagonal norm and operator kernels
\begin{subequations}
\label{kernelsA}
\begin{align}
\mathcal{N}(\varphi)  &\equiv \langle \Psi^{\text{A}}_{0} | \Phi(\varphi) \rangle \, , \label{kernelsA1} \\
\mathcal{O}(\varphi) &\equiv \langle \Psi^{\text{A}}_{0} |  O | \Phi(\varphi) \rangle  \, , \label{kernelsA3}
\end{align}
\end{subequations}
where $| \Phi(\varphi) \rangle \equiv S(\varphi)| \Phi \rangle$ denotes the gauge-rotated reference state, leads to the working form
\begin{align}
\text{O}^{\text{A}}_{0} &= \frac{\int_{0}^{2\pi} \!d\varphi \, e^{-i\text{A}\varphi}  \, {\cal O}(\varphi)}{\int_{0}^{2\pi} \!d\varphi \, e^{-i\text{A}\varphi}  \, {\cal N}(\varphi)}  \, . \label{projeigenequatkernelsA} 
\end{align}

Equation~\eqref{projeigenequatkernelsA} constitutes the master equation on which the PNP-BMBPT formalism is built. In absence of the projection operator, one recovers BMBPT's master equation under the form
\begin{align}
\text{O}^{\text{A}}_{0} &= \mathcal{O}(0) \, , \label{unprojeigenequatkernelsA} 
\end{align}
where intermediate normalization $\mathcal{N}(0)= \langle \Psi^{\text{A}}_{0} | \Phi \rangle = 1$ with the unrotated Bogoliubov reference state has been used. Equations~\eqref{projeigenequatkernelsA} and~\eqref{unprojeigenequatkernelsA} are obviously equivalent in the exact limit but differ as soon as $\langle \Psi^{\text{A}}_{0} |$ is expanded around $\langle \Phi |$ and truncated.

\subsection{Imaginary-time formalism}
\label{timeformalism}

Introducing the evolution operator in imaginary time\footnote{The time is given in units of MeV$^{-1}$.}
\begin{equation}
{\cal U}(\tau) \equiv e^{-\tau \Omega} \, , \label{evoloperator}
\end{equation}
with $\tau$ real, allows one to write the ground state as\footnote{The result is obtained by inserting a complete set of energy eigenstates in both the numerator and the denominator.}\footnote{The chemical potential $\lambda$ is fixed such that $\Omega^{\text{A}_0}_0$ for the targeted particle number $\text{A}_0$ is the lowest value of all $\Omega^{\text{A}}_\mu$ over Fock space, i.e.,\@ it penalizes systems with larger number of particles such that $\Omega^{\text{A}_0}_0 < \Omega^{\text{A}}_\mu$ for all $\text{A}>\text{A}_0$ while maintaining at the same time that $\Omega^{\text{A}_0}_0 < \Omega^{\text{A}}_\mu$ for all $\text{A}<\text{A}_0$. This is practically achievable only if $E^{\text{A}}_0$ is strictly convex in the neighborhood of $\text{A}_0$, which is generally but not always true for atomic nuclei.}
\begin{align}
| \Psi^{\text{A}}_{0} \rangle &= \lim\limits_{\tau \to \infty} | \Psi (\tau) \rangle  \equiv \lim\limits_{\tau \to \infty}  \frac{{\cal U}(\tau) | \Phi \rangle}{\langle \Phi | {\cal U}(\tau) | \Phi \rangle}  \, , \label{evolstate}
\end{align}
where $\langle \Phi | \Psi (\tau) \rangle=1$ for all $\tau$. With this definition at hand, the off-diagonal kernels entering Eq.~\eqref{projeigenequatkernelsA} read as
\begin{subequations}
\label{kernelslimit}
\begin{align}
{\cal N}(\varphi) &\equiv \frac{N(\varphi)}{N(0)} = \lim\limits_{\tau \to \infty} \frac{\langle \Phi |{\cal U}(\tau)  | \Phi(\varphi) \rangle}{\langle \Phi |{\cal U}(\tau) | \Phi \rangle} \, , \\
{\cal O}(\varphi) &\equiv \frac{O(\varphi)}{N(0)} = \lim\limits_{\tau \to \infty}  \frac{\langle \Phi |{\cal U}(\tau) O | \Phi(\varphi) \rangle}{\langle \Phi |{\cal U}(\tau) | \Phi \rangle} \, .
\end{align}
\end{subequations}
The off-diagonal kernels $N(\varphi)$ and $O(\varphi)$ are the many-body quantities to be approximated via a viable expansion method from which $N(0)$ and $O(0)$ can be obtained as a particular case~\cite{Tichai:2018mll,arthuis18a}.

\subsection{Norm kernel}
\label{normvsop}

The off-diagonal norm kernel plays a particular role given that it does not actually involve a non-trivial operator, which makes its perturbative expansion different from the expansion of an operator kernel~\cite{Duguet:2015yle}. In fact, it can be trivially related to the particle-number operator kernel through
\begin{equation}
\frac{d}{d \varphi} \, {\cal N}(\varphi) = i \, {\cal A}(\varphi) \, . \label{NormkernelODE1} 
\end{equation}
Accessing ${\cal N}(\varphi)$ via the integration of Eq.~\eqref{NormkernelODE1} ensures that Eq.~\eqref{schroed} applied to $O\equiv A$ and rewritten as Eq.~\eqref{projeigenequatkernelsA} delivers the expected result $\text{A}^{\text{A}}_{0} = \text{A}$ \emph{even when} $\mathcal{A}(\varphi)$ is computed approximately through, e.g., perturbation theory. Indeed, as long as Eq.~\eqref{NormkernelODE1} is enforced, one has 
\begin{align}
\frac{\int_{0}^{2\pi} \!d\varphi \, e^{-i\text{A}\varphi}  \, {\cal A}(\varphi)}{\int_{0}^{2\pi} \!d\varphi \, e^{-i\text{A}\varphi}  \, {\cal N}(\varphi)} &= -i \frac{\int_{0}^{2\pi} \!d\varphi \, e^{-i\text{A}\varphi}  \, \frac{d }{d\varphi} {\cal N}(\varphi)}{\int_{0}^{2\pi} \!d\varphi \, e^{-i\text{A}\varphi}  \, {\cal N}(\varphi)} \nonumber \\
&= +i \frac{\int_{0}^{2\pi} \!d\varphi \, [\frac{d }{d\varphi} e^{-i\text{A}\varphi}]  \,  {\cal N}(\varphi)}{\int_{0}^{2\pi} \!d\varphi \, e^{-i\text{A}\varphi}  \, {\cal N}(\varphi)}  \nonumber \\
&= \text{A} \, , \nonumber
\end{align}
which is the required result for the symmetry of present interest to be exactly restored. Further introducing the factorization of an arbitrary operator kernel
\begin{align}
\mathcal{O}(\varphi) &\equiv o(\varphi) \, \mathcal{N}(\varphi) \label{factorization}
\end{align}
where $o(\varphi)$ denotes the connected/linked part of the operator kernel~\cite{Duguet:2015yle}, one arrives at the first-order ordinary differential equation (ODE) fulfilled by the norm kernel
\begin{equation}
\frac{d}{d \varphi} \, \mathcal{N}(\varphi)  - i \, a(\varphi) \, \mathcal{N}(\varphi) = 0 \, ,\label{NormkernelODE2} 
\end{equation}
whose closed-form solution reads as
\begin{equation}
\mathcal{N}(\varphi)  = e^{i \int_{0}^{\varphi} \!d\phi \, a(\phi)} \, . \label{solNormkernelODE} 
\end{equation}
From the computation of $a(\varphi)$, the off-diagonal norm kernel is consistently obtained. Eventually, the connected/linked part $o(\varphi)$ of an operator kernel $\mathcal{O}(\varphi)$ is the sole quantity one needs to effectively focus on in order to implement the complete PNP-BMBPT formalism. As a matter of fact, it is not by chance given that, as will be discussed below, $o(\varphi)$ is size-extensive and properly scales with system size, which translates into the fact that it effectively displays a \emph{connected} expansion.

\section{Perturbation theory}
\label{subs:basicsBMBPT}

\subsection{Partitioning}

The grand potential is split into an unperturbed part $\Omega_{0}$ and a residual part $\Omega_1$
\begin{equation}
\label{split1}
\Omega = \Omega_{0} + \Omega_{1} \ .
\end{equation}
For a given number of interacting fermions, the key is to choose $\Omega_0$ with a low-enough symmetry for its ground state $| \Phi \rangle$ to be non-degenerate with respect to elementary excitations. For open-shell superfluid nuclei, this leads to choosing an operator $\Omega_0$ that breaks particle-number conservation, i.e., while $\Omega$ commutes with $U(1)$ transformations, we are interested in the case where $\Omega_0$, and thus $\Omega_1$, do not. More specifically, the partioning is defined through
\begin{subequations}
\label{split2}
\begin{align}
\Omega_{0} &\equiv \Omega^{00}+\bar{\Omega}^{11} \ , \\
\Omega_{1} &\equiv \Omega^{20} + \breve{\Omega}^{11} + \Omega^{02} \nonumber \\
  &\phantom{\equiv } + \Omega^{40} + \Omega^{31} + \Omega^{22} +  \Omega^{13} + \Omega^{04} \nonumber \\
  &\phantom{\equiv } + \Omega^{60}+ \Omega^{51}+ \Omega^{42}+ \Omega^{33}+ \Omega^{24}+ \Omega^{15}+ \Omega^{06} \ ,
 \label{e:perturbation}
\end{align}
\end{subequations}
with $\breve{\Omega}^{11}\equiv\Omega^{11}- \bar{\Omega}^{11}$ and where the normal-ordered one-body part of $\Omega_{0}$ is diagonal, i.e.,
\begin{equation}
\bar{\Omega}^{11} \equiv \sum_{k} E_k \beta^{\dagger}_k \beta_k \, , \label{onebodypiece}
\end{equation}
with $E_k > 0$ for all $k$. 

Introducing many-body states generated via an even number of quasi-particle excitations of the Bogoliubov vacuum
\begin{equation}
| \Phi^{k_1 k_2\ldots} \rangle \equiv \beta^{\dagger}_{k_1} \, \beta^{\dagger}_{k_2} \, \ldots | \Phi \rangle \, , 
\end{equation}
the unperturbed grand potential $\Omega_{0}$ is fully characterized by its complete set of orthonormal eigenstates in Fock space%
\begin{subequations}%
\begin{align}
\Omega_{0}\, |  \Phi \rangle &= \Omega^{00} \, |  \Phi \rangle \, , \\
\Omega_{0}\, |  \Phi^{k_1 k_2\ldots} \rangle &= \left[\Omega^{00} + \epsilon_{k_1 k_2 \ldots}\right] |  \Phi^{k_1 k_2\ldots} \rangle  \label{phi} \, ,
\end{align}
\end{subequations}
where the strict positivity of unperturbed excitations $\epsilon_{k_1 k_2 \ldots} \equiv E_{k_1} + E_{k_2} +\ldots $ characterizes the lifting of the particle-hole degeneracy authorized by the spontaneous breaking of $U(1)$ symmetry in open-shell nuclei at the mean-field level.

In the particular case where $\ket{\Phi}$ solves the HFB variational problem, one has that $\Omega^{20}=\breve{\Omega}^{11}=\Omega^{02}=0$ such that $\Omega_1$ reduces to $\Omega^{[4]}+\Omega^{[6]}$. This choice defines the \emph{canonical} version of (PNP-)BMBPT and reduces significantly the number of non-zero diagrams to be considered. However, we do not make this \textit{a priori} hypothesis such that the reference state $\ket{\Phi}$ and the corresponding unperturbed grand potential $\Omega_{0}$ can be defined more generally, eventually leading to the appearance of \emph{non-canonical} diagrams involving $\Omega^{20}$, $\breve{\Omega}^{11}$ and $\Omega^{02}$ vertices.

On the basis of the above splitting of $\Omega$, one introduces the interaction representation of operators in the quasi-particle basis, e.g.,
\begin{align}
O^{31}(\tau) &\equiv e^{+\tau \Omega_{0}} O^{31} e^{-\tau \Omega_{0}}  \\
&=\frac{1}{3!}\sum_{k_1 k_2 k_3 k_4}  O^{31}_{k_1 k_2 k_3 k_4}
   \beta^{\dagger}_{k_1}(\tau)\beta^{\dagger}_{k_2}(\tau)\beta^{\dagger}_{k_3}(\tau)\beta_{k_4}(\tau) \, ,\nonumber
\end{align}
where
\begin{subequations}
\label{aalphatau}
\begin{align}
\beta_{k} (\tau)  &\equiv e^{+\tau \Omega_{0}} \, \beta_{k} \, e^{-\tau \Omega_{0}}=e^{-\tau E_k} \, \beta_{k} \label{aalphatau2} \ , \\
\beta_{k}^{\dagger}(\tau)  &\equiv e^{+\tau \Omega_{0}} \, \beta_{k}^{\dagger} \, e^{-\tau \Omega_{0}}=e^{+\tau E_{k}} \, \beta_{k}^{\dagger} \ . \label{aalphatau1}
\end{align}
\end{subequations}

\subsection{Perturbative expansion}

Expanding the evolution operator in powers of $\Omega_1$~\cite{blaizot86}
\begin{align}
\mathcal{U}(\tau) &\equiv e^{-\tau \Omega} \nonumber \\
&= e^{-\tau \Omega_{0}} \, \textmd{T}e^{-\int_{0}^{\tau}\!\mathrm{d}\tau \Omega_{1}(\tau) }  \, ,
 \label{evol1}
\end{align}
where $\textmd{T}$ denotes the time-ordering operator\footnote{The time-ordering operator orders a product of operators in decreasing order according to their time labels (i.e., larger times to the left) and multiplies the result with the signature of the permutation used to achieve the corresponding reordering.}, one obtains~\cite{Duguet:2015yle} the expansion of interest\footnote{In agreement with Eq.~\eqref{unprojeigenequatkernelsA}, straight BMBPT is recovered from Eq.~\eqref{observableO1} for $\varphi=0$ given that $\text{O}^{\text{A}}_{0} =   {\cal O}(0)= o(0)$ in this formalism.}
\begin{align}
o(\varphi) &\equiv  \lim\limits_{\tau \to \infty} \frac{\langle \Phi |{\cal U}(\tau) O | \Phi(\varphi) \rangle}{\langle \Phi |{\cal U}(\tau) | \Phi(\varphi) \rangle} \nonumber  \\
&= \lim\limits_{\tau \to \infty}  \langle \Phi | \textmd{T}e^{-\int_{0}^{\tau}dt \Omega_{1}\left(t\right)} O |  \Phi(\varphi) \rangle_{c}   \nonumber \\
&= \langle \Phi | O |  \Phi(\varphi) \rangle \nonumber \\
&\phantom{=} -\frac{1}{1!}\int_{0}^{+\infty}d\tau_1 \langle \Phi | \textmd{T}\left[  \Omega_{1}\left(  \tau_1\right)O(0)\right]|  \Phi(\varphi) \rangle_{c}  \nonumber \\
& \phantom{=} +\frac{1}{2!}\int_{0}^{+\infty} \!d\tau_{1} d\tau_{2} \langle \Phi |\textmd{T}\left[  \Omega_{1}\left(  \tau_{1}\right)  \Omega_{1}\left(  \tau_{2}\right) O(0)\right] |  \Phi(\varphi) \rangle_{c} \nonumber \\
& \phantom{=} -...  \ , \label{observableO1}
\end{align}
where the lower index $c$ refers to the restriction to connected terms, i.e., contributions arising from the application of Wick's theorem in which the associated string of contractions necessarily involves all the operators at play in the many-body matrix element under consideration. The time-independent operator $O$ could be inserted at no cost within the time-ordering by providing it with a fictitious and harmless time dependence $t=0$. Indeed, all $\Omega_{1}\left(  \tau_k\right)$ operators appear to the left of $O$ and occur at a larger time given that their corresponding time variables are positive.

Invoking perturbation theory consists of truncating the Taylor expansion of the time-evolution operator in Eq.~\eqref{observableO1}. Gathering all terms up to order $p$, $o(\varphi)$ sums matrix elements of products of up to $p+1$ time-dependent operators\footnote{The expansion starts at order $p=0$ that corresponds to the term containing no $\Omega_{1}$ operator and no time integral in Eq.~\eqref{observableO1}.}. The running time variables are integrated over from $0$ to $\tau \rightarrow +\infty$ whereas the time label attributed to the operator $O$ itself remains fixed at $t=0$, i.e.,\@ contributions of order $p$ contain a $p$-tuple time integral that needs to be performed to generate the end result under the required form. 

Given the off-diagonal character of the kernels, each matrix element in Eq.~\eqref{observableO1} is computed via the application of \emph{off-diagonal} Wick's theorem~\cite{balian69a}, which is applicable to matrix elements of operators between any two (non-orthogonal) left and right product states. As a result, diagrams at play invoke a set of four off-diagonal unperturbed propagators defined in the quasi-particle basis $\{\beta_{k};\beta^{\dagger}_{k}\}$ as
\begin{subequations}
\label{propagatorsA}
\begin{align}
G^{+- (0)}_{k_1k_2}(\tau_1, \tau_2 ; \varphi) &\equiv \frac{\langle \Phi |  \textmd{T}[\beta^{\dagger}_{k_1}(\tau_1) \beta_{k_2}(\tau_2)] | \Phi(\varphi) \rangle}{\langle \Phi | \Phi(\varphi) \rangle}  \, , \label{propagatorsA1} \\
G^{-- (0)}_{k_1k_2}(\tau_1, \tau_2 ; \varphi) &\equiv \frac{\langle \Phi |  \textmd{T}[\beta_{k_1}(\tau_1) \beta_{k_2}(\tau_2)] | \Phi(\varphi) \rangle}{\langle \Phi | \Phi(\varphi) \rangle}  \, , \label{propagatorsA2} \\
G^{++ (0)}_{k_1k_2}(\tau_1, \tau_2 ; \varphi) &\equiv \frac{\langle \Phi |  \textmd{T}[\beta^{\dagger}_{k_1}(\tau_1) \beta^{\dagger}_{k_2}(\tau_2)] | \Phi(\varphi) \rangle}{\langle \Phi | \Phi(\varphi) \rangle}  \, , \label{propagatorsA3} \\
G^{-+ (0)}_{k_1k_2}(\tau_1, \tau_2 ; \varphi) &\equiv \frac{\langle \Phi |  \textmd{T}[\beta_{k_1}(\tau_1) \beta^{\dagger}_{k_2}(\tau_2)] | \Phi(\varphi) \rangle}{\langle \Phi | \Phi(\varphi) \rangle}  \, . \label{propagatorsA4}
\end{align}
\end{subequations}
By virtue of the off-diagonal elementary contractions worked out in \ref{sectioncontractions}, the four off-diagonal propagators are equal to
\begin{subequations}
\label{propagatorsB}
\begin{align}
G^{+- (0)}_{k_1k_2}(\tau_1, \tau_2 ; \varphi) &= - e^{-(\tau_2-\tau_1)E_{k_1}} \, \theta(\tau_2-\tau_1) \, \delta_{k_1k_2} \, , \label{propagatorsB1} \\
G^{-- (0)}_{k_1k_2}(\tau_1, \tau_2 ; \varphi) &= + e^{-\tau_1 E_{k_1}} \, e^{-\tau_2 E_{k_2}} \, R^{--}_{k_1k_2}(\varphi) \, , \label{propagatorsB2} \\
G^{++ (0)}_{k_1k_2}(\tau_1, \tau_2 ; \varphi) &= 0  \, , \label{propagatorsB3} \\
G^{-+ (0)}_{k_1k_2}(\tau_1, \tau_2 ; \varphi) &= + e^{-(\tau_1-\tau_2)E_{k_1}} \, \theta(\tau_1-\tau_2) \, \delta_{k_1k_2}   \label{propagatorsB4} \, , 
\end{align}
\end{subequations}
where both normal propagators are actually related via antisymmetry under the exchange of time and quasi-particle labels. The higher generality and complexity of the off-diagonal BMBPT diagrammatics of present interest compared to the straight BMBPT diagrammatics discussed in Ref.~\cite{arthuis18a} is due to the presence of the anomalous propagator $G^{-- (0)}(\varphi)$ that carries the full gauge-angle dependence. In particular, the possibility to form anomalous propagators significantly increases the combinatorics~\cite{Duguet:2015yle}. Eventually, the two diagrammatics coincide in the limit $\varphi=0$ given that $G^{-- (0)}(0)=0$. All in all, the present extension of the \textbf{\texttt{ADG}} code amounts to dealing with this higher generality and complexity, which itself originates from the presence of different left and right vacua in the off-diagonal kernel $o(\varphi)$ (see Eq.~\eqref{observableO1}).

Equal-time propagators can solely arise from contracting two quasi-particle operators belonging to the same normal-ordered operator displaying creation operators to the left of annihilation ones. As a result, one finds that~\cite{Duguet:2015yle}
\begin{subequations}
\label{propagatorsC}
\begin{align}
G^{+- (0)}_{k_1k_2}(\tau, \tau ; \varphi) &\equiv 0 \, , \label{propagatorsC1} \\
G^{-- (0)}_{k_1k_2}(\tau, \tau ; \varphi) &\equiv + e^{-\tau (E_{k_1}+E_{k_2})} \, R^{--}_{k_1k_2}(\varphi) \, , \label{propagatorsC2} \\
G^{++ (0)}_{k_1k_2}(\tau, \tau ; \varphi) &\equiv 0  \, , \label{propagatorsC3} \\
G^{-+ (0)}_{k_1k_2}(\tau, \tau ; \varphi) &\equiv 0  \label{propagatorsC4} \, ,
\end{align}
\end{subequations}
such that the sole non-zero equal-time contraction, and thus the sole contraction of an interaction vertex onto itself, is of anomalous character. Correspondingly, no contraction of an interaction vertex onto itself can occur in the diagonal case, i.e.,\@ for $\varphi=0$.

\begin{figure}[t!]
$O^{[0]} =$
	\parbox{40pt}{\begin{fmffile}{O00}
	\begin{fmfgraph*}(40,40)
	\fmftop{t1} \fmfbottom{b1}
	\fmf{phantom}{b1,i1}
	\fmf{phantom}{i1,t1}
	\fmfv{label=$O^{00}$,label.angle=-90,d.shape=square,d.filled=full,d.size=2thick}{i1}
	\end{fmfgraph*}
	\end{fmffile}}

\vspace{\baselineskip}

$O^{[2]} =$
	\parbox{40pt}{\begin{fmffile}{O11}
	\begin{fmfgraph*}(40,40)
	\fmfcmd{style_def half_prop expr p =
    draw_plain p;
    shrink(.7);
        cfill (marrow (p, .5))
    endshrink;
	enddef;}
	\fmftop{t1} \fmfbottom{b1}
	\fmf{half_prop}{b1,i1}
	\fmf{half_prop}{i1,t1}
	\fmfv{d.shape=square,d.filled=full,d.size=2thick}{i1}
	\fmflabel{$O^{11}$}{b1}
	\end{fmfgraph*}
	\end{fmffile}}
+
	\parbox{40pt}{\begin{fmffile}{O20}
	\begin{fmfgraph*}(40,40)
	\fmfcmd{style_def half_prop expr p =
    draw_plain p;
    shrink(.7);
        cfill (marrow (p, .5))
    endshrink;
	enddef;}
	\fmftop{t1,t2} \fmfbottom{b1}
	\fmfstraight
	\fmf{half_prop}{i1,t2}
	\fmf{half_prop}{i1,t1}
	\fmf{phantom,tension=2}{b1,i1}
	\fmfv{d.shape=square,d.filled=full,d.size=2thick}{i1}
	\fmflabel{$O^{20}$}{b1}
	\end{fmfgraph*}
	\end{fmffile}}
+
	\parbox{40pt}{\begin{fmffile}{O02}
	\begin{fmfgraph*}(40,40)
	\fmfcmd{style_def half_prop expr p =
    draw_plain p;
    shrink(.7);
        cfill (marrow (p, .5))
    endshrink;
	enddef;}
	\fmftop{t1} \fmfbottom{b1,b2,b3}
	\fmfstraight
	\fmf{half_prop}{b1,i1}
	\fmf{half_prop}{b3,i1}
	\fmf{phantom,tension=2}{i1,t1}
	\fmfv{d.shape=square,d.filled=full,d.size=2thick}{i1}
	\fmflabel{$O^{02}$}{b2}
	\end{fmfgraph*}
	\end{fmffile}}

\vspace{2\baselineskip}

$O^{[4]} =$
	\parbox{40pt}{\begin{fmffile}{O22}
	\begin{fmfgraph*}(40,40)
	\fmfstraight
	\fmfcmd{style_def half_prop expr p =
    draw_plain p;
    shrink(.7);
        cfill (marrow (p, .5))
    endshrink;
	enddef;}
	\fmftop{t1,t2} \fmfbottom{b1,b2,b3}
	\fmf{half_prop}{b1,i1}
	\fmf{half_prop}{b3,i1}
	\fmf{half_prop}{i1,t1}
	\fmf{half_prop}{i1,t2}
	\fmfv{d.shape=square,d.filled=full,d.size=2thick}{i1}
	\fmflabel{$O^{22}$}{b2}
	\end{fmfgraph*}
	\end{fmffile}}
+
	\parbox{40pt}{\begin{fmffile}{O31}
	\begin{fmfgraph*}(40,40)
	\fmfstraight
	\fmfcmd{style_def half_prop expr p =
    draw_plain p;
    shrink(.7);
        cfill (marrow (p, .5))
    endshrink;
	enddef;}
	\fmftop{t1,t2,t3} \fmfbottom{b1}
	\fmf{half_prop}{i1,t2}
	\fmf{half_prop}{i1,t1}
	\fmf{half_prop}{i1,t3}
	\fmf{half_prop,tension=3}{b1,i1}
	\fmfv{d.shape=square,d.filled=full,d.size=2thick}{i1}
	\fmflabel{$O^{31}$}{b1}
	\end{fmfgraph*}
	\end{fmffile}}
+
	\parbox{40pt}{\begin{fmffile}{O13}
	\begin{fmfgraph*}(40,40)
	\fmfstraight
	\fmfcmd{style_def half_prop expr p =
    draw_plain p;
    shrink(.7);
        cfill (marrow (p, .5))
    endshrink;
	enddef;}
	\fmftop{t1} \fmfbottom{b1,b2,b3}
	\fmf{half_prop}{b1,i1}
	\fmf{half_prop}{b2,i1}
	\fmf{half_prop}{b3,i1}
	\fmf{half_prop,tension=3}{i1,t1}
	\fmfv{d.shape=square,d.filled=full,d.size=2thick}{i1}
	\fmflabel{$O^{13}$}{b2}
	\end{fmfgraph*}
	\end{fmffile}}

\vspace{2\baselineskip}	

$\phantom{O^{[4]} =}$
+
	\parbox{40pt}{\begin{fmffile}{O40}
	\begin{fmfgraph*}(40,40)
	\fmfstraight
	\fmfcmd{style_def half_prop expr p =
    draw_plain p;
    shrink(.7);
        cfill (marrow (p, .5))
    endshrink;
	enddef;}
	\fmftop{t1,t2,t3,t4} \fmfbottom{b1}
	\fmf{half_prop}{i1,t1}
	\fmf{half_prop}{i1,t2}
	\fmf{half_prop}{i1,t3}
	\fmf{half_prop}{i1,t4}
	\fmf{phantom,tension=3}{i1,b1}
	\fmfv{d.shape=square,d.filled=full,d.size=2thick}{i1}
	\fmflabel{$O^{40}$}{b1}
	\end{fmfgraph*}
	\end{fmffile}}
+
	\parbox{40pt}{\begin{fmffile}{O04}
	\begin{fmfgraph*}(40,40)
	\fmfstraight
	\fmfcmd{style_def half_prop expr p =
    draw_plain p;
    shrink(.7);
        cfill (marrow (p, .5))
    endshrink;
	enddef;}
	\fmftop{t1} \fmfbottom{b1,b2,b3,b4}
	\fmf{half_prop}{b1,i1}
	\fmf{half_prop}{b2,i1}
	\fmf{half_prop}{b3,i1}
	\fmf{half_prop}{b4,i1}
	\fmf{phantom,tension=3}{i1,t1}
	\fmfv{d.shape=square,d.filled=full,d.size=2thick}{i1}
	\fmffreeze
	\fmf{phantom,label=$O^{04}$}{b2,b3}
	\end{fmfgraph*}
	\end{fmffile}}

\vspace{2\baselineskip}

$O^{[6]} =$
	\parbox{40pt}{\begin{fmffile}{O33}
	\begin{fmfgraph*}(40,40)
	\fmfstraight
	\fmfcmd{style_def half_prop expr p =
    draw_plain p;
    shrink(.7);
        cfill (marrow (p, .5))
    endshrink;
	enddef;}
	\fmftop{t1,t2,t3} \fmfbottom{b1,b2,b3}
	\fmf{half_prop}{b1,i1}
	\fmf{half_prop}{b2,i1}
	\fmf{half_prop}{b3,i1}
	\fmf{half_prop}{i1,t1}
	\fmf{half_prop}{i1,t2}
	\fmf{half_prop}{i1,t3}
	\fmfv{d.shape=square,d.filled=full,d.size=2thick}{i1}
	\fmflabel{$O^{33}$}{b2}
	\end{fmfgraph*}
	\end{fmffile}}
+
	\parbox{40pt}{\begin{fmffile}{O42}
	\begin{fmfgraph*}(40,40)
	\fmfstraight
	\fmfcmd{style_def half_prop expr p =
    draw_plain p;
    shrink(.7);
        cfill (marrow (p, .5))
    endshrink;
	enddef;}
	\fmftop{t1,t2,t3,t4} \fmfbottom{b1,b2,b3}
	\fmf{half_prop}{i1,t2}
	\fmf{half_prop}{i1,t1}
	\fmf{half_prop}{i1,t3}
	\fmf{half_prop}{i1,t4}
	\fmf{half_prop,tension=2}{b3,i1}
	\fmf{half_prop,tension=2}{b1,i1}
	\fmfv{d.shape=square,d.filled=full,d.size=2thick}{i1}
	\fmflabel{$O^{42}$}{b2}
	\end{fmfgraph*}
	\end{fmffile}}
+
	\parbox{40pt}{\begin{fmffile}{O24}
	\begin{fmfgraph*}(40,40)
	\fmfstraight
	\fmfcmd{style_def half_prop expr p =
    draw_plain p;
    shrink(.7);
        cfill (marrow (p, .5))
    endshrink;
	enddef;}
	\fmftop{t1,t2} \fmfbottom{b1,b2,b3,b4}
	\fmf{half_prop}{b1,i1}
	\fmf{half_prop}{b2,i1}
	\fmf{half_prop}{b3,i1}
	\fmf{half_prop}{b4,i1}
	\fmf{half_prop,tension=2}{i1,t2}
	\fmf{half_prop,tension=2}{i1,t1}
	\fmfv{d.shape=square,d.filled=full,d.size=2thick}{i1}
	\fmffreeze
	\fmf{phantom,label=$O^{24}$}{b2,b3}
	\end{fmfgraph*}
	\end{fmffile}}
+
	\parbox{40pt}{\begin{fmffile}{O51}
	\begin{fmfgraph*}(40,40)
	\fmfstraight
	\fmfcmd{style_def half_prop expr p =
    draw_plain p;
    shrink(.7);
        cfill (marrow (p, .5))
    endshrink;
	enddef;}
	\fmftop{t1,t2,t3,t4,t5} \fmfbottom{b1}
	\fmf{half_prop}{i1,t2}
	\fmf{half_prop}{i1,t1}
	\fmf{half_prop}{i1,t3}
	\fmf{half_prop}{i1,t4}
	\fmf{half_prop}{i1,t5}
	\fmf{half_prop,tension=5}{b1,i1}
	\fmfv{d.shape=square,d.filled=full,d.size=2thick}{i1}
	\fmflabel{$O^{51}$}{b1}
	\end{fmfgraph*}
	\end{fmffile}}
	
\vspace{2\baselineskip}	

$\phantom{O^{[6]} =}$
+
	\parbox{40pt}{\begin{fmffile}{O15}
	\begin{fmfgraph*}(40,40)
	\fmfstraight
	\fmfcmd{style_def half_prop expr p =
    draw_plain p;
    shrink(.7);
        cfill (marrow (p, .5))
    endshrink;
	enddef;}
	\fmftop{t1} \fmfbottom{b1,b2,b3,b4,b5}
	\fmf{half_prop}{b1,i1}
	\fmf{half_prop}{b2,i1}
	\fmf{half_prop}{b3,i1}
	\fmf{half_prop}{b4,i1}
	\fmf{half_prop}{b5,i1}
	\fmf{half_prop,tension=5}{i1,t1}
	\fmfv{d.shape=square,d.filled=full,d.size=2thick}{i1}
	\fmflabel{$O^{15}$}{b3}
	\end{fmfgraph*}
	\end{fmffile}}
+
	\parbox{40pt}{\begin{fmffile}{O60}
	\begin{fmfgraph*}(40,40)
	\fmfstraight
	\fmfcmd{style_def half_prop expr p =
    draw_plain p;
    shrink(.7);
        cfill (marrow (p, .5))
    endshrink;
	enddef;}
	\fmftop{t1,t2,t3,t4,t5,t6} \fmfbottom{b1}
	\fmf{half_prop}{i1,t1}
	\fmf{half_prop}{i1,t2}
	\fmf{half_prop}{i1,t3}
	\fmf{half_prop}{i1,t4}
	\fmf{half_prop}{i1,t5}
	\fmf{half_prop}{i1,t6}
	\fmf{phantom,tension=6}{i1,b1}
	\fmfv{d.shape=square,d.filled=full,d.size=2thick}{i1}
	\fmflabel{$O^{60}$}{b1}
	\end{fmfgraph*}
	\end{fmffile}}
+
	\parbox{40pt}{\begin{fmffile}{O06}
	\begin{fmfgraph*}(40,40)
	\fmfstraight
	\fmfcmd{style_def half_prop expr p =
    draw_plain p;
    shrink(.7);
        cfill (marrow (p, .5))
    endshrink;
	enddef;}
	\fmftop{t1} \fmfbottom{b1,b2,b3,b4,b5,b6}
	\fmf{half_prop}{b1,i1}
	\fmf{half_prop}{b2,i1}
	\fmf{half_prop}{b3,i1}
	\fmf{half_prop}{b4,i1}
	\fmf{half_prop}{b5,i1}
	\fmf{half_prop}{b6,i1}
	\fmf{phantom,tension=6}{i1,t1}
	\fmfv{d.shape=square,d.filled=full,d.size=2thick}{i1}
	\fmffreeze
	\fmf{phantom,label=$O^{06}$}{b3,b4}
	\end{fmfgraph*}
	\end{fmffile}}

\vspace{\baselineskip}

\caption{
\label{f:verticesO}
Canonical diagrammatic representation of normal-ordered contributions to the operator $O$ in the Schr\"odinger representation.}
\end{figure}
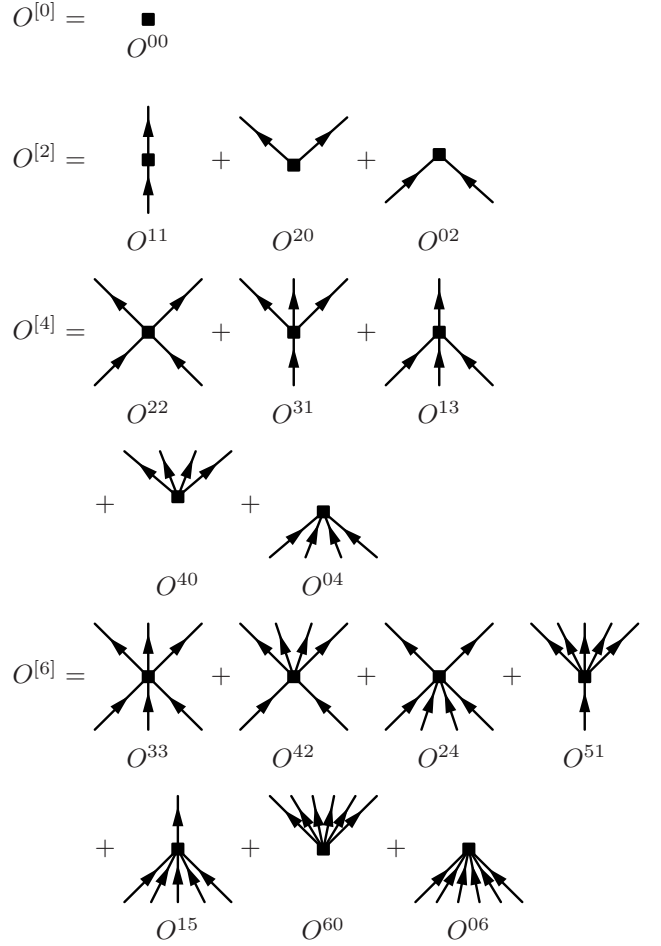

\subsection{Diagrammatic representation}
\label{subs:prod_bmbpt_diags}

The pedestrian application of the off-diagonal Wick's theorem becomes quickly cumbersome as the order $p$ increases. Furthermore, it leads to computing independently many contributions that are in fact identical. By identifying the corresponding pattern, one can design a diagrammatic representation of the various contributions and evaluate their algebraic expressions such that a single diagram captures all identical contributions  at once. In order to achieve this goal, one must first introduce the diagrammatic representation of the building blocks. 

The operator $O$ expressed in the quasi-particle basis is displayed in the Schrödinger representation in Fig.~\ref{f:verticesO} as a sum of Hugenholtz vertices denoting its various normal-ordered contributions $O^{ij}$. The antisymmetrized matrix element $O^{ij}_{k_1 \ldots k_i k_{i+1} \ldots k_{i+j}}$ must be assigned to the corresponding square vertex, where $i$ ($j$) denotes the number of lines traveling out of (into) the vertex and representing quasi-particle creation (annihilation) operators. The operator $O(\tau)$ in the interaction representation possesses the same diagrammatic except that a time $\tau$ is attributed to each of the vertices, i.e., to each of the lines coming in or out of them.

\begin{figure}[t!]
\begin{center}
\parbox{40pt}{\begin{fmffile}{OrientRule}
	\begin{fmfgraph*}(40,40)
	\fmfstraight
	\fmfcmd{style_def half_prop expr p =
    draw_plain p;
    shrink(.7);
        cfill (marrow (p, .5))
    endshrink;
	enddef;}
	\fmftop{t1,t2} \fmfbottom{b1,b2}
	\fmfv{label=$+O^{22}_{k_1 k_2 k_3 k_4}$,label.angle=20,label.dist=15pt}{i1}
	\fmf{half_prop}{b1,i1}
	\fmf{half_prop}{b2,i1}
	\fmf{half_prop}{i1,t1}
	\fmf{half_prop}{i1,t2}
	\fmfv{d.shape=square,d.filled=full,d.size=2thick}{i1}
	\fmfv{l=$k_1$,l.d=0.05w}{t1}
	\fmfv{l=$k_2$,l.d=0.05w}{t2}
	\fmfv{l=$k_3$,l.d=0.03w}{b1}
	\fmfv{l=$k_4$,l.d=0.05w}{b2}
	\fmffreeze
	\fmfleft{l1}
	\fmfright{r1}
	\fmf{plain,fore=red}{l1,i1}
	\fmf{dashes,fore=red}{i1,r1}
	\end{fmfgraph*}
	\end{fmffile}}
\hspace{40pt} = \hspace{5pt}
	\parbox{40pt}{\begin{fmffile}{OrientRule_1}
	\begin{fmfgraph*}(40,40)
	\fmfcmd{style_def half_prop expr p =
    draw_plain p;
    shrink(.7);
        cfill (marrow (p, .5))
    endshrink;
	enddef;}
	\fmftop{t1} \fmfbottom{b1,b2,b3,b4}
	\fmfv{label=$+O^{22}_{k_1 k_2 k_3 k_4}$,label.angle=90}{i1}
	\fmf{half_prop}{b1,i1}
	\fmf{half_prop}{b2,i1}
	\fmf{half_prop}{i1,b3}
	\fmf{half_prop}{i1,b4}
	\fmf{phantom,tension=3}{i1,t1}
	\fmfv{d.shape=square,d.filled=full,d.size=2thick}{i1}
	\fmfv{l=$k_3$,l.d=0.03w}{b1}
	\fmfv{l=$k_4$,l.d=0.05w}{b2}
	\fmfv{l=$k_2$,l.d=0.05w}{b3}
	\fmfv{l=$k_1$,l.d=0.05w}{b4}
	\end{fmfgraph*}
	\end{fmffile}}
\hspace{5pt} = \hspace{10pt}
	\parbox{40pt}{\begin{fmffile}{OrientRule_2}
	\begin{fmfgraph*}(40,40)
	\fmfcmd{style_def half_prop expr p =
    draw_plain p;
    shrink(.7);
        cfill (marrow (p, .5))
    endshrink;
	enddef;}
	\fmftop{t1} \fmfbottom{b1,b2,b3,b4}
	\fmfv{label=$-O^{22}_{k_1 k_2 k_3 k_4}$,label.angle=90}{i1}
	\fmf{half_prop}{b1,i1}
	\fmf{half_prop}{i1,b2}
	\fmf{half_prop}{b3,i1}
	\fmf{half_prop}{i1,b4}
	\fmf{phantom,tension=3}{i1,t1}
	\fmfv{d.shape=square,d.filled=full,d.size=2thick}{i1}
	\fmfv{l=$k_3$,l.d=0.03w}{b1}
	\fmfv{l=$k_2$,l.d=0.05w}{b2}
	\fmfv{l=$k_4$,l.d=0.05w}{b3}
	\fmfv{l=$k_1$,l.d=0.05w}{b4}
	\end{fmfgraph*}
	\end{fmffile}}
\end{center}
\caption{
\label{variousvertices3}
Rules to apply when departing from the canonical diagrammatic representation of a normal-ordered operator. Oriented lines can be rotated through the dashed line but not through the full line.}
\end{figure}
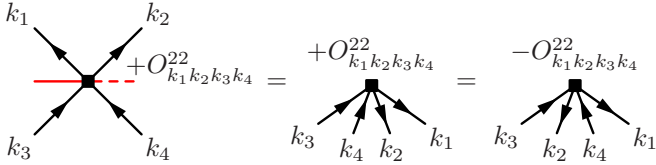

In the canonical representation used in Fig.~\ref{f:verticesO}, all oriented lines go up, i.e., lines representing quasi-particle creation (annihilation) operators appear above (below) the vertex. Accordingly, indices $k_1 \ldots k_i$ must be assigned consecutively from the leftmost to the rightmost line above the vertex, while $k_{i+1} \ldots k_{i+j}$ must be similarly assigned consecutively to lines below the vertex. In the diagrammatic representation of the observable  $\text{O}^{\text{A}}_0$, it is however possible for a line to propagate downwards. This can be obtained unambiguously starting from the canonical representation of Fig.~\ref{f:verticesO}  at the price of adding a specific rule. As illustrated in Fig.~\ref{variousvertices3} for the diagram representing $O^{22}$, lines must only be rotated through the right of the diagram, i.e., going through the dashed line, while it is forbidden to rotate them through the full line. Additionally, a minus sign must be added to the amplitude $O^{ij}_{k_1 \ldots k_i k_{i+1} \ldots k_{i+j}}$ associated with the canonical diagram each time two lines cross as illustrated in Fig.~\ref{variousvertices3}.

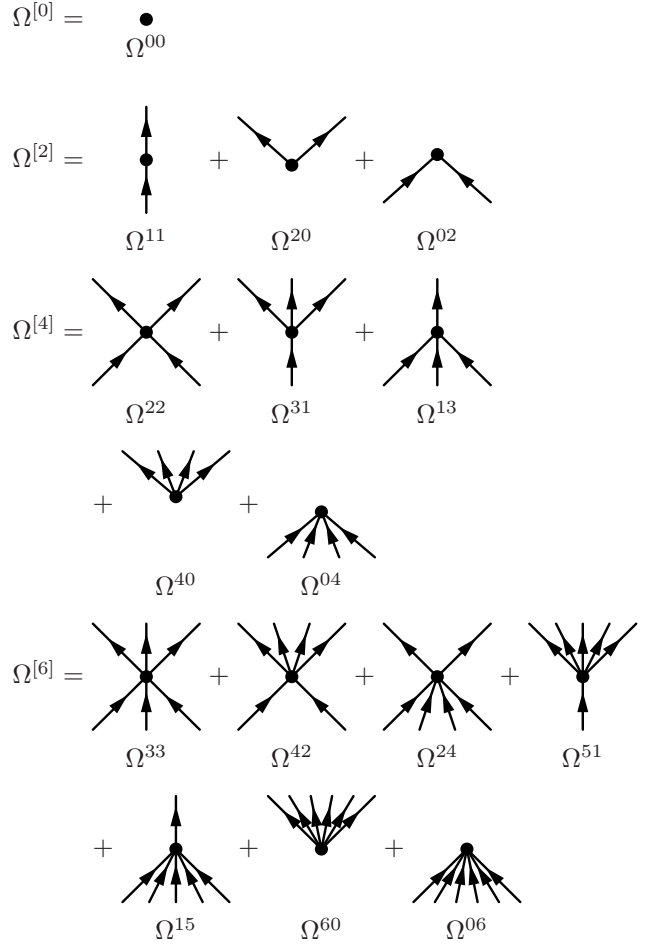
\begin{figure}[t!]
$\Omega^{[0]} =$
	\parbox{40pt}{\begin{fmffile}{Omega00}
	\begin{fmfgraph*}(40,40)
	\fmftop{t1} \fmfbottom{b1}
	\fmf{phantom}{b1,i1}
	\fmf{phantom}{i1,t1}
	\fmfv{label=$\Omega^{00}$,label.angle=-90,d.shape=circle,d.filled=full,d.size=2thick}{i1}
	\end{fmfgraph*}
	\end{fmffile}}

\vspace{\baselineskip}

$\Omega^{[2]} =$
	\parbox{40pt}{\begin{fmffile}{Omega11}
	\begin{fmfgraph*}(40,40)
	\fmfcmd{style_def half_prop expr p =
    draw_plain p;
    shrink(.7);
        cfill (marrow (p, .5))
    endshrink;
	enddef;}
	\fmftop{t1} \fmfbottom{b1}
	\fmf{half_prop}{b1,i1}
	\fmf{half_prop}{i1,t1}
	\fmfv{d.shape=circle,d.filled=full,d.size=2thick}{i1}
	\fmflabel{$\Omega^{11}$}{b1}
	\end{fmfgraph*}
	\end{fmffile}}
+
	\parbox{40pt}{\begin{fmffile}{Omega20}
	\begin{fmfgraph*}(40,40)
	\fmfcmd{style_def half_prop expr p =
    draw_plain p;
    shrink(.7);
        cfill (marrow (p, .5))
    endshrink;
	enddef;}
	\fmftop{t1,t2} \fmfbottom{b1}
	\fmfstraight
	\fmf{half_prop}{i1,t2}
	\fmf{half_prop}{i1,t1}
	\fmf{phantom,tension=2}{b1,i1}
	\fmfv{d.shape=circle,d.filled=full,d.size=2thick}{i1}
	\fmflabel{$\Omega^{20}$}{b1}
	\end{fmfgraph*}
	\end{fmffile}}
+
	\parbox{40pt}{\begin{fmffile}{Omega02}
	\begin{fmfgraph*}(40,40)
	\fmfcmd{style_def half_prop expr p =
    draw_plain p;
    shrink(.7);
        cfill (marrow (p, .5))
    endshrink;
	enddef;}
	\fmftop{t1} \fmfbottom{b1,b2,b3}
	\fmfstraight
	\fmf{half_prop}{b1,i1}
	\fmf{half_prop}{b3,i1}
	\fmf{phantom,tension=2}{i1,t1}
	\fmfv{d.shape=circle,d.filled=full,d.size=2thick}{i1}
	\fmflabel{$\Omega^{02}$}{b2}
	\end{fmfgraph*}
	\end{fmffile}}

\vspace{2\baselineskip}

$\Omega^{[4]} =$
	\parbox{40pt}{\begin{fmffile}{Omega22}
	\begin{fmfgraph*}(40,40)
	\fmfcmd{style_def half_prop expr p =
    draw_plain p;
    shrink(.7);
        cfill (marrow (p, .5))
    endshrink;
	enddef;}
	\fmfstraight
	\fmftop{t1,t2} \fmfbottom{b1,b2,b3}
	\fmf{half_prop}{b1,i1}
	\fmf{half_prop}{b3,i1}
	\fmf{half_prop}{i1,t1}
	\fmf{half_prop}{i1,t2}
	\fmfv{d.shape=circle,d.filled=full,d.size=2thick}{i1}
	\fmflabel{$\Omega^{22}$}{b2}
	\end{fmfgraph*}
	\end{fmffile}}
+
	\parbox{40pt}{\begin{fmffile}{Omega31}
	\begin{fmfgraph*}(40,40)
	\fmfcmd{style_def half_prop expr p =
    draw_plain p;
    shrink(.7);
        cfill (marrow (p, .5))
    endshrink;
	enddef;}
	\fmfstraight
	\fmftop{t1,t2,t3} \fmfbottom{b1}
	\fmf{half_prop}{i1,t2}
	\fmf{half_prop}{i1,t1}
	\fmf{half_prop}{i1,t3}
	\fmf{half_prop,tension=3}{b1,i1}
	\fmfv{d.shape=circle,d.filled=full,d.size=2thick}{i1}
	\fmflabel{$\Omega^{31}$}{b1}
	\end{fmfgraph*}
	\end{fmffile}}
+
	\parbox{40pt}{\begin{fmffile}{Omega13}
	\begin{fmfgraph*}(40,40)
	\fmfcmd{style_def half_prop expr p =
    draw_plain p;
    shrink(.7);
        cfill (marrow (p, .5))
    endshrink;
	enddef;}
	\fmfstraight
	\fmftop{t1} \fmfbottom{b1,b2,b3}
	\fmf{half_prop}{b1,i1}
	\fmf{half_prop}{b2,i1}
	\fmf{half_prop}{b3,i1}
	\fmf{half_prop,tension=3}{i1,t1}
	\fmfv{d.shape=circle,d.filled=full,d.size=2thick}{i1}
	\fmflabel{$\Omega^{13}$}{b2}
	\end{fmfgraph*}
	\end{fmffile}}
	
\vspace{2\baselineskip}	

$\phantom{\Omega^{[4]} =}$
+
	\parbox{40pt}{\begin{fmffile}{Omega40}
	\begin{fmfgraph*}(40,40)
	\fmfcmd{style_def half_prop expr p =
    draw_plain p;
    shrink(.7);
        cfill (marrow (p, .5))
    endshrink;
	enddef;}
	\fmfstraight
	\fmftop{t1,t2,t3,t4} \fmfbottom{b1}
	\fmf{half_prop}{i1,t1}
	\fmf{half_prop}{i1,t2}
	\fmf{half_prop}{i1,t3}
	\fmf{half_prop}{i1,t4}
	\fmf{phantom,tension=3}{i1,b1}
	\fmfv{d.shape=circle,d.filled=full,d.size=2thick}{i1}
	\fmflabel{$\Omega^{40}$}{b1}
	\end{fmfgraph*}
	\end{fmffile}}
+
	\parbox{40pt}{\begin{fmffile}{Omega04}
	\begin{fmfgraph*}(40,40)
	\fmfcmd{style_def half_prop expr p =
    draw_plain p;
    shrink(.7);
        cfill (marrow (p, .5))
    endshrink;
	enddef;}
	\fmfstraight
	\fmftop{t1} \fmfbottom{b1,b2,b3,b4}
	\fmf{half_prop}{b1,i1}
	\fmf{half_prop}{b2,i1}
	\fmf{half_prop}{b3,i1}
	\fmf{half_prop}{b4,i1}
	\fmf{phantom,tension=3}{i1,t1}
	\fmfv{d.shape=circle,d.filled=full,d.size=2thick}{i1}
	\fmffreeze
	\fmf{phantom,label=$\Omega^{04}$}{b2,b3}
	\end{fmfgraph*}
	\end{fmffile}}

\vspace{2\baselineskip}

$\Omega^{[6]} =$
	\parbox{40pt}{\begin{fmffile}{Omega33}
	\begin{fmfgraph*}(40,40)
	\fmfcmd{style_def half_prop expr p =
    draw_plain p;
    shrink(.7);
        cfill (marrow (p, .5))
    endshrink;
	enddef;}
	\fmfstraight
	\fmftop{t1,t2,t3} \fmfbottom{b1,b2,b3}
	\fmf{half_prop}{b1,i1}
	\fmf{half_prop}{b2,i1}
	\fmf{half_prop}{b3,i1}
	\fmf{half_prop}{i1,t1}
	\fmf{half_prop}{i1,t2}
	\fmf{half_prop}{i1,t3}
	\fmfv{d.shape=circle,d.filled=full,d.size=2thick}{i1}
	\fmflabel{$\Omega^{33}$}{b2}
	\end{fmfgraph*}
	\end{fmffile}}
+
	\parbox{40pt}{\begin{fmffile}{Omega42}
	\begin{fmfgraph*}(40,40)
	\fmfcmd{style_def half_prop expr p =
    draw_plain p;
    shrink(.7);
        cfill (marrow (p, .5))
    endshrink;
	enddef;}
	\fmfstraight
	\fmftop{t1,t2,t3,t4} \fmfbottom{b1,b2,b3}
	\fmf{half_prop}{i1,t2}
	\fmf{half_prop}{i1,t1}
	\fmf{half_prop}{i1,t3}
	\fmf{half_prop}{i1,t4}
	\fmf{half_prop,tension=2}{b3,i1}
	\fmf{half_prop,tension=2}{b1,i1}
	\fmfv{d.shape=circle,d.filled=full,d.size=2thick}{i1}
	\fmflabel{$\Omega^{42}$}{b2}
	\end{fmfgraph*}
	\end{fmffile}}
+
	\parbox{40pt}{\begin{fmffile}{Omega24}
	\begin{fmfgraph*}(40,40)
	\fmfcmd{style_def half_prop expr p =
    draw_plain p;
    shrink(.7);
        cfill (marrow (p, .5))
    endshrink;
	enddef;}
	\fmfstraight
	\fmftop{t1,t2} \fmfbottom{b1,b2,b3,b4}
	\fmf{half_prop}{b1,i1}
	\fmf{half_prop}{b2,i1}
	\fmf{half_prop}{b3,i1}
	\fmf{half_prop}{b4,i1}
	\fmf{half_prop,tension=2}{i1,t2}
	\fmf{half_prop,tension=2}{i1,t1}
	\fmfv{d.shape=circle,d.filled=full,d.size=2thick}{i1}
	\fmffreeze
	\fmf{phantom,label=$\Omega^{24}$}{b2,b3}
	\end{fmfgraph*}
	\end{fmffile}}
+
	\parbox{40pt}{\begin{fmffile}{Omega51}
	\begin{fmfgraph*}(40,40)
	\fmfcmd{style_def half_prop expr p =
    draw_plain p;
    shrink(.7);
        cfill (marrow (p, .5))
    endshrink;
	enddef;}
	\fmfstraight
	\fmftop{t1,t2,t3,t4,t5} \fmfbottom{b1}
	\fmf{half_prop}{i1,t2}
	\fmf{half_prop}{i1,t1}
	\fmf{half_prop}{i1,t3}
	\fmf{half_prop}{i1,t4}
	\fmf{half_prop}{i1,t5}
	\fmf{half_prop,tension=5}{b1,i1}
	\fmfv{d.shape=circle,d.filled=full,d.size=2thick}{i1}
	\fmflabel{$\Omega^{51}$}{b1}
	\end{fmfgraph*}
	\end{fmffile}}
	
\vspace{2\baselineskip}	

$\phantom{\Omega^{[6]} =}$
+
	\parbox{40pt}{\begin{fmffile}{Omega15}
	\begin{fmfgraph*}(40,40)
	\fmfcmd{style_def half_prop expr p =
    draw_plain p;
    shrink(.7);
        cfill (marrow (p, .5))
    endshrink;
	enddef;}
	\fmfstraight
	\fmftop{t1} \fmfbottom{b1,b2,b3,b4,b5}
	\fmf{half_prop}{b1,i1}
	\fmf{half_prop}{b2,i1}
	\fmf{half_prop}{b3,i1}
	\fmf{half_prop}{b4,i1}
	\fmf{half_prop}{b5,i1}
	\fmf{half_prop,tension=5}{i1,t1}
	\fmfv{d.shape=circle,d.filled=full,d.size=2thick}{i1}
	\fmflabel{$\Omega^{15}$}{b3}
	\end{fmfgraph*}
	\end{fmffile}}
+
	\parbox{40pt}{\begin{fmffile}{Omega60}
	\begin{fmfgraph*}(40,40)
	\fmfcmd{style_def half_prop expr p =
    draw_plain p;
    shrink(.7);
        cfill (marrow (p, .5))
    endshrink;
	enddef;}
	\fmfstraight
	\fmftop{t1,t2,t3,t4,t5,t6} \fmfbottom{b1}
	\fmf{half_prop}{i1,t1}
	\fmf{half_prop}{i1,t2}
	\fmf{half_prop}{i1,t3}
	\fmf{half_prop}{i1,t4}
	\fmf{half_prop}{i1,t5}
	\fmf{half_prop}{i1,t6}
	\fmf{phantom,tension=6}{i1,b1}
	\fmfv{d.shape=circle,d.filled=full,d.size=2thick}{i1}
	\fmflabel{$\Omega^{60}$}{b1}
	\end{fmfgraph*}
	\end{fmffile}}
+
	\parbox{40pt}{\begin{fmffile}{Omega06}
	\begin{fmfgraph*}(40,40)
	\fmfcmd{style_def half_prop expr p =
    draw_plain p;
    shrink(.7);
        cfill (marrow (p, .5))
    endshrink;
	enddef;}
	\fmfstraight
	\fmftop{t1} \fmfbottom{b1,b2,b3,b4,b5,b6}
	\fmf{half_prop}{b1,i1}
	\fmf{half_prop}{b2,i1}
	\fmf{half_prop}{b3,i1}
	\fmf{half_prop}{b4,i1}
	\fmf{half_prop}{b5,i1}
	\fmf{half_prop}{b6,i1}
	\fmf{phantom,tension=6}{i1,t1}
	\fmfv{d.shape=circle,d.filled=full,d.size=2thick}{i1}
	\fmffreeze
	\fmf{phantom,label=$\Omega^{06}$}{b3,b4}
	\end{fmfgraph*}
	\end{fmffile}}

\vspace{\baselineskip}

\caption{
\label{f:verticesOmega}
Canonical diagrammatic representation of normal-ordered contributions to the grand potential operator $\Omega$ in the Schr\"odinger representation.}
\end{figure}

Since the grand canonical potential $\Omega$ is involved in the evaluation of any observable $\text{O}^{\text{A}}_0$, its own diagrammatic representation is needed and displayed in Fig.~\ref{f:verticesOmega}. The only difference with  Fig.~\ref{f:verticesO} relates to the use of dots rather than square symbols to represent the vertices. The same is easily done for other operators of interest, i.e.,\@ $H$ and $A$. It is to be noted that $\Omega_{1}$ has the same diagrammatic representation as $\Omega$ except that $\Omega^{00}$ must be omitted and $\Omega^{11}$ replaced by $\breve{\Omega}^{11}$, which requires to use a different symbol for that particular vertex\footnote{We omit to use a different symbol for $\breve{\Omega}^{11}$ in the following although it must be clear that the vertex with one line coming in and one line coming out does represent $\breve{\Omega}^{11}$ whenever it originates from the perturbative expansion of the evolution operator. This may be confusing whenever $O=\Omega$ since in this case there can also be a vertex $\Omega^{11}$ at fixed time $t=0$.}.

\begin{figure*}[t!]
\begin{center}
\parbox{60pt}{\begin{fmffile}{Gplumotest}
\begin{fmfgraph*}(40,40)
\fmfcmd{style_def prop_pm expr p =
    draw_plain p;
    shrink(.7);
        cfill (marrow (p, .25));
        cfill (marrow (p, .75))
    endshrink;
enddef;
}
\fmftop{t1} \fmfbottom{b1}
\fmflabel{\small $k_2 \ \tau_2$}{b1}
\fmflabel{\small $k_1 \ \tau_1$}{t1}
\fmf{prop_pm}{t1,b1}
\end{fmfgraph*}
\end{fmffile}}
\hbox{\quad}
\parbox{60pt}{\begin{fmffile}{Gmomotest}
\begin{fmfgraph*}(40,40)
\fmfcmd{style_def prop_mm expr p =
        draw_plain p;
        shrink(.7);
            cfill (marrow (p, .75));
            cfill (marrow (reverse p, .75))
        endshrink;
        enddef;}
\fmftop{t1} \fmfbottom{b1}
\fmflabel{\small $k_2 \ \tau_2$}{b1}
\fmflabel{\small $k_1 \ \tau_1$}{t1}
\fmf{prop_mm}{b1,t1}
\end{fmfgraph*}
\end{fmffile}}
\hbox{\quad}
\parbox{60pt}{\begin{fmffile}{Gpluplutest}
\begin{fmfgraph*}(40,40)
\fmfcmd{style_def prop_pp expr p =
    cdraw p;
    shrink(.7);
        cfill (marrow (p, .25));
        cfill (marrow (reverse p, .25))
    endshrink;
enddef;}
\fmftop{t1} \fmfbottom{b1}
\fmflabel{\small $k_2 \ \tau_2$}{b1}
\fmflabel{\small $k_1 \ \tau_1$}{t1}
\fmf{prop_pp}{b1,t1}
\end{fmfgraph*}
\end{fmffile}}
\hbox{\quad}
\parbox{60pt}{\begin{fmffile}{Gmoplutest}
\begin{fmfgraph*}(40,40)
\fmfcmd{style_def prop_mp expr p =
    draw_plain p;
    shrink(.7);
        cfill (marrow (reverse p, .25));
        cfill (marrow (reverse p, .75))
    endshrink;
enddef;
}
\fmftop{t1} \fmfbottom{b1}
\fmflabel{\small $k_2 \ \tau_2$}{b1}
\fmflabel{\small $k_1 \ \tau_1$}{t1}
\fmf{prop_mp}{t1,b1}
\end{fmfgraph*}
\end{fmffile}}

\vspace{2\baselineskip}

\parbox{60pt}{$G^{+- (0)}_{k_1k_2}(\tau_1, \tau_2;\varphi)$} \hbox{\quad}
\parbox{60pt}{$G^{-- (0)}_{k_1k_2}(\tau_1, \tau_2;\varphi)$} \hbox{\quad}
\parbox{60pt}{$G^{++ (0)}_{k_1k_2}(\tau_1, \tau_2;\varphi)$} \hbox{\quad}
\parbox{60pt}{$G^{-+ (0)}_{k_1k_2}(\tau_1, \tau_2;\varphi)$}
\end{center}
\caption{
\label{f:prop}
Diagrammatic representation of the four unperturbed elementary one-body propagators $G^{gg' (0)}(\varphi)$. The convention is that the left-to-right reading of a matrix element corresponds to the up-down reading of the diagram.}
\end{figure*}
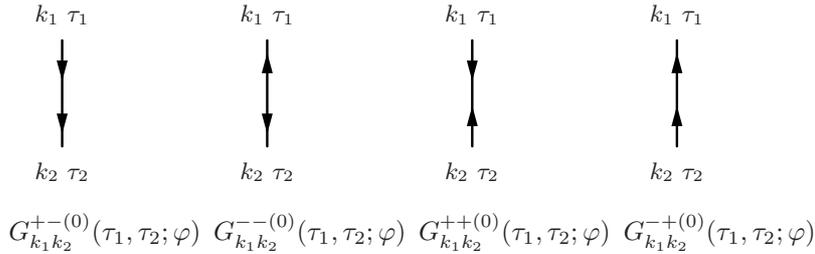

As the off-diagonal Wick theorem contracts pairs of quasi-particle operators together, the lines entering the diagrammatic representation of operators are eventually connected in the computation of the kernel $o(\varphi)$, thus, forming elementary contractions. Consequently, the four unperturbed propagators at play also need to be represented diagrammatically, which is done in Fig.~\ref{f:prop}. Here, the convention is that the left-to-right reading of a matrix element corresponds to the up-down reading of the diagram.

\subsection{Diagrams generation}
\label{subs:diag_gene}

With the building blocks at hand, off-diagonal BMBPT Feynman diagrams representing the contributions to $o(\varphi)$ are generated by assembling them according to a set of topological rules~\cite{Duguet:2015yle}
\begin{enumerate}
\item A Feynman diagram of order $p$ consists of $p$ vertices $\Omega^{i_kj_k}(\tau_k)$, $i_k+j_k=2,4$ or $6$, along with one vertex $O^{mn}(0)$, $m+n=0,2,4$ or $6$, that are connected by fermionic quasi-particle lines, i.e., via non-zero propagators $G^{+- (0)}$, $G^{-+ (0)}$ or $G^{-- (0)}$.
\item Each vertex is labeled by a time variable while each line is labeled by two time labels associated with the two vertices the line is attached to. 
\item Generating all contributions to Eq.~\eqref{observableO1} requires to form all possible diagrams, i.e., contract quasi-particle lines attached to the vertices in all possible ways while fulfilling the following restrictions.
\begin{enumerate}
\item Restrict equal-time propagators starting and ending at the same vertex to anomalous propagators. In the diagonal case, i.e.,\@ for $\varphi=0$, no such self-contraction may occur.
\item Restrict the set to \emph{connected} diagrams, i.e., omit diagrams containing parts that are not connected to each other by either propagators or vertices. This implies in particular that the vertex $O^{00}$ with no line can only appear at order $p=0$.
\item Because of the time-ordering relations carried by the propagators (see Eq.~\eqref{propagatorsB}), normal lines linking a set of vertices must not form an oriented loop. For two given vertices $\Omega^{i_kj_k}(\tau_k)$ and $\Omega^{i_{k'}j_{k'}}(\tau_{k'})$, it means that normal lines must propagate between them in the same direction. Correspondingly, normal lines connected to the generic operator $O$ at fixed time $0$ must go out of it, i.e.,\@ upwards in time. Anomalous lines do not carry time-ordering relations and are, thus, not concerned by these restrictions. In the diagonal limit, i.e.,\@ $\varphi=0$, where no anomalous line may be formed, the above constraint imposes that contributing vertices $O^{mn}(0)$ can only have lines going out, i.e., one necessarily has $n=0$.
\item Restrict the set to \emph{vacuum-to-vacuum} diagrams forming a set of closed loops with no external, i.e., unpaired, lines. This condition, together with the fact that $G^{++ (0)}(\varphi)$ is identically zero, strongly constrains which normal-ordered parts $\Omega^{i_kj_k}(\tau_k)$ and $O^{mn}(0)$ of the $p+1$ involved operators can be combined, i.e., the condition 
\begin{equation}
n_a \equiv \sum_{k=1}^{p}(j_k-i_k) +n-m \geq 0 \, , \nonumber
\end{equation}
must be fulfilled. The number $n_a$ corresponds to the number of anomalous propagators $G^{-- (0)}(\varphi)$ in the diagram. In the diagonal limit for which $G^{-- (0)}(0)=0$, the set of combined operators are further reduced to $n_a=0$.
\item Restrict the set to \emph{topologically distinct} time-unlabelled diagrams, i.e., time-unlabelled diagrams that cannot be obtained from one another via a mere displacement, i.e., translation, of the vertices.
\end{enumerate}
\end{enumerate}

\subsection{Diagram evaluation}
\label{subs:diag_eval}

\subsubsection{Feynman expression}
\label{subs:feyndiag_eval}

\begin{figure}[t!]
\begin{center}
  \parbox{100pt}{\begin{fmffile}{OrientRule_22}
	\begin{fmfgraph*}(100,100)
	\fmfcmd{style_def half_prop expr p =
    draw_plain p;
    shrink(.7);
        cfill (marrow (p, .5))
    endshrink;
	enddef;}
	\fmfstraight
	\fmftop{t1,t2,t3,t4,t5} \fmfbottom{b1,b2,b3,b4,b5}
  \fmf{phantom}{b1,i1}
  \fmf{phantom}{b2,i2}
  \fmf{phantom}{b3,i3}
  \fmf{phantom}{b4,i4}
  \fmf{phantom}{b5,i5}
  \fmf{phantom}{i1,j1}
  \fmf{phantom}{i2,j2}
  \fmf{phantom}{i3,j3}
  \fmf{phantom}{i4,j4}
  \fmf{phantom}{i5,j5}
  \fmf{phantom}{j1,k1}
  \fmf{phantom}{j2,k2}
  \fmf{phantom}{j3,k3}
  \fmf{phantom}{j4,k4}
  \fmf{phantom}{j5,k5}
  \fmf{phantom}{k1,t1}
  \fmf{phantom}{k2,t2}
  \fmf{phantom}{k3,t3}
  \fmf{phantom}{k4,t4}
  \fmf{phantom}{k5,t5}
  \fmffreeze
	\fmf{dashes,tag=1}{j3,t2}
	\fmf{dashes,tag=2}{b2,j3}
  \fmf{half_prop,left,tag=3}{j5,j3}
  \fmf{half_prop,right,tag=4}{j5,j3}
  \fmf{dashes,fore=red,left}{j2,j4}
  \fmf{dashes,fore=red,left}{k3,i3}
  \fmf{phantom_arrow,fore=red,width=0.2}{i4,i2}
	\fmfv{d.shape=circle,d.filled=full,d.size=3thick,label=$\tau$}{j3}
	\fmf{plain,fore=red,width=1.2}{j1,j3}
  \fmfposition
  \fmfipath{p[]}
  \fmfiset{p1}{vpath1(__j3,__t2)}
  \fmfiset{p2}{vpath2(__b2,__j3)}
  \fmfiset{p3}{vpath3(__j5,__j3)}
  \fmfiset{p4}{vpath4(__j5,__j3)}
  \fmfi{half_prop}{point 0 of p1 -- point length(p1)*0.60 of p1}
  \fmfi{half_prop}{point length(p2)*0.35 of p2 -- point length(p2) of p2}
  \fmfiv{label=$k_1$,l.angle=-130,l.dist=0.02w}{point length(p1)*0.27 of p1}
  \fmfiv{label=$k_2$,l.angle=22,l.dist=0.05w}{point length(p2)*0.65 of p2}
  \fmfiv{label=$k_3$,l.angle=-90,l.dist=0.05w}{point length(p3)/2 of p3}
  \fmfiv{label=$k_4$,l.angle=90,l.dist=0.05w}{point length(p4)/2 of p4}
  \fmfiv{label=$G^{--(0)}_{k_4k_3}(\tau ,,\tau ; \varphi)$}{point 0 of p3}
  \fmfiv{label=$O^{13}_{k_1k_2k_3k_4}$,l.angle=-150,l.dist=0.1w}{point 0 of p1}
	\end{fmfgraph*}
	\end{fmffile}}
\end{center}
\caption{
\label{fig:anomalousself}
Convention to draw and read anomalous self-contractions. The example is given for a vertex $\Omega^{13}$ displaying a self-contraction.}
\end{figure}
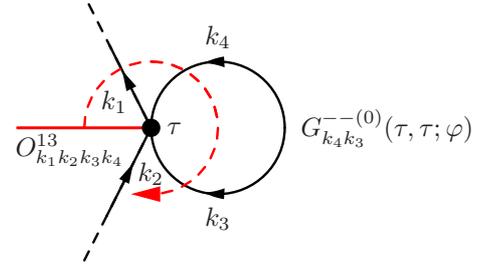

The way to translate off-diagonal BMBPT Feynman diagrams into their mathematical expressions follows a set of algebraic rules
\begin{enumerate}
\item Each of the $p+1$ vertices contributes a factor, e.g., $\Omega^{ij}_{k_1 \ldots k_i k_{i+1} \ldots k_{i+j}}$ with the sign convention detailed in Sec.~\ref{subs:prod_bmbpt_diags}. 
\item Each of the 
\begin{equation}
n_b\equiv \left(\sum_{k=1}^{p}\left(j_k+i_k\right) +n+m\right)/2 \, , \nonumber
\end{equation}
lines contributes a factor $G^{gg' (0)}_{k_1k_2}(\tau_k, \tau_{k'})$, where $g$ and $g'$ characterize the type of elementary propagator the line corresponds to. According to Eq.~\eqref{propagatorsB}, each of the $n_a$ anomalous propagators carries an exponential function of the two time labels and an anomalous contraction $R^{--}_{k_1k_2}(\varphi)$ while each of the $n_b-n_a$ normal propagators carries an exponential function and a step function of the two time labels. 
\item A normal line can be interpreted as $G^{-+ (0)}$ or $G^{+- (0)}$ depending on the ascendant or descendant reading of the diagram. Similarly, the ordering of quasi-particle and time labels of a propagator depends on the ascendant or descendant reading of the diagram. \emph{All} the lines involved in a given diagram must be interpreted in the \emph{same} way, i.e., sticking to an ascendant or descendant way of reading the diagram all throughout. In the present work, the chosen convention corresponds to reading diagrams from top to bottom, which further relates to reading the many-body matrix element it originates from in Eq.~\eqref{observableO1} in a left-right fashion. It is the convention employed to represent the four propagators in Fig.~\ref{f:prop}.
\item The reading of an anomalous line linking two different vertices is unambiguous as long as one stick to the up-down convention displayed in Fig.~\ref{f:prop}. However, the up-down reading of a self-contraction is potentially ambiguous depending on the way the line is actually drawn. As illustrated in Fig.~\ref{fig:anomalousself}, one must further fix a convention based on the insertion of a fictitious semi-straight, e.g., horizontal line originating from the vertex that the self-contraction is forbidden to cross. Taking the semi-straight line as a reference point, the quasi-particle indices must be attributed to the equal-time propagator in the order the lines are crossed when going around the vertex in a clockwise fashion.
\item All quasi-particle labels must be summed over while all running time variables must be integrated over from $0$ to $\tau\rightarrow +\infty$.
\item A sign factor $(-1)^{p+n_c}$, where $p$ denotes the order of the diagram and $n_c$ denotes the number of crossing lines in the diagram, must be considered\footnote{In case a line is drawn such that it crosses itself, the crossing(s) must be omitted when evaluating $p$.}. The overall sign results from multiplying this factor with the sign associated with each matrix element.
\item Each diagram comes with a numerical prefactor obtained from the following combination
\begin{enumerate}
\item A factor $1/(n_e)!$ must be considered for \emph{each} group of $n_e$ equivalent lines. Equivalent lines must all begin and end at the same vertices (or vertex, for anomalous propagators starting and ending at the same vertex), and must correspond to the same type of contractions, i.e., they must all correspond to propagators characterized by the same superscripts $g$ and $g'$ in addition to having identical time labels.
\item Given the previous rule, an extra factor $1/2$ must be considered for \emph{each} anomalous propagator that starts and ends at the same vertex.
\item A symmetry factor $1/n_s$ must be considered in connection with exchanging the time labels of the vertices in all possible ways, counting the identity as one. The factor $n_s$ corresponds to the number of ways exchanging the time labels provides a time-labelled diagram that is topologically equivalent to the original one.
\end{enumerate} 
\end{enumerate} 

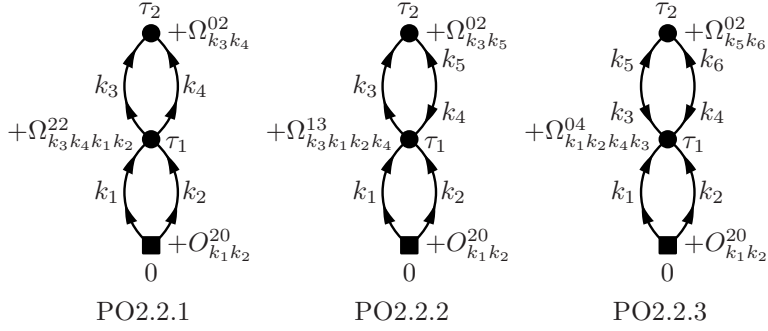
\begin{figure*}[h!]
\begin{center}

  \vspace{\baselineskip}

  \parbox{80pt}{\begin{fmffile}{PNP-BMBPT2_2_1p}
    \begin{fmfgraph*}(80,80)
    \fmfcmd{style_def prop_pm expr p =
    draw_plain p;
    shrink(.7);
        cfill (marrow (p, .25));
        cfill (marrow (p, .75))
    endshrink;
enddef;
}
      \fmftop{t1}
      \fmfbottom{b1}
      \fmf{prop_pm,left=0.5,tag=1,label=$k_1$,l.dist=0.02w}{b1,i1}
      \fmf{prop_pm,right=0.5,tag=2,label=$k_2$,l.dist=0.02w}{b1,i1}
      \fmf{prop_pm,left=0.5,tag=3,label=$k_3$,l.dist=0.02w}{i1,t1}
      \fmf{prop_pm,right=0.5,tag=4,label=$k_4$,l.dist=0.02w}{i1,t1}
      \fmfv{d.shape=circle,d.filled=full,d.size=3thick,l=$+\Omega^{02}_{k_3k_4}$,l.angle=0}{t1}
      \fmfv{d.shape=circle,d.filled=full,d.size=3thick,l=$+\Omega^{22}_{k_3k_4k_1k_2}$,l.angle=160}{i1}
      \fmfv{d.shape=square,d.filled=full,d.size=3thick,l=$+O^{20}_{k_1k_2}$,l.angle=0}{b1}
      \fmfposition
      \fmfipath{p[]}
      \fmfiset{p1}{vpath1(__b1,__i1)}
      \fmfiset{p2}{vpath2(__b1,__i1)}
      \fmfiset{p3}{vpath3(__i1,__t1)}
      \fmfiset{p4}{vpath4(__i1,__t1)}
      \fmfiv{label=$0$,l.angle=-73,l.dist=0.13w}{point 0 of p1}
      \fmfiv{label=$\tau_1$,l.angle=0,l.dist=0.04w}{point length(p2) of p2}
      \fmfiv{label=$\tau_2$,l.angle=75,l.dist=0.10w}{point length(p3) of p3}
      \fmfiv{label=$\text{PO2.2.1}$,l.angle=-90,l.dist=0.30w}{point 0 of p1}
    \end{fmfgraph*}
  \end{fmffile}}
  \hbox{\quad}
  \parbox{80pt}{\begin{fmffile}{PNP-BMBPT2_2_2p}
    \begin{fmfgraph*}(80,80)
    \fmfcmd{style_def prop_pm expr p =
    draw_plain p;
    shrink(.7);
        cfill (marrow (p, .25));
        cfill (marrow (p, .75))
    endshrink;
enddef;
}
\fmfcmd{style_def prop_mm expr p =
        draw_plain p;
        shrink(.7);
            cfill (marrow (p, .75));
            cfill (marrow (reverse p, .75))
        endshrink;
        enddef;}
      \fmftop{t1}
      \fmfbottom{b1}
      \fmf{prop_pm,left=0.5,tag=1,label=$k_1$,l.dist=0.02w}{b1,i1}
      \fmf{prop_pm,right=0.5,tag=2,label=$k_2$,l.dist=0.02w}{b1,i1}
      \fmf{prop_pm,left=0.5,tag=3,label=$k_3$,l.dist=0.02w}{i1,t1}
      \fmf{prop_mm,right=0.5,tag=4}{i1,t1}
      \fmfv{d.shape=circle,d.filled=full,d.size=3thick,l=$+\Omega^{02}_{k_3k_5}$,l.angle=0}{t1}
      \fmfv{d.shape=circle,d.filled=full,d.size=3thick,l=$+\Omega^{13}_{k_3k_1k_2k_4}$,l.angle=160}{i1}
      \fmfv{d.shape=square,d.filled=full,d.size=3thick,l=$+O^{20}_{k_1k_2}$,l.angle=0}{b1}
      \fmfposition
      \fmfipath{p[]}
      \fmfiset{p1}{vpath1(__b1,__i1)}
      \fmfiset{p2}{vpath2(__b1,__i1)}
      \fmfiset{p3}{vpath3(__i1,__t1)}
      \fmfiset{p4}{vpath4(__i1,__t1)}
      \fmfiv{label=$k_4$,l.angle=0,l.dist=0.05w}{point length(p4)/4 of p4}
      \fmfiv{label=$k_5$,l.angle=0,l.dist=0.05w}{point 3*length(p4)/4 of p4}
      \fmfiv{label=$0$,l.angle=-73,l.dist=0.13w}{point 0 of p1}
      \fmfiv{label=$\tau_1$,l.angle=0,l.dist=0.04w}{point length(p2) of p2}
      \fmfiv{label=$\tau_2$,l.angle=75,l.dist=0.10w}{point length(p3) of p3}
      \fmfiv{label=$\text{PO2.2.2}$,l.angle=-90,l.dist=0.30w}{point 0 of p1}
    \end{fmfgraph*}
  \end{fmffile}}
  \hbox{\quad}
  \parbox{80pt}{\begin{fmffile}{PNP-BMBPT2_2_3p}
    \begin{fmfgraph*}(80,80)
    \fmfcmd{style_def prop_pm expr p =
    draw_plain p;
    shrink(.7);
        cfill (marrow (p, .25));
        cfill (marrow (p, .75))
    endshrink;
enddef;
}
\fmfcmd{style_def prop_mm expr p =
        draw_plain p;
        shrink(.7);
            cfill (marrow (p, .75));
            cfill (marrow (reverse p, .75))
        endshrink;
        enddef;}
      \fmftop{t1}
      \fmfbottom{b1}
      \fmf{prop_pm,left=0.5,tag=1,label=$k_1$,l.dist=0.02w}{b1,i1}
      \fmf{prop_pm,right=0.5,tag=2,label=$k_2$,l.dist=0.02w}{b1,i1}
      \fmf{prop_mm,left=0.5,tag=3}{i1,t1}
      \fmf{prop_mm,right=0.5,tag=4}{i1,t1}
      \fmfv{d.shape=circle,d.filled=full,d.size=3thick,l=$+\Omega^{02}_{k_5k_6}$,l.angle=0}{t1}
      \fmfv{d.shape=circle,d.filled=full,d.size=3thick,l=$+\Omega^{04}_{k_1k_2k_4k_3}$,l.angle=160}{i1}
      \fmfv{d.shape=square,d.filled=full,d.size=3thick,l=$+O^{20}_{k_1k_2}$,l.angle=0}{b1}
      \fmfposition
      \fmfipath{p[]}
      \fmfiset{p1}{vpath1(__b1,__i1)}
      \fmfiset{p2}{vpath2(__b1,__i1)}
      \fmfiset{p3}{vpath3(__i1,__t1)}
      \fmfiset{p4}{vpath4(__i1,__t1)}
      \fmfiv{label=$k_3$,l.angle=180,l.dist=0.05w}{point length(p3)/4 of p3}
      \fmfiv{label=$k_5$,l.angle=180,l.dist=0.05w}{point 3*length(p3)/4 of p3}
      \fmfiv{label=$k_4$,l.angle=0,l.dist=0.05w}{point length(p4)/4 of p4}
      \fmfiv{label=$k_6$,l.angle=0,l.dist=0.05w}{point 3*length(p4)/4 of p4}
      \fmfiv{label=$0$,l.angle=-73,l.dist=0.13w}{point 0 of p1}
      \fmfiv{label=$\tau_1$,l.angle=0,l.dist=0.04w}{point length(p2) of p2}
      \fmfiv{label=$\tau_2$,l.angle=75,l.dist=0.10w}{point length(p3) of p3}
      \fmfiv{label=$\text{PO2.2.3}$,l.angle=-90,l.dist=0.30w}{point 0 of p1}
    \end{fmfgraph*}
  \end{fmffile}}

  \vspace{2\baselineskip}
  \caption{\label{f:ex_offbmbpt_diag}
  Selected second-order off-diagonal Feynman BMBPT diagrams.}
\end{center}
\end{figure*}

In order to illustrate the typical expression of off-diagonal Feynman BMBPT diagrams and to anticipate several key characteristics, let us compute the three second-order diagrams displayed in Fig.~\ref{f:ex_offbmbpt_diag}, i.e., 
\begin{subequations}
\label{e:ex_bmbpt_feynman}
\begin{align}
    \text{PO2.2.1}
    &= \frac{1}{4} \sum_{k_i} \Omega^{02}_{k_3k_4} \Omega^{22}_{k_3k_4k_1k_2} O^{20}_{k_1k_2} \notag \\
    &\hphantom{=} \times \lim_{\tau\rightarrow\infty} \int_0^{\tau} \mathrm{d}\tau_1 \mathrm{d}\tau_2 \, \theta(\tau_2-\tau_1) \\
    &\hphantom{=} \hspace{1.5cm} \times  \mathrm{e}^{-\tau_2\epsilon_{k_3k_4}} \mathrm{e}^{-\tau_1\epsilon_{k_1k_2}^{k_3k_4}} \notag\\
    \text{PO2.2.2}
    &= \frac{1}{2} \sum_{k_i} \Omega^{02}_{k_3k_5} \Omega^{13}_{k_3k_1k_2k_4} O^{20}_{k_1k_2} R^{--}_{k_5k_4}(\varphi) \notag \\
    &\hphantom{=} \times \lim_{\tau\rightarrow\infty} \int_0^{\tau} \mathrm{d}\tau_1 \mathrm{d}\tau_2 \, \theta(\tau_2-\tau_1) \\
    &\hphantom{=} \hspace{1.5cm} \times  \mathrm{e}^{-\tau_2\epsilon_{k_3k_5}} \mathrm{e}^{-\tau_1\epsilon_{k_1k_2k_4}^{k_3}} \notag\\
    \text{PO2.2.3}
    &= \frac{1}{4} \sum_{k_i} \Omega^{02}_{k_5k_6} \Omega^{04}_{k_1k_2k_4k_3} O^{20}_{k_1k_2} R^{--}_{k_6k_4}(\varphi) R^{--}_{k_5k_3}(\varphi) \notag \\
    &\hphantom{=} \times \lim_{\tau\rightarrow\infty} \int_0^{\tau} \mathrm{d}\tau_1 \mathrm{d}\tau_2  \\
    &\hphantom{=} \hspace{1.5cm} \times \mathrm{e}^{-\tau_2\epsilon_{k_5k_6}} \mathrm{e}^{-\tau_1\epsilon_{k_1k_2k_3k_4}} \, , \notag
\end{align}
\end{subequations}
where the extended notation 
\begin{equation}
\epsilon^{k_{a}k_{b}\dots}_{k_{i}k_{j}\dots} \equiv  E_{k_{i}} +  E_{k_{j}} + \ldots - E_{k_{a}} -  E_{k_{b}} - \ldots  \, ,
\end{equation}
was introduced. In each case, the sign, the combinatorial factors and the three matrix elements directly reflect Feynman's algebraic rules listed above and are easy to interpret. Eventually, the final form of the integrand originates from expliciting the $n_b=4$ propagators via Eq.~\eqref{propagatorsB}, which induces the presence of one off-diagonal elementary contractions per anomalous propagator.  

The three chosen diagrams display the same overall topology\footnote{The number of quasi-particle indices on which summation is performed increases by one per anomalous propagator due to the fact that the matrix $R^{--}(\varphi)$ is not diagonal in quasi-particle space.}, i.e., while the vertex $O^{20}$ is at fixed time $0$, the vertex $\Omega^{22}$/$\Omega^{13}$/$\Omega^{04}$ is at running time $\tau_1$ and the $\Omega^{02}$ vertex is at running time $\tau_2$. However, the three diagrams differ in their number of anomalous lines and, as such, clearly illustrate key consequences of going from diagonal to off-diagonal BMBPT. The first diagram, PO2.2.1, contains no anomalous line ($n_a=0$) and already occurs in straight, i.e., diagonal, BMBPT\footnote{This diagram is the one denoted as PO2.2 in Fig.~6 of Ref.~\cite{arthuis18a}.}. By turning the second vertex $\Omega^{22}$ into $\Omega^{13}$ ($\Omega^{04}$), PO2.2.2 (PO2.2.3) contains $n_a=1$ ($n_a=2$) anomalous line(s) between the second and the third vertices. As a consequence, the integrands display typical structures that need to be scrutinized for the following. 
\begin{itemize}
\item The fact that the two running variables $\tau_1$ and $\tau_2$ are positive is directly encoded into the boundary of the double integral.
\item In PO2.2.1, the explicit step function characterizes the  time ordering induced between $\Omega^{22}$ and $\Omega^{02}$ vertices by the two normal propagators connecting them. This step function, i.e., time ordering, remains at play in PO2.2.2 given than one normal line still connects the second and third vertices. Contrarily, the absence of step function in PO2.2.3 characterizes the fact that $\Omega^{04}$ and $\Omega^{02}$ are solely connected via anomalous propagators that do not induce any time-ordering relation between them. While in the first two cases the integral over $\tau_1$ depends on the integral over $\tau_2$, both integrals are independent from each other in PO2.2.3. 
\item Grouping appropriately the exponential functions coming from the four propagators, the integrand displays one exponential factor per running time, i.e., per $\Omega^{i_kj_k}(\tau_k)$ vertex. The relevant energy factor $\epsilon^{k_{a}k_{b}\dots}_{k_{i}k_{j}\dots} $ multiplying the variable $\tau_k$ in this exponential function denotes the sum/difference of quasi-particle energies associated with the lines entering/leaving the corresponding vertex.
\end{itemize}
The three diagrams exemplify the fact that the off-diagonal BMBPT diagrammatics differentiates itself by the presence of anomalous lines that, depending on the situation, may change the time-ordering structure between the vertices compared to the diagonal BMBPT diagram displaying the same overall topology.

\subsubsection{Time-integrated expression}
\label{subs:feynman_to_goldstone}

The expression obtained via the application of Feynman's algebraic rules does not yet constitute the form needed for the numerical implementation of the formalism. While the sign, the combinatorial factor and the matrix elements will remain untouched, the $p$-tuple time integral must be performed in order to obtain the needed expression. 

A major part of Ref.~\cite{arthuis18a} was dedicated to the automated computation of the $p$-tuple time integrals via the introduction of the so-called time-structure diagram (TSD) underlying any given BMBPT diagram of arbitrary order and topology. We thus refer to Ref.~\cite{arthuis18a} for the general theory of TSDs and will come back later on to its implementation for more general off-diagonal BMBPT diagrams. For now, it is sufficient to focus on the main consequence of the above analysis, i.e.,\@ while the presence of one anomalous line in PO2.2.2 does not change its time structure compared to PO2.2.1, turning the other propagator connecting the second and third vertices into an anomalous line does modify it. Consequently, while the TSD associated to diagrams PO2.2.1 and PO2.2.2 is T2.1 (see Fig.~\ref{diagTSD0123}), it becomes T2.2 for PO2.2.3. Generically denoting as $a_k$ the energy factor multiplying the time label $\tau_k$ in the integrand, the integrals associated with the examples given in Eq.~\eqref{e:ex_bmbpt_feynman} are
\begin{subequations}
\label{examplesTSD}
\begin{align}
\text{T2.1} &= \lim\limits_{\tau \to \infty} \int_0^{\tau} d\tau_1 d\tau_2 \, \theta(\tau_2 - \tau_1) e^{-a_1\tau_1} e^{-a_2\tau_2} \nonumber \\
&= \frac{1}{a_2(a_1+a_2)} \ , \\
\text{T2.2} &= \lim\limits_{\tau \to \infty} \int_0^{\tau} d\tau_1 d\tau_2  \, e^{-a_1\tau_1} e^{-a_2\tau_2} \nonumber \\
&= \frac{1}{a_1a_2} \ ,
\end{align}
\end{subequations}
the first (second) of which applies to PO2.2.1 and PO2.2.2 (PO2.2.3).

In order to obtain the final, i.e. time-integrated, expression of each of the three diagrams, the factors $a_1$ and $a_2$ must be expressed back in terms of quasi-particle energies. As discussed in Ref.~\cite{arthuis18a} for diagonal BMBPT diagrams, and as generalized to off-diagonal BMBPT diagrams below, the specific combinations of these factors emerging from the TSDs correspond necessarily to positive sums of quasi-particle energies that can be straightforwardly extracted from the diagram itself. Combining Eqs.~\eqref{e:ex_bmbpt_feynman} and~\eqref{examplesTSD} before inserting the appropriate combinations of quasi-particle energies, one eventually obtains the desired expressions under the form
\begin{subequations}
\label{e:ex_bmbpt_goldstone}
\begin{align}
    \text{PO2.2.1}
    &= \frac{1}{4} \sum_{k_i} \frac{\Omega^{02}_{k_3k_4} \Omega^{22}_{k_3k_4k_1k_2} O^{20}_{k_1k_2}}{\epsilon_{k_1 k_2}\epsilon_{k_3 k_4}} \, , \notag\\
    \text{PO2.2.2}
    &= \frac{1}{2} \sum_{k_i} \frac{\Omega^{02}_{k_3k_5} \Omega^{13}_{k_3k_1k_2k_4} O^{20}_{k_1k_2}}{\epsilon_{k_1 k_2 k_4 k_5} \epsilon_{k_3 k_5}} R^{--}_{k_5k_4}(\varphi) \, ,\notag\\
    \text{PO2.2.3}
    &= \frac{1}{4} \sum_{k_i} \frac{\Omega^{02}_{k_5k_6} \Omega^{04}_{k_1k_2k_4k_3} O^{20}_{k_1k_2}}{\epsilon_{k_1 k_2 k_3 k_4}\epsilon_{k_5 k_6}} R^{--}_{k_6k_4}(\varphi) R^{--}_{k_5k_3}(\varphi) \, . \notag
\end{align}
\end{subequations}

\begin{figure*}[h!]
  
  \vspace{-0.5\baselineskip}

  \begin{center}
    \parbox{40pt}{\begin{fmffile}{PNP-BMBPT0_1_1_1na0}
      \begin{fmfgraph*}(40,40)
        \fmftop{t1}
        \fmfbottom{b1}
        \fmf{phantom,tag=1}{b1,t1}
        \fmffreeze
        \fmfv{d.shape=square,d.filled=full,d.size=3thick,l=$O^{00}$,l.angle=0}{b1}
        \fmfposition
        \fmfipath{p[]}
        \fmfiset{p1}{vpath1(__b1,__t1)}
        \fmfiv{label=$\text{PO0.1.1}^{(1)}$,l.angle=-90,l.dist=0.50w}{point 0 of p1}
        \fmfiv{label=$\mathbf{n_a = 0}$,l.angle=0,l.dist=0.00w}{point length(p1)*0.7 of p1}
        \fmfiv{label=$0$,l.angle=-90,l.dist=0.22w}{point 0 of p1}
      \end{fmfgraph*}
    \end{fmffile}}
    \hbox{\qquad\qquad}
    \parbox{40pt}{\begin{fmffile}{PNP-BMBPT0_1_1_2na1}
      \begin{fmfgraph*}(40,40)
      \fmfcmd{style_def half_prop expr p =
    draw_plain p;
    shrink(.7);
        cfill (marrow (p, .5))
    endshrink;
	enddef;}
        \fmftop{t1}
        \fmfbottom{bmm,bm,b0,b1,b2,b3,b4}
        \fmf{phantom,tag=1}{b1,t1}
        \fmffreeze
        \fmf{half_prop,left}{b3,b1}
        \fmf{half_prop,right}{b3,b1}
        \fmfv{d.shape=square,d.filled=full,d.size=3thick,l=$O^{02}$,l.angle=0,l.dist=0.40w}{b1}
        \fmfposition
        \fmfipath{p[]}
        \fmfiset{p1}{vpath1(__b1,__t1)}
        \fmfiv{label=$\text{PO0.1.1}^{(2)}$,l.angle=-90,l.dist=0.50w}{point 0 of p1}
        \fmfiv{label=$\mathbf{n_a = 1}$,l.angle=0,l.dist=0.00w}{point length(p1)*0.7 of p1}
        \fmfiv{label=$0$,l.angle=-90,l.dist=0.22w}{point 0 of p1}
      \end{fmfgraph*}
    \end{fmffile}}
    \hbox{\qquad\qquad}
    \parbox{40pt}{\begin{fmffile}{PNP-BMBPT0_1_1_3na2}
      \begin{fmfgraph*}(40,40)
      \fmfcmd{style_def half_prop expr p =
    draw_plain p;
    shrink(.7);
        cfill (marrow (p, .5))
    endshrink;
	enddef;}
        \fmftop{t1}
        \fmfbottom{bmm,bm,b0,b1,b2,b3,b4}
        \fmf{phantom,tag=1}{b1,t1}
        \fmffreeze
        \fmf{half_prop,left}{b3,b1}
        \fmf{half_prop,right}{b3,b1}
        \fmf{half_prop,left}{bm,b1}
        \fmf{half_prop,right}{bm,b1}
        \fmfv{d.shape=square,d.filled=full,d.size=3thick,l=$O^{04}$,l.angle=0,l.dist=0.40w}{b1}
        \fmfposition
        \fmfipath{p[]}
        \fmfiset{p1}{vpath1(__b1,__t1)}
        \fmfiv{label=$\text{PO0.1.1}^{(3)}$,l.angle=-90,l.dist=0.50w}{point 0 of p1}
        \fmfiv{label=$\mathbf{n_a = 2}$,l.angle=0,l.dist=0.00w}{point length(p1)*0.7 of p1}
        \fmfiv{label=$0$,l.angle=-90,l.dist=0.22w}{point 0 of p1}
      \end{fmfgraph*}
    \end{fmffile}}
    \hbox{\qquad\qquad}
    \parbox{40pt}{\begin{fmffile}{empty40ptna3}
      \begin{fmfgraph*}(40,40)
        \fmftop{t1}
        \fmfbottom{b1}
        \fmf{phantom,tag=1}{b1,t1}
        \fmffreeze
        \fmfposition
        \fmfipath{p[]}
        \fmfiset{p1}{vpath1(__b1,__t1)}
        \fmfiv{label=$\mathbf{n_a = 3}$,l.angle=0,l.dist=0.00w}{point length(p1)*0.7 of p1}
      \end{fmfgraph*}
    \end{fmffile}}
    \hbox{\qquad\qquad}
    \parbox{40pt}{\begin{fmffile}{empty40ptna4}
      \begin{fmfgraph*}(40,40)
        \fmftop{t1}
        \fmfbottom{b1}
        \fmf{phantom,tag=1}{b1,t1}
        \fmffreeze
        \fmfposition
        \fmfipath{p[]}
        \fmfiset{p1}{vpath1(__b1,__t1)}
        \fmfiv{label=$\mathbf{n_a = 4}$,l.angle=0,l.dist=0.00w}{point length(p1)*0.7 of p1}
      \end{fmfgraph*}
    \end{fmffile}}

    \vspace{4\baselineskip}

    \parbox{40pt}{\begin{fmffile}{PNP-BMBPT1_1_1_1p}
      \begin{fmfgraph*}(40,40)
      \fmfcmd{style_def prop_pm expr p =
    draw_plain p;
    shrink(.7);
        cfill (marrow (p, .25));
        cfill (marrow (p, .75))
    endshrink;
enddef;
}
        \fmftop{t1}
        \fmfbottom{b1}
        \fmf{prop_pm,left=0.5,tag=1}{b1,t1}
        \fmf{prop_pm,right=0.5,tag=2}{b1,t1}
        \fmfv{d.shape=circle,d.filled=full,d.size=3thick,l=$\Omega^{02}$,l.dist=0.18w,l.angle=20}{t1}
        \fmfv{d.shape=square,d.filled=full,d.size=3thick,l=$O^{20}$,l.angle=0}{b1}
        \fmfposition
        \fmfipath{p[]}
        \fmfiset{p1}{vpath1(__b1,__t1)}
        \fmfiv{label=$\text{PO1.1.1}^{(1)}$,l.angle=-90,l.dist=0.50w}{point 0 of p1}
        \fmfiv{label=$0$,l.angle=-70,l.dist=0.22w}{point 0 of p1}
        \fmfiv{label=$\tau_1$,l.angle=70,l.dist=0.18w}{point length(p1) of p1}
      \end{fmfgraph*}
    \end{fmffile}}
    \hbox{\qquad\qquad}
    \parbox{40pt}{\begin{fmffile}{PNP-BMBPT1_1_1_2p}
      \begin{fmfgraph*}(40,40)
      \fmfcmd{style_def prop_pm expr p =
    draw_plain p;
    shrink(.7);
        cfill (marrow (p, .25));
        cfill (marrow (p, .75))
    endshrink;
enddef;
}
\fmfcmd{style_def prop_mm expr p =
        draw_plain p;
        shrink(.7);
            cfill (marrow (p, .75));
            cfill (marrow (reverse p, .75))
        endshrink;
        enddef;}
        \fmftop{t1}
        \fmfbottom{b1}
        \fmf{prop_pm,left=0.5,tag=1}{b1,t1}
        \fmf{prop_mm,right=0.5,tag=2}{b1,t1}
        \fmfv{d.shape=circle,d.filled=full,d.size=3thick,l=$\Omega^{02}$,l.dist=0.18w,l.angle=20}{t1}
        \fmfv{d.shape=square,d.filled=full,d.size=3thick,l=$O^{11}$,l.angle=0}{b1}
        \fmfposition
        \fmfipath{p[]}
        \fmfiset{p1}{vpath1(__b1,__t1)}
        \fmfiv{label=$\text{PO1.1.1}^{(2)}$,l.angle=-90,l.dist=0.50w}{point 0 of p1}
        \fmfiv{label=$0$,l.angle=-70,l.dist=0.22w}{point 0 of p1}
        \fmfiv{label=$\tau_1$,l.angle=70,l.dist=0.18w}{point length(p1) of p1}
      \end{fmfgraph*}
    \end{fmffile}}
    \hbox{\qquad\qquad}
    \parbox{40pt}{\begin{fmffile}{PNP-BMBPT1_1_1_3p}
      \begin{fmfgraph*}(40,40)
      \fmfcmd{style_def prop_mm expr p =
        draw_plain p;
        shrink(.7);
            cfill (marrow (p, .75));
            cfill (marrow (reverse p, .75))
        endshrink;
        enddef;}
        \fmftop{t1}
        \fmfbottom{b1}
        \fmf{prop_mm,left=0.5,tag=1}{b1,t1}
        \fmf{prop_mm,right=0.5,tag=2}{b1,t1}
        \fmfv{d.shape=circle,d.filled=full,d.size=3thick,l=$\Omega^{02}$,l.dist=0.18w,l.angle=20}{t1}
        \fmfv{d.shape=square,d.filled=full,d.size=3thick,l=$O^{02}$,l.angle=0}{b1}
        \fmfposition
        \fmfipath{p[]}
        \fmfiset{p1}{vpath1(__b1,__t1)}
        \fmfiv{label=$\text{PO1.1.1}^{(3)}$,l.angle=-90,l.dist=0.50w}{point 0 of p1}
        \fmfiv{label=$0$,l.angle=-70,l.dist=0.22w}{point 0 of p1}
        \fmfiv{label=$\tau_1$,l.angle=70,l.dist=0.18w}{point length(p1) of p1}
      \end{fmfgraph*}
    \end{fmffile}}
    \hbox{\qquad\qquad}
    \parbox{40pt}{\begin{fmffile}{empty40pt}
      \begin{fmfgraph*}(40,40)
      \end{fmfgraph*}
    \end{fmffile}}
    \hbox{\qquad\qquad}
    \parbox{40pt}{\begin{fmffile}{empty40pt}
      \begin{fmfgraph*}(40,40)
      \end{fmfgraph*}
    \end{fmffile}}

    \vspace{4\baselineskip}

    \parbox{40pt}{\begin{fmffile}{empty40pt}
      \begin{fmfgraph*}(40,40)
      \end{fmfgraph*}
    \end{fmffile}}
    \hbox{\qquad\qquad}
    \parbox{40pt}{\begin{fmffile}{PNP-BMBPT1_1_1_4p}
      \begin{fmfgraph*}(40,40)
      \fmfcmd{style_def half_prop expr p =
    draw_plain p;
    shrink(.7);
        cfill (marrow (p, .5))
    endshrink;
	enddef;}
	\fmfcmd{style_def prop_pm expr p =
    draw_plain p;
    shrink(.7);
        cfill (marrow (p, .25));
        cfill (marrow (p, .75))
    endshrink;
enddef;
}
        \fmftop{t1}
        \fmfbottom{bm,b0,b1,b2,b3}
        \fmf{prop_pm,left=0.5,tag=1}{b1,t1}
        \fmf{prop_pm,right=0.5,tag=2}{b1,t1}
        \fmf{half_prop,left}{b2,b1}
        \fmf{half_prop,right}{b2,b1}
        \fmfv{d.shape=circle,d.filled=full,d.size=3thick,l=$\Omega^{02}$,l.angle=0}{t1}
        \fmfv{d.shape=square,d.filled=full,d.size=3thick,l=$O^{22}$,l.dist=0.30w,l.angle=0}{b1}
        \fmfposition
        \fmfipath{p[]}
        \fmfiset{p1}{vpath1(__b1,__t1)}
        \fmfiv{label=$\text{PO1.1.1}^{(4)}$,l.angle=-90,l.dist=0.50w}{point 0 of p1}
        \fmfiv{label=$0$,l.angle=-70,l.dist=0.22w}{point 0 of p1}
        \fmfiv{label=$\tau_1$,l.angle=70,l.dist=0.18w}{point length(p1) of p1}
      \end{fmfgraph*}
    \end{fmffile}}
    \hbox{\qquad\qquad}
    \parbox{40pt}{\begin{fmffile}{PNP-BMBPT1_1_1_5p}
      \begin{fmfgraph*}(40,40)
      \fmfcmd{style_def half_prop expr p =
    draw_plain p;
    shrink(.7);
        cfill (marrow (p, .5))
    endshrink;
	enddef;}
	\fmfcmd{style_def prop_pm expr p =
    draw_plain p;
    shrink(.7);
        cfill (marrow (p, .25));
        cfill (marrow (p, .75))
    endshrink;
enddef;
}
\fmfcmd{style_def prop_mm expr p =
        draw_plain p;
        shrink(.7);
            cfill (marrow (p, .75));
            cfill (marrow (reverse p, .75))
        endshrink;
        enddef;}
        \fmftop{t1}
        \fmfbottom{bm,b0,b1,b2,b3}
        \fmf{prop_pm,left=0.5,tag=1}{b1,t1}
        \fmf{prop_mm,right=0.5,tag=2}{b1,t1}
        \fmf{half_prop,left}{b2,b1}
        \fmf{half_prop,right}{b2,b1}
        \fmfv{d.shape=circle,d.filled=full,d.size=3thick,l=$\Omega^{02}$,l.dist=0.18w,l.angle=20}{t1}
        \fmfv{d.shape=square,d.filled=full,d.size=3thick,l=$O^{13}$,l.dist=0.30w,l.angle=0}{b1}
        \fmfposition
        \fmfipath{p[]}
        \fmfiset{p1}{vpath1(__b1,__t1)}
        \fmfiv{label=$\text{PO1.1.1}^{(5)}$,l.angle=-90,l.dist=0.50w}{point 0 of p1}
        \fmfiv{label=$0$,l.angle=-70,l.dist=0.22w}{point 0 of p1}
        \fmfiv{label=$\tau_1$,l.angle=70,l.dist=0.18w}{point length(p1) of p1}
      \end{fmfgraph*}
    \end{fmffile}}
    \hbox{\qquad\qquad}
    \parbox{40pt}{\begin{fmffile}{PNP-BMBPT1_1_1_6p}
      \begin{fmfgraph*}(40,40)
      \fmfcmd{style_def half_prop expr p =
    draw_plain p;
    shrink(.7);
        cfill (marrow (p, .5))
    endshrink;
	enddef;}
	\fmfcmd{style_def prop_mm expr p =
        draw_plain p;
        shrink(.7);
            cfill (marrow (p, .75));
            cfill (marrow (reverse p, .75))
        endshrink;
        enddef;}
        \fmftop{t1}
        \fmfbottom{bm,b0,b1,b2,b3}
        \fmf{prop_mm,left=0.5,tag=1}{b1,t1}
        \fmf{prop_mm,right=0.5,tag=2}{b1,t1}
        \fmf{half_prop,left}{b2,b1}
        \fmf{half_prop,right}{b2,b1}
        \fmfv{d.shape=circle,d.filled=full,d.size=3thick,l=$\Omega^{02}$,l.dist=0.18w,l.angle=20}{t1}
        \fmfv{d.shape=square,d.filled=full,d.size=3thick,l=$O^{04}$,l.dist=0.30w,l.angle=0}{b1}
        \fmfposition
        \fmfipath{p[]}
        \fmfiset{p1}{vpath1(__b1,__t1)}
        \fmfiv{label=$\text{PO1.1.1}^{(6)}$,l.angle=-90,l.dist=0.50w}{point 0 of p1}
        \fmfiv{label=$0$,l.angle=-70,l.dist=0.22w}{point 0 of p1}
        \fmfiv{label=$\tau_1$,l.angle=70,l.dist=0.18w}{point length(p1) of p1}
      \end{fmfgraph*}
    \end{fmffile}}
    \hbox{\qquad\qquad}
    \parbox{40pt}{\begin{fmffile}{empty40pt}
      \begin{fmfgraph*}(40,40)
      \end{fmfgraph*}
    \end{fmffile}}

    \vspace{4\baselineskip}

    \parbox{40pt}{\begin{fmffile}{empty40pt}
      \begin{fmfgraph*}(40,40)
      \end{fmfgraph*}
    \end{fmffile}}
    \hbox{\qquad\qquad}
    \parbox{40pt}{\begin{fmffile}{PNP-BMBPT1_1_2_1p}
      \begin{fmfgraph*}(40,40)
      \fmfcmd{style_def half_prop expr p =
    draw_plain p;
    shrink(.7);
        cfill (marrow (p, .5))
    endshrink;
	enddef;}
	\fmfcmd{style_def prop_pm expr p =
    draw_plain p;
    shrink(.7);
        cfill (marrow (p, .25));
        cfill (marrow (p, .75))
    endshrink;
enddef;
}
        \fmftop{tm,t0,t1,t2,t3}
        \fmfbottom{b1}
        \fmf{prop_pm,left=0.5,tag=1}{b1,t1}
        \fmf{prop_pm,right=0.5}{b1,t1}
        \fmf{half_prop,left}{t2,t1}
        \fmf{half_prop,right}{t2,t1}
        \fmfv{d.shape=circle,d.filled=full,d.size=3thick,l=$\Omega^{04}$,l.dist=0.35w,l.angle=5}{t1}
        \fmfv{d.shape=square,d.filled=full,d.size=3thick,l=$O^{20}$,l.angle=0}{b1}
        \fmfposition
        \fmfipath{p[]}
        \fmfiset{p1}{vpath1(__b1,__t1)}
        \fmfiv{label=$\text{PO1.1.2}^{(1)}$,l.angle=-90,l.dist=0.50w}{point 0 of p1}
        \fmfiv{label=$0$,l.angle=-70,l.dist=0.22w}{point 0 of p1}
        \fmfiv{label=$\tau_1$,l.angle=90,l.dist=0.18w}{point length(p1) of p1}
      \end{fmfgraph*}
    \end{fmffile}}
    \hbox{\qquad\qquad}
    \parbox{40pt}{\begin{fmffile}{PNP-BMBPT1_1_2_2p}
      \begin{fmfgraph*}(40,40)
      \fmfcmd{style_def half_prop expr p =
    draw_plain p;
    shrink(.7);
        cfill (marrow (p, .5))
    endshrink;
	enddef;}
	\fmfcmd{style_def prop_pm expr p =
    draw_plain p;
    shrink(.7);
        cfill (marrow (p, .25));
        cfill (marrow (p, .75))
    endshrink;
enddef;
}
\fmfcmd{style_def prop_mm expr p =
        draw_plain p;
        shrink(.7);
            cfill (marrow (p, .75));
            cfill (marrow (reverse p, .75))
        endshrink;
        enddef;}
        \fmftop{tm,t0,t1,t2,t3}
        \fmfbottom{b1}
        \fmf{prop_pm,left=0.5,tag=1}{b1,t1}
        \fmf{prop_mm,right=0.5,tag=2}{b1,t1}
        \fmf{half_prop,left}{t2,t1}
        \fmf{half_prop,right}{t2,t1}
        \fmfv{d.shape=circle,d.filled=full,d.size=3thick,l=$\Omega^{04}$,l.dist=0.35w,l.angle=5}{t1}
        \fmfv{d.shape=square,d.filled=full,d.size=3thick,l=$O^{11}$,l.angle=0}{b1}
        \fmfposition
        \fmfipath{p[]}
        \fmfiset{p1}{vpath1(__b1,__t1)}
        \fmfiv{label=$\text{PO1.1.2}^{(2)}$,l.angle=-90,l.dist=0.50w}{point 0 of p1}
        \fmfiv{label=$0$,l.angle=-70,l.dist=0.22w}{point 0 of p1}
        \fmfiv{label=$\tau_1$,l.angle=90,l.dist=0.18w}{point length(p1) of p1}
      \end{fmfgraph*}
    \end{fmffile}}
    \hbox{\qquad\qquad}
    \parbox{40pt}{\begin{fmffile}{PNP-BMBPT1_1_2_3p}
      \begin{fmfgraph*}(40,40)
      \fmfcmd{style_def half_prop expr p =
    draw_plain p;
    shrink(.7);
        cfill (marrow (p, .5))
    endshrink;
	enddef;}
	\fmfcmd{style_def prop_mm expr p =
        draw_plain p;
        shrink(.7);
            cfill (marrow (p, .75));
            cfill (marrow (reverse p, .75))
        endshrink;
        enddef;}
        \fmftop{tm,t0,t1,t2,t3}
        \fmfbottom{b1}
        \fmf{prop_mm,left=0.5,tag=1}{b1,t1}
        \fmf{prop_mm,right=0.5,tag=2}{b1,t1}
        \fmf{half_prop,left}{t2,t1}
        \fmf{half_prop,right}{t2,t1}
        \fmfv{d.shape=circle,d.filled=full,d.size=3thick,l=$\Omega^{04}$,l.dist=0.35w,l.angle=5}{t1}
        \fmfv{d.shape=square,d.filled=full,d.size=3thick,l=$O^{02}$,l.angle=0}{b1}
        \fmfposition
        \fmfipath{p[]}
        \fmfiset{p1}{vpath1(__b1,__t1)}
        \fmfiv{label=$\text{PO1.1.2}^{(3)}$,l.angle=-90,l.dist=0.50w}{point 0 of p1}
        \fmfiv{label=$0$,l.angle=-70,l.dist=0.22w}{point 0 of p1}
        \fmfiv{label=$\tau_1$,l.angle=90,l.dist=0.18w}{point length(p1) of p1}
      \end{fmfgraph*}
    \end{fmffile}}
    \hbox{\qquad\qquad}
    \parbox{40pt}{\begin{fmffile}{empty40pt}
      \begin{fmfgraph*}(40,40)
      \end{fmfgraph*}
    \end{fmffile}}

    \vspace{4\baselineskip}

    \parbox{40pt}{\begin{fmffile}{empty40pt}
      \begin{fmfgraph*}(40,40)
      \end{fmfgraph*}
    \end{fmffile}}
    \hbox{\qquad\qquad}
    \parbox{40pt}{\begin{fmffile}{empty40pt}
      \begin{fmfgraph*}(40,40)
      \end{fmfgraph*}
    \end{fmffile}}
    \hbox{\qquad\qquad}
    \parbox{40pt}{\begin{fmffile}{PNP-BMBPT1_1_2_4p}
      \begin{fmfgraph*}(40,40)
      \fmfcmd{style_def half_prop expr p =
    draw_plain p;
    shrink(.7);
        cfill (marrow (p, .5))
    endshrink;
	enddef;}
	\fmfcmd{style_def prop_pm expr p =
    draw_plain p;
    shrink(.7);
        cfill (marrow (p, .25));
        cfill (marrow (p, .75))
    endshrink;
enddef;
}
        \fmftop{tm,t0,t1,t2,t3}
        \fmfbottom{bm,b0,b1,b2,b3}
        \fmf{prop_pm,left=0.5,tag=1}{b1,t1}
        \fmf{prop_pm,right=0.5,tag=2}{b1,t1}
        \fmf{half_prop,left}{b2,b1}
        \fmf{half_prop,right}{b2,b1}
        \fmf{half_prop,left}{t2,t1}
        \fmf{half_prop,right}{t2,t1}
        \fmfv{d.shape=circle,d.filled=full,d.size=3thick,l=$\Omega^{04}$,l.dist=0.35w,l.angle=5}{t1}
        \fmfv{d.shape=square,d.filled=full,d.size=3thick,l=$O^{22}$,l.dist=0.30w,l.angle=0}{b1}
        \fmfposition
        \fmfipath{p[]}
        \fmfiset{p1}{vpath1(__b1,__t1)}
        \fmfiv{label=$\text{PO1.1.2}^{(4)}$,l.angle=-90,l.dist=0.50w}{point 0 of p1}
        \fmfiv{label=$0$,l.angle=-70,l.dist=0.22w}{point 0 of p1}
        \fmfiv{label=$\tau_1$,l.angle=90,l.dist=0.18w}{point length(p1) of p1}
      \end{fmfgraph*}
    \end{fmffile}}
    \hbox{\qquad\qquad}
    \parbox{40pt}{\begin{fmffile}{PNP-BMBPT1_1_2_5p}
      \begin{fmfgraph*}(40,40)
      \fmfcmd{style_def half_prop expr p =
    draw_plain p;
    shrink(.7);
        cfill (marrow (p, .5))
    endshrink;
	enddef;}
	\fmfcmd{style_def prop_pm expr p =
    draw_plain p;
    shrink(.7);
        cfill (marrow (p, .25));
        cfill (marrow (p, .75))
    endshrink;
enddef;
}
\fmfcmd{style_def prop_mm expr p =
        draw_plain p;
        shrink(.7);
            cfill (marrow (p, .75));
            cfill (marrow (reverse p, .75))
        endshrink;
        enddef;}
        \fmftop{tm,t0,t1,t2,t3}
        \fmfbottom{bm,b0,b1,b2,b3}
        \fmf{prop_pm,left=0.5,tag=1}{b1,t1}
        \fmf{prop_mm,right=0.5,tag=2}{b1,t1}
        \fmf{half_prop,left}{b2,b1}
        \fmf{half_prop,right}{b2,b1}
        \fmf{half_prop,left}{t2,t1}
        \fmf{half_prop,right}{t2,t1}
        \fmfv{d.shape=circle,d.filled=full,d.size=3thick,l=$\Omega^{04}$,l.dist=0.35w,l.angle=5}{t1}
        \fmfv{d.shape=square,d.filled=full,d.size=3thick,l=$O^{13}$,l.dist=0.30w,l.angle=0}{b1}
        \fmfposition
        \fmfipath{p[]}
        \fmfiset{p1}{vpath1(__b1,__t1)}
        \fmfiv{label=$\text{PO1.1.2}^{(5)}$,l.angle=-90,l.dist=0.50w}{point 0 of p1}
        \fmfiv{label=$0$,l.angle=-70,l.dist=0.22w}{point 0 of p1}
        \fmfiv{label=$\tau_1$,l.angle=90,l.dist=0.18w}{point length(p1) of p1}
      \end{fmfgraph*}
    \end{fmffile}}
    \hbox{\qquad\qquad}
    \parbox{40pt}{\begin{fmffile}{PNP-BMBPT1_1_2_6p}
      \begin{fmfgraph*}(40,40)
      \fmfcmd{style_def half_prop expr p =
    draw_plain p;
    shrink(.7);
        cfill (marrow (p, .5))
    endshrink;
	enddef;}
	\fmfcmd{style_def prop_mm expr p =
        draw_plain p;
        shrink(.7);
            cfill (marrow (p, .75));
            cfill (marrow (reverse p, .75))
        endshrink;
        enddef;}
        \fmftop{tm,t0,t1,t2,t3}
        \fmfbottom{bm,b0,b1,b2,b3}
        \fmf{prop_mm,left=0.5,tag=1}{b1,t1}
        \fmf{prop_mm,right=0.5,tag=2}{b1,t1}
        \fmf{half_prop,left}{b2,b1}
        \fmf{half_prop,right}{b2,b1}
        \fmf{half_prop,left}{t2,t1}
        \fmf{half_prop,right}{t2,t1}
        \fmfv{d.shape=circle,d.filled=full,d.size=3thick,l=$\Omega^{04}$,l.dist=0.35w,l.angle=5}{t1}
        \fmfv{d.shape=square,d.filled=full,d.size=3thick,l=$O^{04}$,l.dist=0.30w,l.angle=0}{b1}
        \fmfposition
        \fmfipath{p[]}
        \fmfiset{p1}{vpath1(__b1,__t1)}
        \fmfiv{label=$\text{PO1.1.2}^{(6)}$,l.angle=-90,l.dist=0.50w}{point 0 of p1}
        \fmfiv{label=$0$,l.angle=-70,l.dist=0.22w}{point 0 of p1}
        \fmfiv{label=$\tau_1$,l.angle=90,l.dist=0.18w}{point length(p1) of p1}
      \end{fmfgraph*}
    \end{fmffile}}

    \vspace{4\baselineskip}

    \parbox{40pt}{\begin{fmffile}{PNP-BMBPT1_2_1_1p}
      \begin{fmfgraph*}(40,40)
      \fmfcmd{style_def prop_pm expr p =
    draw_plain p;
    shrink(.7);
        cfill (marrow (p, .25));
        cfill (marrow (p, .75))
    endshrink;
enddef;
}
        \fmftop{t1}
        \fmfbottom{b1}
        \fmf{prop_pm,left=0.8,tag=1}{b1,t1}
        \fmf{prop_pm,left=0.3,tag=2}{b1,t1}
        \fmf{prop_pm,right=0.3,tag=3}{b1,t1}
        \fmf{prop_pm,right=0.8,tag=4}{b1,t1}
        \fmfv{d.shape=circle,d.filled=full,d.size=3thick,l=$\Omega^{04}$,l.dist=0.18w,l.angle=22}{t1}
        \fmfv{d.shape=square,d.filled=full,d.size=3thick,l=$O^{40}$,l.angle=-10}{b1}
        \fmfposition
        \fmfipath{p[]}
        \fmfiset{p1}{vpath1(__b1,__t1)}
        \fmfiv{label=$\text{PO1.2.1}^{(1)}$,l.angle=-90,l.dist=0.50w}{point 0 of p1}
        \fmfiv{label=$0$,l.angle=-70,l.dist=0.19w}{point 0 of p1}
        \fmfiv{label=$\tau_1$,l.angle=70,l.dist=0.15w}{point length(p1) of p1}
      \end{fmfgraph*}
    \end{fmffile}}
    \hbox{\qquad\qquad}
    \parbox{40pt}{\begin{fmffile}{PNP-BMBPT1_2_1_2p}
      \begin{fmfgraph*}(40,40)
      \fmfcmd{style_def prop_pm expr p =
    draw_plain p;
    shrink(.7);
        cfill (marrow (p, .25));
        cfill (marrow (p, .75))
    endshrink;
enddef;
}
\fmfcmd{style_def prop_mm expr p =
        draw_plain p;
        shrink(.7);
            cfill (marrow (p, .75));
            cfill (marrow (reverse p, .75))
        endshrink;
        enddef;}
        \fmftop{t1}
        \fmfbottom{b1}
        \fmf{prop_pm,left=0.8,tag=1}{b1,t1}
        \fmf{prop_pm,left=0.3,tag=2}{b1,t1}
        \fmf{prop_pm,right=0.3,tag=3}{b1,t1}
        \fmf{prop_mm,right=0.8,tag=4}{b1,t1}
        \fmfv{d.shape=circle,d.filled=full,d.size=3thick,l=$\Omega^{04}$,l.dist=0.18w,l.angle=22}{t1}
        \fmfv{d.shape=square,d.filled=full,d.size=3thick,l=$O^{31}$,l.angle=-10}{b1}
        \fmfposition
        \fmfipath{p[]}
        \fmfiset{p1}{vpath1(__b1,__t1)}
        \fmfiv{label=$\text{PO1.2.1}^{(2)}$,l.angle=-90,l.dist=0.50w}{point 0 of p1}
        \fmfiv{label=$0$,l.angle=-70,l.dist=0.19w}{point 0 of p1}
        \fmfiv{label=$\tau_1$,l.angle=70,l.dist=0.15w}{point length(p1) of p1}
      \end{fmfgraph*}
    \end{fmffile}}
    \hbox{\qquad\qquad}
    \parbox{40pt}{\begin{fmffile}{PNP-BMBPT1_2_1_3p}
      \begin{fmfgraph*}(40,40)
      \fmfcmd{style_def prop_pm expr p =
    draw_plain p;
    shrink(.7);
        cfill (marrow (p, .25));
        cfill (marrow (p, .75))
    endshrink;
enddef;
}
\fmfcmd{style_def prop_mm expr p =
        draw_plain p;
        shrink(.7);
            cfill (marrow (p, .75));
            cfill (marrow (reverse p, .75))
        endshrink;
        enddef;}
        \fmftop{t1}
        \fmfbottom{b1}
        \fmf{prop_pm,left=0.8,tag=1}{b1,t1}
        \fmf{prop_pm,left=0.3,tag=2}{b1,t1}
        \fmf{prop_mm,right=0.3,tag=3}{b1,t1}
        \fmf{prop_mm,right=0.8,tag=4}{b1,t1}
        \fmfv{d.shape=circle,d.filled=full,d.size=3thick,l=$\Omega^{04}$,l.dist=0.18w,l.angle=22}{t1}
        \fmfv{d.shape=square,d.filled=full,d.size=3thick,l=$O^{22}$,l.angle=-10}{b1}
        \fmfposition
        \fmfipath{p[]}
        \fmfiset{p1}{vpath1(__b1,__t1)}
        \fmfiv{label=$\text{PO1.2.1}^{(3)}$,l.angle=-90,l.dist=0.50w}{point 0 of p1}
        \fmfiv{label=$0$,l.angle=-70,l.dist=0.19w}{point 0 of p1}
        \fmfiv{label=$\tau_1$,l.angle=70,l.dist=0.15w}{point length(p1) of p1}
      \end{fmfgraph*}
    \end{fmffile}}
    \hbox{\qquad\qquad}
    \parbox{40pt}{\begin{fmffile}{PNP-BMBPT1_2_1_4p}
      \begin{fmfgraph*}(40,40)
      \fmfcmd{style_def prop_pm expr p =
    draw_plain p;
    shrink(.7);
        cfill (marrow (p, .25));
        cfill (marrow (p, .75))
    endshrink;
enddef;
}
\fmfcmd{style_def prop_mm expr p =
        draw_plain p;
        shrink(.7);
            cfill (marrow (p, .75));
            cfill (marrow (reverse p, .75))
        endshrink;
        enddef;}
        \fmftop{t1}
        \fmfbottom{b1}
        \fmf{prop_pm,left=0.8,tag=1}{b1,t1}
        \fmf{prop_mm,left=0.3,tag=2}{b1,t1}
        \fmf{prop_mm,right=0.3,tag=3}{b1,t1}
        \fmf{prop_mm,right=0.8,tag=4}{b1,t1}
        \fmfv{d.shape=circle,d.filled=full,d.size=3thick,l=$\Omega^{04}$,l.dist=0.18w,l.angle=22}{t1}
        \fmfv{d.shape=square,d.filled=full,d.size=3thick,l=$O^{13}$,l.angle=-10}{b1}
        \fmfposition
        \fmfipath{p[]}
        \fmfiset{p1}{vpath1(__b1,__t1)}
        \fmfiv{label=$\text{PO1.2.1}^{(4)}$,l.angle=-90,l.dist=0.50w}{point 0 of p1}
        \fmfiv{label=$0$,l.angle=-70,l.dist=0.19w}{point 0 of p1}
        \fmfiv{label=$\tau_1$,l.angle=70,l.dist=0.15w}{point length(p1) of p1}
      \end{fmfgraph*}
    \end{fmffile}}
    \hbox{\qquad\qquad}
    \parbox{40pt}{\begin{fmffile}{PNP-BMBPT1_2_1_5p}
      \begin{fmfgraph*}(40,40)
      \fmfcmd{style_def prop_mm expr p =
        draw_plain p;
        shrink(.7);
            cfill (marrow (p, .75));
            cfill (marrow (reverse p, .75))
        endshrink;
        enddef;}
        \fmftop{t1}
        \fmfbottom{b1}
        \fmf{prop_mm,left=0.8,tag=1}{b1,t1}
        \fmf{prop_mm,left=0.3,tag=2}{b1,t1}
        \fmf{prop_mm,right=0.3,tag=3}{b1,t1}
        \fmf{prop_mm,right=0.8,tag=4}{b1,t1}
        \fmfv{d.shape=circle,d.filled=full,d.size=3thick,l=$\Omega^{04}$,l.dist=0.18w,l.angle=22}{t1}
        \fmfv{d.shape=square,d.filled=full,d.size=3thick,l=$O^{04}$,l.angle=-10}{b1}
        \fmfposition
        \fmfipath{p[]}
        \fmfiset{p1}{vpath1(__b1,__t1)}
        \fmfiv{label=$\text{PO1.2.1}^{(5)}$,l.angle=-90,l.dist=0.50w}{point 0 of p1}
        \fmfiv{label=$0$,l.angle=-70,l.dist=0.19w}{point 0 of p1}
        \fmfiv{label=$\tau_1$,l.angle=70,l.dist=0.15w}{point length(p1) of p1}
      \end{fmfgraph*}
    \end{fmffile}}
    \vspace{1.5\baselineskip}
  \end{center}
  \caption{\label{offdiagBMBPT01}
  Zero- and first-order \emph{off-diagonal} Feynman BMBPT diagrams. Diagrams are grouped vertically according to the number $n_a$ of anomalous lines they contain.}
\end{figure*}
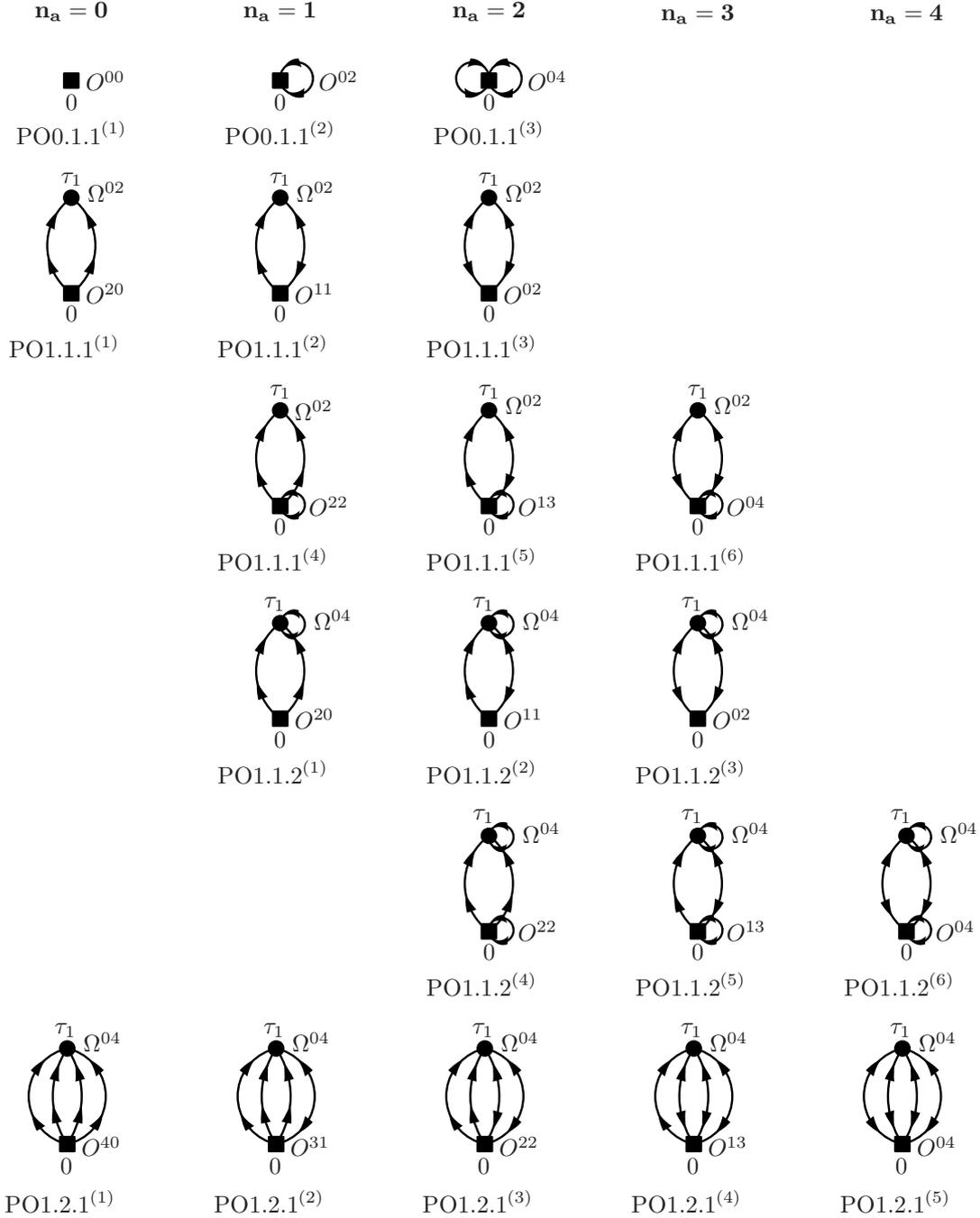

\subsection{Towards higher orders}

Off-diagonal BMBPT diagrams of order $p=0$ and $1$ have been generated and evaluated manually for two-body operators, i.e., operators summing $O^{[k]}$ with $k\leq 4$~\cite{Duguet:2015yle}. The twenty corresponding diagrams are displayed in Fig.~\ref{offdiagBMBPT01} for illustration. Among these twenty diagrams, only the three diagrams in the first column ($n_a=0$) remain in straight, i.e.,\@ diagonal, BMBPT that has been dealt with in the first version of the \textbf{\texttt{ADG}} code~\cite{arthuis18a}.

While it was already challenging to automatically generate and evaluate diagonal BMBPT diagrams of arbitrary orders and topologies, off-diagonal BMBPT obviously reaches yet another level of complexity related to the proliferation of diagrams, itself increasing with the perturbative order, associated with the possibility to form off-diagonal propagators. In this context, the step accomplished in Ref.~\cite{arthuis18a} will however happen to be of tremendous help to achieve the automatic generation and evaluation of the off-diagonal BMBPT diagrams.

\section{Generation of off-diagonal BMBPT diagrams}
\label{secautomaticgeneration}

The automated generation of diagonal BMBPT Feynman diagrams via elements of graph theory was explained at length in Ref.~\cite{arthuis18a}. We do not repeat it here and refer the reader to Ref.~\cite{arthuis18a} for details. As a matter of fact, the strategy presently employed is not to follow a similar method to generate off-diagonal diagrams from scratch but rather to take advantage of having already done so for the diagonal ones, i.e.,\@ to start from the order $p$ diagonal BMBPT diagrams to systematically produce their off-diagonal partners.

\subsection{Basic analysis}

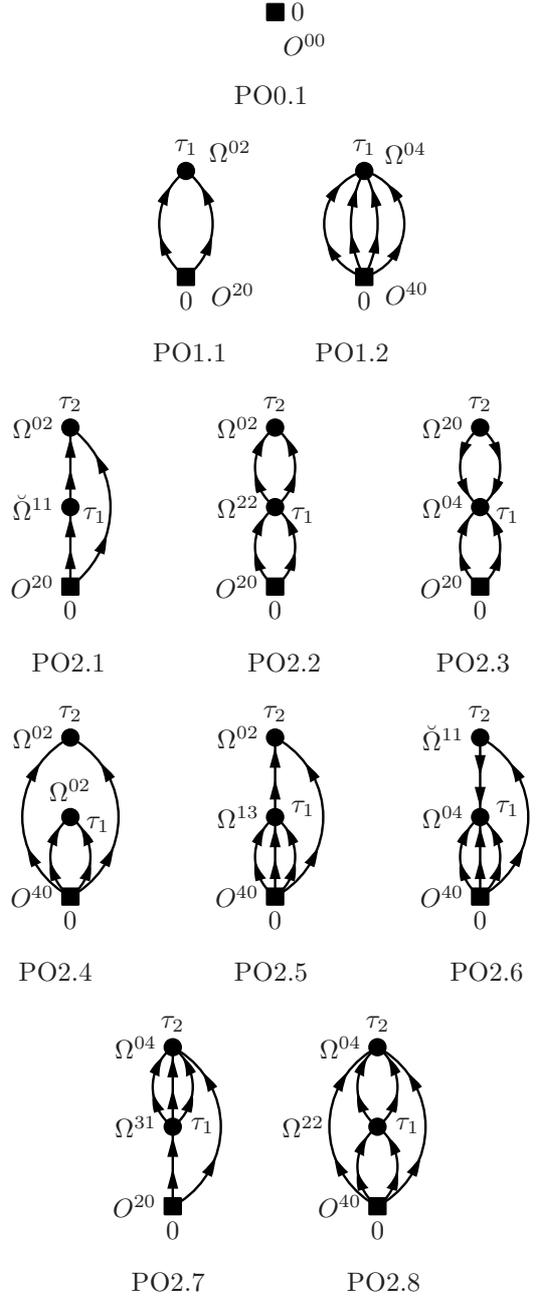
\begin{figure}[t!]
\begin{center}
\parbox{30pt}{\begin{fmffile}{PO0_1}
    \begin{fmfgraph*}(30,30)
    \fmftop{t1} \fmfbottom{b1}
    \fmf{phantom,tag=1}{b1,i1}
    \fmf{phantom}{i1,t1}
    \fmfv{d.shape=square,d.filled=full,d.size=3thick}{i1}
    \fmflabel{$0$}{i1}
    \fmfposition
    \fmfipath{p[]}
    \fmfiset{p1}{vpath1(__b1,__i1)}
    \fmfiv{label=$ O^{00}$,l.angle=-60}{point length(p1) of p1}
    \end{fmfgraph*}
    \end{fmffile}}

\vspace{\baselineskip}
PO0.1
\vspace{2\baselineskip}

\parbox{40pt}{\begin{fmffile}{PO1_1}
    \begin{fmfgraph*}(40,40)
    \fmfcmd{style_def prop_pm expr p =
    draw_plain p;
    shrink(.7);
        cfill (marrow (p, .25));
        cfill (marrow (p, .75))
    endshrink;
enddef;
}
    \fmftop{t1} \fmfbottom{b1}
    \fmf{prop_pm,left=0.5,tag=1}{b1,t1}
    \fmf{prop_pm,right=0.5,tag=2}{b1,t1}
    \fmfv{d.shape=square,d.filled=full,d.size=3thick}{b1}
    \fmfv{d.shape=circle,d.filled=full,d.size=3thick}{t1}
    \fmflabel{$0$}{b1}
    \fmflabel{$\tau_1$}{t1}
    \fmfposition
    \fmfipath{p[]}
    \fmfiset{p1}{vpath1(__b1,__t1)}
    \fmfiset{p2}{vpath2(__b1,__t1)}
    \fmfiv{label=$\ O^{20}$,l.angle=-60}{point 0 of p2}
    \fmfiv{label=$\ \Omega^{02}$,l.angle=60}{point length(p2) of p2}
    \end{fmfgraph*}
    \end{fmffile}}
\hbox{\quad\quad}
\parbox{40pt}{\begin{fmffile}{PO1_2}
    \begin{fmfgraph*}(40,40)
    \fmfcmd{style_def prop_pm expr p =
    draw_plain p;
    shrink(.7);
        cfill (marrow (p, .25));
        cfill (marrow (p, .75))
    endshrink;
enddef;
}
    \fmftop{t1} \fmfbottom{b1}
    \fmf{prop_pm,left=0.75,tag=1}{b1,t1}
    \fmf{prop_pm,right=0.75,tag=2}{b1,t1}
    \fmf{prop_pm,left=0.25,tag=3}{b1,t1}
    \fmf{prop_pm,right=0.25,tag=4}{b1,t1}
    \fmfv{d.shape=square,d.filled=full,d.size=3thick}{b1}
    \fmfv{d.shape=circle,d.filled=full,d.size=3thick}{t1}
    \fmflabel{$0$}{b1}
    \fmflabel{$\tau_1$}{t1}
    \fmfposition
    \fmfipath{p[]}
    \fmfiset{p1}{vpath1(__b1,__t1)}
    \fmfiset{p2}{vpath2(__b1,__t1)}
    \fmfiset{p3}{vpath3(__b1,__t1)}
    \fmfiset{p4}{vpath4(__b1,__t1)}
    \fmfiv{label=$\ O^{40}$,l.angle=-60}{point 0 of p4}
    \fmfiv{label=$\ \Omega^{04}$,l.angle=60}{point length(p4) of p4}
    \end{fmfgraph*}
    \end{fmffile}}

\vspace{2\baselineskip}
PO1.1 \quad\quad \quad PO1.2
\vspace{2\baselineskip}

\parbox{60pt}{\begin{fmffile}{PO2_1}
    \begin{fmfgraph*}(60,60)
    \fmfcmd{style_def prop_pm expr p =
    draw_plain p;
    shrink(.7);
        cfill (marrow (p, .25));
        cfill (marrow (p, .75))
    endshrink;
enddef;
}
    \fmftop{t1} \fmfbottom{b1}
    \fmf{prop_pm,tag=1}{b1,i1}
    \fmf{prop_pm,tag=2}{i1,t1}
    \fmffreeze
    \fmf{prop_pm,right=0.5,tag=3}{b1,t1}
    \fmfv{d.shape=square,d.filled=full,d.size=3thick,l=$O^{20}$,l.angle=180}{b1}
    \fmfv{d.shape=circle,d.filled=full,d.size=3thick,l=$\breve{\Omega}^{11}$,l.angle=180}{i1}
    \fmfv{d.shape=circle,d.filled=full,d.size=3thick,l=$\Omega^{02}$,l.angle=180}{t1}
    \fmfposition
    \fmfipath{p[]}
    \fmfiset{p1}{vpath1(__b1,__i1)}
    \fmfiset{p2}{vpath2(__i1,__t1)}
    \fmfiset{p3}{vpath3(__b1,__t1)}
    \fmfiv{label=$0$,l.angle=-90,l.dist=0.15w}{point 0 of p1}
    \fmfiv{label=$\tau_1$,l.angle=0,l.dist=0.08w}{point length(p1) of p1}
    \fmfiv{label=$\tau_2$,l.angle=90,l.dist=0.15w}{point length(p2) of p2}
    \end{fmfgraph*}
    \end{fmffile}}
\hbox{\quad}
\parbox{60pt}{\begin{fmffile}{PO2_2}
    \begin{fmfgraph*}(60,60)
    \fmfcmd{style_def prop_pm expr p =
    draw_plain p;
    shrink(.7);
        cfill (marrow (p, .25));
        cfill (marrow (p, .75))
    endshrink;
enddef;
}
    \fmftop{t1} \fmfbottom{b1}
    \fmf{prop_pm,left=0.5,tag=1}{b1,i1}
    \fmf{prop_pm,right=0.5,tag=2}{b1,i1}
    \fmf{prop_pm,left=0.5,tag=3}{i1,t1}
    \fmf{prop_pm,right=0.5,tag=4}{i1,t1}
    \fmf{phantom,tag=5}{i1,t1}
    \fmf{phantom,tag=6}{b1,i1}
    \fmfv{d.shape=square,d.filled=full,d.size=3thick,l=$O^{20}$,l.angle=180}{b1}
    \fmfv{d.shape=circle,d.filled=full,d.size=3thick,l=$\Omega^{22}$,l.angle=180}{i1}
    \fmfv{d.shape=circle,d.filled=full,d.size=3thick,l=$\Omega^{02}$,l.angle=180}{t1}
    \fmfposition
    \fmfipath{p[]}
    \fmfiset{p1}{vpath1(__b1,__i1)}
    \fmfiset{p2}{vpath2(__b1,__i1)}
    \fmfiset{p3}{vpath3(__i1,__t1)}
    \fmfiset{p4}{vpath4(__i1,__t1)}
    \fmfiset{p5}{vpath5(__i1,__t1)}
    \fmfiset{p6}{vpath6(__b1,__i1)}
    \fmfiv{label=$0$,l.angle=-90,l.dist=0.15w}{point 0 of p6}
    \fmfiv{label=$\tau_1$,l.angle=0,l.dist=0.1w}{point length(p6) of p6}
    \fmfiv{label=$\tau_2$,l.angle=90,l.dist=0.15w}{point length(p5) of p5}
    \end{fmfgraph*}
    \end{fmffile}}
\hbox{\quad}
\parbox{60pt}{\begin{fmffile}{PO2_3}
    \begin{fmfgraph*}(60,60)
    \fmfcmd{style_def prop_pm expr p =
    draw_plain p;
    shrink(.7);
        cfill (marrow (p, .25));
        cfill (marrow (p, .75))
    endshrink;
enddef;
}
\fmfcmd{style_def prop_mp expr p =
    draw_plain p;
    shrink(.7);
        cfill (marrow (reverse p, .25));
        cfill (marrow (reverse p, .75))
    endshrink;
enddef;
}
    \fmftop{t1} \fmfbottom{b1}
    \fmf{prop_pm,left=0.5,tag=1}{b1,i1}
    \fmf{prop_pm,right=0.5,tag=2}{b1,i1}
    \fmf{prop_mp,left=0.5,tag=3}{i1,t1}
    \fmf{prop_mp,right=0.5,tag=4}{i1,t1}
    \fmf{phantom,tag=5}{i1,t1}
    \fmf{phantom,tag=6}{b1,i1}
    \fmfv{d.shape=square,d.filled=full,d.size=3thick,l=$O^{20}$,l.angle=180}{b1}
    \fmfv{d.shape=circle,d.filled=full,d.size=3thick,l=$\Omega^{04}$,l.angle=180}{i1}
    \fmfv{d.shape=circle,d.filled=full,d.size=3thick,l=$\Omega^{20}$,l.angle=180}{t1}
    \fmfposition
    \fmfipath{p[]}
    \fmfiset{p1}{vpath1(__b1,__i1)}
    \fmfiset{p2}{vpath2(__b1,__i1)}
    \fmfiset{p3}{vpath3(__i1,__t1)}
    \fmfiset{p4}{vpath4(__i1,__t1)}
    \fmfiset{p5}{vpath5(__i1,__t1)}
    \fmfiset{p6}{vpath6(__b1,__i1)}
    \fmfiv{label=$0$,l.angle=-90,l.dist=0.15w}{point 0 of p6}
    \fmfiv{label=$\tau_1$,l.angle=0,l.dist=0.1w}{point length(p6) of p6}
    \fmfiv{label=$\tau_2$,l.angle=90,l.dist=0.15w}{point length(p5) of p5}
    \end{fmfgraph*}
    \end{fmffile}}

\vspace{2\baselineskip}
PO2.1 \quad\quad\quad\quad\quad PO2.2 \quad\quad\quad\quad PO2.3
\vspace{2\baselineskip}

\parbox{60pt}{\begin{fmffile}{PO2_4}
    \begin{fmfgraph*}(60,60)
    \fmfcmd{style_def prop_pm expr p =
    draw_plain p;
    shrink(.7);
        cfill (marrow (p, .25));
        cfill (marrow (p, .75))
    endshrink;
enddef;
}
    \fmftop{t1} \fmfbottom{b1}
    \fmf{phantom,tag=5}{i1,t1}
    \fmf{phantom,tag=6}{i1,b1}
    \fmffreeze
    \fmf{prop_pm,left=0.5,tag=1}{b1,i1}
    \fmf{prop_pm,right=0.5,tag=2}{b1,i1}
    \fmf{prop_pm,left=0.6,tag=3}{b1,t1}
    \fmf{prop_pm,right=0.6,tag=4}{b1,t1}
    \fmfv{d.shape=square,d.filled=full,d.size=3thick,l=$O^{40}$,l.angle=180}{b1}
    \fmfv{d.shape=circle,d.filled=full,d.size=3thick,l=$\Omega^{02}$,l.angle=90}{i1}
    \fmfv{d.shape=circle,d.filled=full,d.size=3thick,l=$\Omega^{02}$,l.angle=180}{t1}
    \fmfposition
    \fmfipath{p[]}
    \fmfiset{p1}{vpath1(__b1,__i1)}
    \fmfiset{p2}{vpath2(__b1,__i1)}
    \fmfiset{p3}{vpath3(__b1,__t1)}
    \fmfiset{p4}{vpath4(__b1,__t1)}
    \fmfiset{p5}{vpath5(__i1,__t1)}
    \fmfiset{p6}{vpath6(__i1,__b1)}
    \fmfiv{label=$0$,l.angle=-90,l.dist=0.15w}{point length(p6) of p6}
    \fmfiv{label=$\tau_1$,l.angle=0,l.dist=0.1w}{point 0 of p6}
    \fmfiv{label=$\tau_2$,l.angle=90,l.dist=0.15w}{point length(p5) of p5}
    \end{fmfgraph*}
    \end{fmffile}}
\hbox{\quad}
\parbox{60pt}{\begin{fmffile}{PO2_5}
    \begin{fmfgraph*}(60,60)
    \fmfcmd{style_def prop_pm expr p =
    draw_plain p;
    shrink(.7);
        cfill (marrow (p, .25));
        cfill (marrow (p, .75))
    endshrink;
enddef;
}
    \fmftop{t1} \fmfbottom{b1}
    \fmf{prop_pm,tag=1}{b1,i1}
    \fmf{prop_pm,tag=2}{i1,t1}
    \fmffreeze
    \fmf{prop_pm,left=0.5,tag=3}{b1,i1}
    \fmf{prop_pm,right=0.5,tag=4}{b1,i1}
    \fmf{prop_pm,right=0.6,tag=5}{b1,t1}
    \fmfv{d.shape=square,d.filled=full,d.size=3thick,l=$O^{40}$,l.angle=180}{b1}
    \fmfv{d.shape=circle,d.filled=full,d.size=3thick,l=$\Omega^{13}$,l.angle=180}{i1}
    \fmfv{d.shape=circle,d.filled=full,d.size=3thick,l=$\Omega^{02}$,l.angle=180}{t1}
    \fmfposition
    \fmfipath{p[]}
    \fmfiset{p1}{vpath1(__b1,__i1)}
    \fmfiset{p2}{vpath2(__i1,__t1)}
    \fmfiset{p3}{vpath3(__b1,__i1)}
    \fmfiset{p4}{vpath4(__b1,__i1)}
    \fmfiset{p5}{vpath5(__b1,__t1)}
    \fmfiv{label=$0$,l.angle=-90,l.dist=0.15w}{point 0 of p1}
    \fmfiv{label=$\tau_1$,l.angle=0,l.dist=0.1w}{point 0 of p2}
    \fmfiv{label=$\tau_2$,l.angle=90,l.dist=0.15w}{point length(p2) of p2}
    \end{fmfgraph*}
    \end{fmffile}}
\hbox{\quad}
\parbox{60pt}{\begin{fmffile}{PO2_6}
    \begin{fmfgraph*}(60,60)
    \fmfcmd{style_def prop_pm expr p =
    draw_plain p;
    shrink(.7);
        cfill (marrow (p, .25));
        cfill (marrow (p, .75))
    endshrink;
enddef;
}
\fmfcmd{style_def prop_mp expr p =
    draw_plain p;
    shrink(.7);
        cfill (marrow (reverse p, .25));
        cfill (marrow (reverse p, .75))
    endshrink;
enddef;
}
    \fmftop{t1} \fmfbottom{b1}
    \fmf{prop_pm,tag=1}{b1,i1}
    \fmf{prop_mp,tag=2}{i1,t1}
    \fmffreeze
    \fmf{prop_pm,left=0.5,tag=3}{b1,i1}
    \fmf{prop_pm,right=0.5,tag=4}{b1,i1}
    \fmf{prop_pm,right=0.6,tag=5}{b1,t1}
    \fmfv{d.shape=square,d.filled=full,d.size=3thick,l=$O^{40}$,l.angle=180}{b1}
    \fmfv{d.shape=circle,d.filled=full,d.size=3thick,l=$\Omega^{04}$,l.angle=180}{i1}
    \fmfv{d.shape=circle,d.filled=full,d.size=3thick,l=$\breve{\Omega}^{11}$,l.angle=180}{t1}
    \fmfposition
    \fmfipath{p[]}
    \fmfiset{p1}{vpath1(__b1,__i1)}
    \fmfiset{p2}{vpath2(__i1,__t1)}
    \fmfiset{p3}{vpath3(__b1,__i1)}
    \fmfiset{p4}{vpath4(__b1,__i1)}
    \fmfiset{p5}{vpath5(__b1,__t1)}
    \fmfiv{label=$0$,l.angle=-90,l.dist=0.15w}{point 0 of p1}
    \fmfiv{label=$\tau_1$,l.angle=0,l.dist=0.1w}{point 0 of p2}
    \fmfiv{label=$\tau_2$,l.angle=90,l.dist=0.15w}{point length(p2) of p2}
    \end{fmfgraph*}
    \end{fmffile}}

\vspace{2\baselineskip}
PO2.4 \quad\quad\quad \quad\quad PO2.5 \quad\quad\quad \quad\quad PO2.6
\vspace{2\baselineskip}

\parbox{60pt}{\begin{fmffile}{PO2_7}
    \begin{fmfgraph*}(60,60)
    \fmfcmd{style_def prop_pm expr p =
    draw_plain p;
    shrink(.7);
        cfill (marrow (p, .25));
        cfill (marrow (p, .75))
    endshrink;
enddef;
}
    \fmftop{t1} \fmfbottom{b1}
    \fmf{prop_pm,tag=1}{b1,i1}
    \fmf{prop_pm,tag=2}{i1,t1}
    \fmffreeze
    \fmf{prop_pm,left=0.5,tag=3}{i1,t1}
    \fmf{prop_pm,right=0.5,tag=4}{i1,t1}
    \fmf{prop_pm,right=0.6,tag=5}{b1,t1}
    \fmfv{d.shape=square,d.filled=full,d.size=3thick,l=$O^{20}$,l.angle=180}{b1}
    \fmfv{d.shape=circle,d.filled=full,d.size=3thick,l=$\Omega^{31}$,l.angle=180}{i1}
    \fmfv{d.shape=circle,d.filled=full,d.size=3thick,l=$\Omega^{04}$,l.angle=180}{t1}
    \fmfposition
    \fmfipath{p[]}
    \fmfiset{p1}{vpath1(__b1,__i1)}
    \fmfiset{p2}{vpath2(__i1,__t1)}
    \fmfiset{p3}{vpath3(__i1,__t1)}
    \fmfiset{p4}{vpath4(__i1,__t1)}
    \fmfiset{p5}{vpath5(__b1,__t1)}
    \fmfiv{label=$0$,l.angle=-90,l.dist=0.15w}{point 0 of p1}
    \fmfiv{label=$\tau_1$,l.angle=-20,l.dist=0.12w}{point 0 of p2}
    \fmfiv{label=$\tau_2$,l.angle=90,l.dist=0.15w}{point length(p2) of p2}
    \end{fmfgraph*}
    \end{fmffile}}
\hbox{\quad}
\parbox{60pt}{\begin{fmffile}{PO2_8}
    \begin{fmfgraph*}(60,60)
    \fmfcmd{style_def prop_pm expr p =
    draw_plain p;
    shrink(.7);
        cfill (marrow (p, .25));
        cfill (marrow (p, .75))
    endshrink;
enddef;
}
    \fmftop{t1} \fmfbottom{b1}
    \fmf{prop_pm,left=0.5,tag=1}{b1,i1}
    \fmf{prop_pm,right=0.5,tag=2}{b1,i1}
    \fmf{prop_pm,left=0.5,tag=3}{i1,t1}
    \fmf{prop_pm,right=0.5,tag=4}{i1,t1}
    \fmffreeze
    \fmf{prop_pm,left=0.6,tag=5}{b1,t1}
    \fmf{prop_pm,right=0.6,tag=6}{b1,t1}
    \fmf{phantom,tag=7}{i1,t1}
    \fmf{phantom,tag=8}{b1,i1}
    \fmfv{d.shape=square,d.filled=full,d.size=3thick,l=$O^{40}$,l.angle=180}{b1}
    \fmfv{d.shape=circle,d.filled=full,d.size=3thick,l=$\Omega^{22}$,l.angle=180,l.dist=20pt}{i1}
    \fmfv{d.shape=circle,d.filled=full,d.size=3thick,l=$\Omega^{04}$,l.angle=180}{t1}
    \fmfposition
    \fmfipath{p[]}
    \fmfiset{p1}{vpath1(__b1,__i1)}
    \fmfiset{p2}{vpath2(__b1,__i1)}
    \fmfiset{p3}{vpath3(__i1,__t1)}
    \fmfiset{p4}{vpath4(__i1,__t1)}
    \fmfiset{p5}{vpath5(__b1,__t1)}
    \fmfiset{p6}{vpath6(__b1,__t1)}
    \fmfiset{p7}{vpath7(__i1,__t1)}
    \fmfiset{p8}{vpath8(__b1,__i1)}
    \fmfiv{label=$0$,l.angle=-90,l.dist=0.15w}{point 0 of p8}
    \fmfiv{label=$\tau_1$,l.angle=-20,l.dist=0.12w}{point 0 of p7}
    \fmfiv{label=$\tau_2$,l.angle=90,l.dist=0.15w}{point length(p7) of p7}
    \end{fmfgraph*}
    \end{fmffile}}

\vspace{2\baselineskip}
PO2.7  \quad\quad\quad \quad\quad  PO2.8

\end{center}
\caption{Zero-, first- and second-order \emph{diagonal} Feynman BMBPT diagrams generated from operator vertices containing four legs at most, i.e., with \texttt{deg\_max}~$=4$.}
\label{diagBMBPT012}
\end{figure}

Given that diagonal BMBPT diagrams constitute the base line for generating the off-diagonal ones, the eleven zero-, first- and second-order diagonal BMBPT diagrams generated from operator vertices containing four legs at most are displayed in Fig.~\ref{diagBMBPT012} for reference. One recognizes in particular the three zero- and first-order diagrams PO0.1, PO1.1 and PO1.2 already appearing in Fig.~\ref{offdiagBMBPT01} with a slightly different denomination whose aim is to group all diagrams originating from the same diagonal diagram.

Diagonal and off-diagonal diagrams of order $p$ derive from the same many-body matrix element in Eq.~\eqref{observableO1}, the emergence of both categories at once being authorized by the fact that the ket is gauge rotated. The latter feature leads to the necessity to consider contractions between pairs of quasi-particle annihilation operators in addition to only contracting one creation and one annihilation operators in the strict diagonal limit. Starting from diagonal BMBPT diagrams of order $p$, the complete set of off-diagonal diagrams can thus be obtained via the application of two basic operations
\begin{enumerate}
\item adding self-contractions to each vertex, while changing the nature of the vertex accordingly, until the sum of lines entering and leaving the vertex is equal to the rank $\texttt{deg\_max}$ of the associated operator.
\item incrementally changing normal propagators linking two vertices into anomalous ones. This is achieved by turning the arrow associated with an outgoing line in the original propagator into an incoming line, thus modifying the concerned vertex accordingly.
\end{enumerate}
Let us now exemplified the two above operations that must eventually be applied systematically. 

Considering the zero-order diagonal diagram PO0.1.1$^{(1)}$ in Fig.~\ref{offdiagBMBPT01} (i.e.,\@ PO0.1 in Fig.~\ref{diagBMBPT012}), and a two-body operator $O$ ($\texttt{deg\_max}=4$), the vertex $O^{00}$ has no line entering or leaving it. Replacing it by $O^{02}$ and $O^{04}$, one generates two valid off-diagonal diagrams containing one and two anomalous contractions denoted as $\text{PO0.1.1}^{(2)}$ and $\text{PO0.1.1}^{(3)}$ in Fig.~\ref{offdiagBMBPT01}. One can proceed similarly starting from the first-order diagram denoted as PO1.1.1$^{(1)}$ in Fig.~\ref{offdiagBMBPT01} (i.e.,\@ PO1.1 in Fig.~\ref{diagBMBPT012}). Adding a self-contraction to each of the two vertices provides three additional off-diagonal diagrams containing one or two anomalous lines and denoted as PO1.1.1$^{(4)}$, PO1.1.2$^{(1)}$ and PO1.1.2$^{(4)}$ in Fig.~\ref{offdiagBMBPT01}.

To illustrate the second operation, let us consider the first-order diagonal diagram PO1.2.1$^{(1)}$ in Fig.~\ref{offdiagBMBPT01} (i.e.,\@ PO1.2 in Fig.~\ref{diagBMBPT012}). This diagram contains four normal lines between the two vertices, each of which can be transformed into an anomalous line. Doing so generates four additional topologically-distinct off-diagonal diagrams denoted as PO1.2.1$^{(2)}$, PO1.2.1$^{(3)}$, PO1.2.1$^{(4)}$ and PO1.2.1$^{(5)}$ in Fig.~\ref{offdiagBMBPT01}.

Combining the transformation of normal lines into anomalous lines and the addition of self-contractions, one obtains the other topologically distinct off-diagonal diagrams displayed in Fig.~\ref{offdiagBMBPT01}.

\subsection{Similarity-transformed operator}

An important feature is that the bottom vertex $O^{m0}$ appearing in diagonal BMBPT diagrams is always at fixed time zero. Consequently, off-diagonal diagrams generated by adding self-contractions to it and/or by transforming a normal line leaving it into an anomalous line entering it possess the same time structure as the diagonal diagram it derives from. Indeed, a self-contraction carries no time dependence and thus cannot impact the time structure of the diagram. Furthermore, the fact that all $\Omega^{ij}$ vertices are at higher times than $O^{m0}$ remains true even if all the lines attached to the bottom vertex are changed into anomalous ones such that the time structure is invariant under this transformation.

A key consequence of the above observations is that all diagrams differing only by the number of self-contractions onto the bottom vertex and/or the number of anomalous propagators connected to it can be grouped into a single diagram in which the bottom vertex is replaced by its similarity transformed partner at gauge angle $\varphi$~\cite{Duguet:2015yle}
\begin{align}
  \tilde{O}(\varphi) &\equiv e^{-Z(\varphi)} O e^{Z(\varphi)} \, , \label{transformedop}
\end{align}
where $Z(\varphi)$ is the Thouless operator, see \ref{SecThouless}. As explained at length in \ref{transformME}, the transformed operator $\tilde{O}(\varphi)$ possesses the same formal structure as the initial operator $O$. As such, it is decomposed as a sum of terms $\tilde{O}^{mn}(\varphi)$ with the same overall rank as $O$, i.e., $m+n\leq \texttt{deg\_max}$. The only difference relates to the definition of the (gauge-dependent) matrix elements entering each term $\tilde{O}^{mn}(\varphi)$. The expression of these matrix elements in terms of the original ones were provided in Ref.~\cite{Duguet:2015yle} for $\texttt{deg\_max}=4$.

\begin{figure}[t!]

  \vspace{\baselineskip}

  \parbox{40pt}{\begin{fmffile}{PNP-BMBPT0_1_1}
    \begin{fmfgraph*}(40,40)
      \fmftop{t1}
      \fmfbottom{b1}
      \fmf{phantom,tag=1}{b1,t1}
      \fmffreeze
      \fmfv{d.shape=square,d.filled=full,d.size=3thick,l=$\tilde{O}^{00}(\varphi)$,l.angle=0}{b1}
      \fmfposition
      \fmfipath{p[]}
      \fmfiset{p1}{vpath1(__b1,__t1)}
      \fmfiv{label=$\text{PO0.1.1}$,l.angle=-90,l.dist=0.30w}{point 0 of p1}
    \end{fmfgraph*}
  \end{fmffile}}
  \hbox{\quad}
  \parbox{40pt}{\begin{fmffile}{PNP-BMBPT1_1_1}
    \begin{fmfgraph*}(40,40)
    \fmfcmd{style_def prop_pm expr p =
    draw_plain p;
    shrink(.7);
        cfill (marrow (p, .25));
        cfill (marrow (p, .75))
    endshrink;
enddef;
}
      \fmftop{t1}
      \fmfbottom{b1}
      \fmf{prop_pm,left=0.5,tag=1}{b1,t1}
      \fmf{prop_pm,right=0.5,tag=2}{b1,t1}
      \fmfv{d.shape=circle,d.filled=full,d.size=3thick,l=$\Omega^{02}$,l.angle=0}{t1}
      \fmfv{d.shape=square,d.filled=full,d.size=3thick,l=$\tilde{O}^{20}(\varphi)$,l.angle=0}{b1}
      \fmfposition
      \fmfipath{p[]}
      \fmfiset{p1}{vpath1(__b1,__t1)}
      \fmfiv{label=$\text{PO1.1.1}$,l.angle=-90,l.dist=0.30w}{point 0 of p1}
    \end{fmfgraph*}
  \end{fmffile}}
  \hbox{\quad}
  \parbox{40pt}{\begin{fmffile}{PNP-BMBPT1_1_2}
    \begin{fmfgraph*}(40,40)
    \fmfcmd{style_def half_prop expr p =
    draw_plain p;
    shrink(.7);
        cfill (marrow (p, .5))
    endshrink;
	enddef;}
	\fmfcmd{style_def prop_pm expr p =
    draw_plain p;
    shrink(.7);
        cfill (marrow (p, .25));
        cfill (marrow (p, .75))
    endshrink;
enddef;
}
      \fmftop{tm,t0,t1,t2,t3}
      \fmfbottom{b1}
      \fmf{prop_pm,left=0.5,tag=1}{b1,t1}
      \fmf{prop_pm,right=0.5}{b1,t1}
      \fmf{half_prop,left}{t2,t1}
      \fmf{half_prop,right}{t2,t1}
      \fmfv{d.shape=circle,d.filled=full,d.size=3thick,l=$\Omega^{04}$,l.dist=0.30w,l.angle=0}{t1}
      \fmfv{d.shape=square,d.filled=full,d.size=3thick,l=$\tilde{O}^{20}(\varphi)$,l.angle=0}{b1}
      \fmfposition
      \fmfipath{p[]}
      \fmfiset{p1}{vpath1(__b1,__t1)}
      \fmfiv{label=$\text{PO1.1.2}$,l.angle=-90,l.dist=0.30w}{point 0 of p1}
    \end{fmfgraph*}
  \end{fmffile}}
  \hbox{\quad}
  \parbox{40pt}{\begin{fmffile}{PNP-BMBPT1_2_1}
    \begin{fmfgraph*}(40,40)
    \fmfcmd{style_def prop_pm expr p =
    draw_plain p;
    shrink(.7);
        cfill (marrow (p, .25));
        cfill (marrow (p, .75))
    endshrink;
enddef;
}
      \fmftop{t1}
      \fmfbottom{b1}
      \fmf{prop_pm,left=0.6,tag=1}{b1,t1}
      \fmf{prop_pm,left=0.3,tag=2}{b1,t1}
      \fmf{prop_pm,right=0.3,tag=3}{b1,t1}
      \fmf{prop_pm,right=0.6,tag=4}{b1,t1}
      \fmfv{d.shape=circle,d.filled=full,d.size=3thick,l=$\Omega^{04}$,l.dist=0.15w,l.angle=0}{t1}
      \fmfv{d.shape=square,d.filled=full,d.size=3thick,l=$\tilde{O}^{40}(\varphi)$,l.angle=0}{b1}
      \fmfposition
      \fmfipath{p[]}
      \fmfiset{p1}{vpath1(__b1,__t1)}
      \fmfiv{label=$\text{PO1.2.1}$,l.angle=-90,l.dist=0.30w}{point 0 of p1}
    \end{fmfgraph*}
  \end{fmffile}}

  \vspace{1.5\baselineskip}

  \caption{\label{fig:offdiagrecastedpo01}
  Zero- and first-order effective off-diagonal BMBPT diagrams recasting the twenty displayed in Fig.~\ref{offdiagBMBPT01}.}
\end{figure}
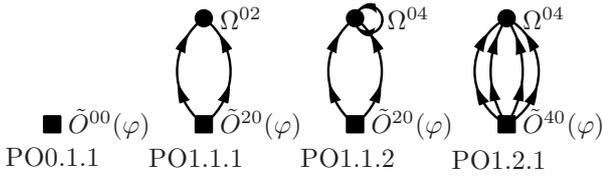

Exploiting this key observation, one can reduce drastically the number of {\it effective} diagrams at play. Employing the transformed operator for the bottom vertex and forbidding any anomalous line to connect to it, the twenty off-diagonal diagrams displayed in Fig.~\ref{offdiagBMBPT01} are recasted into the four effective off-diagonal diagrams displayed in Fig.~\ref{fig:offdiagrecastedpo01}. This feature being generic, the recasting procedure extends to any order $p$.

\subsection{Systematic scheme}

The analysis provided above puts us in position to state the systematic rules used to generate all effective off-diagonal BMBPT diagrams of order $p$ from the diagonal ones. Starting from a diagonal BMBPT diagram of order $p$
\begin{enumerate}
\item replace the bottom vertex $O^{m0}$ by its transformed partner $\tilde{O}^{m0}(\varphi)$,
\item for each energy vertex $\Omega^{ij}$
\begin{enumerate}
\item transform $l$ outgoing arrows into incoming arrows to form $l$ anomalous lines while turning the vertex into $\Omega^{i-lj+l}$, with $l\in \mathbb{N}$, until $i-l=0$,
\item add $k$ self-contractions while turning the vertex into $\Omega^{ij+2k}$, with $k\in \mathbb{N}$, until $i+j+2k=\texttt{deg\_max}$,
\end{enumerate}
  \item retain only topologically distinct diagrams.
\end{enumerate}

While the method is straightforward, it is indeed important to discard topologically equivalent diagrams generated through this brute-force procedure. Anticipating it, one can actually reduce the need to check for all of them, which is particularly beneficial given that the corresponding test scales factorially with the number of vertices in the diagrams. 
The first step in our algorithm being to generate off-diagonal diagrams from a given straight BMBPT one, one avoids producing topologically equivalent diagrams in the subset of children in the following way:
\begin{itemize}
\item As normal propagators are turned anomalous, the list of initial equivalent lines is recorded taking into account possible vertex exchanges. Doing so, only unique permutations of normal propagators are turned into anomalous ones.
\item Once this first set of diagrams is generated, adding self-contractions on them can only create topologically equivalent diagrams when doing so on topologically equivalent vertices. As such, one first looks out for such vertices, and makes sure to add self-contractions on unique combinations of vertices.
\end{itemize}

Once order-$p$ diagrams are generated in this way, checks for topologically equivalent diagrams are still required. This is due to the fact that by turning normal propagators anomalous, one creates new topologies that might in fact arise from several straight BMBPT diagrams. Some diagrams can be excluded from this check though, starting from the ones containing only normal propagators, i.e.,\@ straight BMBPT ones, which have been tested beforehand. This exclusion can be extended to diagrams in which the only anomalous lines are self-contractions, i.e.,\@ straight BMBPT diagrams with extra anomalous self-contractions, as the structure of propagators exchanged between vertices has not been affected. Applying the discussed method eventually results in the number of diagrams displayed in Tab.~\ref{TabDiagNumbers}.

\begin{table*}
\begin{center}
\begin{tabular}{|l|l|c|c|c|c|c|}
\hline  Order & & 0 & 1 & 2 & 3 & 4  \\
  \hline\hline Straight BMBPT & \texttt{deg\_max}~$=4$ & 1 & 2 & 8 & 59 & 568 \\
  \hline                        	  & \texttt{deg\_max}~$=6$ & 1 & 3 & 23 & 396 & 10 716 \\
  \hline\hline Off-diagonal BMBPT & \texttt{deg\_max}~$=4$ & 1 & 3 & 33 & 602 & 14 977 \\
  \hline 							  & \texttt{deg\_max}~$=6$ & 1 & 6 & 189 & 13 046 & \dots \\
  \hline
\end{tabular}
\end{center}
\caption{Number of \emph{diagonal} and effective \emph{off-diagonal} BMBPT diagrams generated from operators containing at most four (\texttt{deg\_max}~$=4$) or six (\texttt{deg\_max}~$=6$) legs.}
\label{TabDiagNumbers}
\end{table*}

\subsection{Drawing associated BMBPT diagrams}

Once the off-diagonal BMBPT diagrams have been produced, it is important to be able to represent them graphically. As in Ref.~\cite{arthuis18a}, the BMBPT diagrams are created as objects generated via the Python package \emph{NetworkX}~\cite{SciPyProceedings_11}, allowing for an efficient and easy to handle storage of the necessary information. It is then easy to read these objects and extract the information necessary to produce the drawing instructions in the form of \emph{FeynMP}~\cite{Ohl:1995kr} commands. The routines designed in Ref.~\cite{arthuis18a} only had to be adapted to allow the drawing of anomalous propagators and self-contractions. As an example, the output displaying the drawing instructions of the off-diagonal BMBPT diagram displayed in Fig.~\ref{f:off_tsd_labelling} is given in Fig.~\ref{f:feynmp_inst}.

\begin{figure}[t]
{\small
\begin{verbatim}
\begin{fmffile}{offdiag_ex}
\begin{fmfgraph*}(80,80)
\fmfcmd{style_def prop_pm expr p =
    draw_plain p;
    shrink(.7);
        cfill (marrow (p, .25));
        cfill (marrow (p, .75))
    endshrink;
	enddef;}
\fmfcmd{style_def prop_mm expr p =
        draw_plain p;
        shrink(.7);
            cfill (marrow (p, .75));
            cfill (marrow (reverse p, .75))
        endshrink;
        enddef;}
\fmfcmd{style_def half_prop expr p =
    draw_plain p;
    shrink(.7);
        cfill (marrow (p, .5))
    endshrink;
	enddef;}
\fmfstraight
\fmftop{d1,d2,v2,d3,d4}\fmfbottom{v0}
\fmf{phantom}{v0,v1}
\fmfv{d.shape=square,d.filled=full,d.size=3thick}{v0}
\fmf{phantom}{v1,v2}
\fmfv{d.shape=circle,d.filled=full,d.size=3thick}{v1}
\fmfv{d.shape=circle,d.filled=full,d.size=3thick}{v2}
\fmffreeze
\fmf{prop_pm,left=0.5}{v0,v1}
\fmf{prop_pm,right=0.5}{v0,v1}
\fmf{prop_pm,left=0.5}{v1,v2}
\fmf{prop_mm,right=0.5}{v1,v2}
\fmf{half_prop,right}{d3,v2}
\fmf{half_prop,left}{d3,v2}
\end{fmfgraph*}
\end{fmffile}
\end{verbatim}
}

\caption{\emph{FeynMP} instructions to draw the off-diagonal BMBPT diagram displayed in Fig.~\ref{f:off_tsd_labelling}.}
\label{f:feynmp_inst}
\end{figure}

\section{Evaluation of off-diagonal BMBPT diagrams}
\label{secautomaticevaluation}

Having the capacity to generate all off-diagonal BMBPT Feynman diagrams of order $p$, the next challenge is to systematically derive their expression. Doing so on the basis of Feynman's algebraic rules is rather straightforward. However, it leaves the $p$-tuple time integral to perform in order to obtain the time-integrated expression of interest. In Ref.~\cite{arthuis18a}, an algorithm was found to overcome this challenge for \emph{diagonal} BMBPT diagrams at play in straight BMBPT without prior knowledge of the perturbative order or of the topology of the diagram. This eventually led to the identification of a novel diagrammatic rule. The present section details how the method only needs to be slightly generalized in order to realize the same objective for \emph{off-diagonal} BMBPT diagrams at play in PNP-BMBPT.

\subsection{Time-structure diagrams}
\label{subsubs:timediags}

Obtaining the result of $p$-tuple time integrals in an automatic fashion was made possible via the introduction of the time-structure diagram underlying any given diagonal BMBPT diagram of arbitrary order and topology. We refer to Ref.~\cite{arthuis18a} for the general theory of TSDs and only comment here on the specificities encountered when dealing with more general off-diagonal BMBPT diagrams.

The key point was already alluded to in Sec.~\ref{subs:diag_eval} and relates to the impact anomalous lines may have on the TSD attributed to a given off-diagonal BMBPT diagram. The main features are
\begin{itemize}
\item The running time labels $(\tau_1, \ldots, \tau_p)$ are positive such that each $\Omega$ vertex entertains at least an ordering relation with the bottom vertex $\tilde{O}(\varphi)$ independently of the network of lines running through the BMBPT diagram. Consequently, the TSD remains necessarily connected, independently of its topology.
\item Contrarily to normal lines,  anomalous lines do not induce any time ordering relation. This means that, while two $\Omega$ vertices connected by at least one normal line are time ordered, it is not the case if they are solely connected via anomalous propagators. Consequently, a link connecting two $\Omega$ vertices in the TSD associated to a diagonal BMBPT diagram will disappear when the two vertices become only connected via anomalous propagators in an off-diagonal partner diagram\footnote{As a TSD stores only the minimal information associated with time-ordering relations, such a disappearance may parallel the occurence of one or several new links with respect to the TSD of the diagonal BMBPT diagram. Hence TSDs must always be produced starting from a given off-diagonal BMBPT diagram and \emph{not} from the TSD of its parent BMBPT diagram.}. Whenever an $\Omega$ vertex ends up entertaining no time relation with any other due to the replacement of normal lines by anomalous ones, it becomes directly linked to the bottom vertex $\tilde{O}(\varphi)$ in the associated TSD.
\item The addition of a self contraction to any given $\Omega$ vertex does not change the time structure of the diagram and thus the associated TSD.
\end{itemize}
In conclusion, the presence of anomalous lines may, depending on the situation, change the TSD associated to an off-diagonal BMBPT diagram compared to the diagonal diagram displaying the same topology. Eventually, the TSD associated to an off-diagonal BMBPT diagram can be obtained through the following steps
\begin{enumerate}
\item copy the off-diagonal BMBPT diagram,
\item remove all the anomalous propagators,
\item replace the normal propagators by links,
\item add a link between the bottom vertex at time 0 and every other vertex if such a link does not exist,
\item for each pair of vertices, consider all paths linking them and only retain the longest one,
\item match the label $a_q$ associated to a given vertex in the TSD diagram to the sum/difference of quasi-particle energies associated with the lines entering/leaving the corresponding vertex in the BMBPT diagram.
\end{enumerate}
The only difference with the procedure followed for strictly diagonal BMBPT diagrams~\cite{arthuis18a} relates to step 2 that trivially stipulates to strip off anomalous propagators if any.

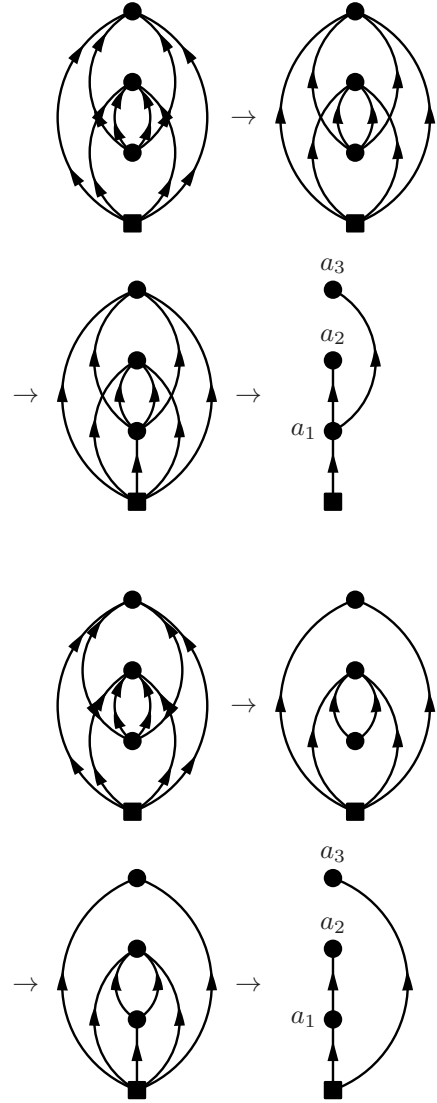
\begin{figure}[t!]
\begin{center}
$\phantom{\rightarrow}$
\hspace{-10pt}
\parbox{80pt}{\begin{fmffile}{diag_tsd_const_ex0}
\begin{fmfgraph*}(80,80)
\fmfcmd{style_def prop_pm expr p =
    draw_plain p;
    shrink(.7);
        cfill (marrow (p, .25));
        cfill (marrow (p, .75))
    endshrink;
	enddef;}
\fmftop{v3}\fmfbottom{v0}
\fmf{phantom}{v0,v1}
\fmfv{d.shape=square,d.filled=full,d.size=3thick}{v0}
\fmf{phantom}{v1,v2}
\fmfv{d.shape=circle,d.filled=full,d.size=3thick}{v1}
\fmf{phantom}{v2,v3}
\fmfv{d.shape=circle,d.filled=full,d.size=3thick}{v2}
\fmfv{d.shape=circle,d.filled=full,d.size=3thick}{v3}
\fmffreeze
\fmf{prop_pm,left=0.6}{v0,v2}
\fmf{prop_pm,right=0.6}{v0,v2}
\fmf{prop_pm,left=0.7}{v0,v3}
\fmf{prop_pm,right=0.7}{v0,v3}
\fmf{prop_pm,left=0.5}{v1,v2}
\fmf{prop_pm,right=0.5}{v1,v2}
\fmf{prop_pm,left=0.6}{v1,v3}
\fmf{prop_pm,right=0.6}{v1,v3}
\end{fmfgraph*}
\end{fmffile}}
\hspace{-10pt}
$\rightarrow$
\hspace{-10pt}
\parbox{80pt}{\begin{fmffile}{diag_tsd_const_ex1}
\begin{fmfgraph*}(80,80)
\fmfcmd{style_def half_prop expr p =
    draw_plain p;
    shrink(.7);
        cfill (marrow (p, .5))
    endshrink;
	enddef;}
\fmftop{v3}\fmfbottom{v0}
\fmf{phantom}{v0,v1}
\fmfv{d.shape=square,d.filled=full,d.size=3thick}{v0}
\fmf{phantom}{v1,v2}
\fmfv{d.shape=circle,d.filled=full,d.size=3thick}{v1}
\fmf{phantom}{v2,v3}
\fmfv{d.shape=circle,d.filled=full,d.size=3thick}{v2}
\fmfv{d.shape=circle,d.filled=full,d.size=3thick}{v3}
\fmffreeze
\fmf{half_prop,left=0.6}{v0,v2}
\fmf{half_prop,right=0.6}{v0,v2}
\fmf{half_prop,left=0.7}{v0,v3}
\fmf{half_prop,right=0.7}{v0,v3}
\fmf{half_prop,left=0.5}{v1,v2}
\fmf{half_prop,right=0.5}{v1,v2}
\fmf{half_prop,left=0.6}{v1,v3}
\fmf{half_prop,right=0.6}{v1,v3}
\end{fmfgraph*}
\end{fmffile}}
\hspace{-10pt}

\vspace{2\baselineskip}

$\rightarrow$
\hspace{-10pt}
\parbox{80pt}{\begin{fmffile}{diag_tsd_const_ex2}
\begin{fmfgraph*}(80,80)
\fmfcmd{style_def half_prop expr p =
    draw_plain p;
    shrink(.7);
        cfill (marrow (p, .5))
    endshrink;
	enddef;}
\fmftop{v3}\fmfbottom{v0}
\fmf{phantom}{v0,v1}
\fmfv{d.shape=square,d.filled=full,d.size=3thick}{v0}
\fmf{phantom}{v1,v2}
\fmfv{d.shape=circle,d.filled=full,d.size=3thick}{v1}
\fmf{phantom}{v2,v3}
\fmfv{d.shape=circle,d.filled=full,d.size=3thick}{v2}
\fmfv{d.shape=circle,d.filled=full,d.size=3thick}{v3}
\fmffreeze
\fmf{half_prop,left=0.6}{v0,v2}
\fmf{half_prop,right=0.6}{v0,v2}
\fmf{half_prop,left=0.7}{v0,v3}
\fmf{half_prop,right=0.7}{v0,v3}
\fmf{half_prop,left=0.5}{v1,v2}
\fmf{half_prop,right=0.5}{v1,v2}
\fmf{half_prop,left=0.6}{v1,v3}
\fmf{half_prop,right=0.6}{v1,v3}
\fmf{half_prop}{v0,v1}
\end{fmfgraph*}
\end{fmffile}}
\hspace{-10pt}
$\rightarrow$
\hspace{-20pt}
\parbox{80pt}{\begin{fmffile}{tsd_const_ex}
\begin{fmfgraph*}(80,80)
\fmfcmd{style_def half_prop expr p =
    draw_plain p;
    shrink(.7);
        cfill (marrow (p, .5))
    endshrink;
	enddef;}
\fmftop{v3}\fmfbottom{v0}
\fmfv{d.shape=square,d.filled=full,d.size=3thick}{v0}
\fmfv{d.shape=circle,d.filled=full,d.size=3thick,l=$a_1$,l.a=180}{v1}
\fmf{phantom}{v2,v3}
\fmfv{d.shape=circle,d.filled=full,d.size=3thick,l=$a_2$,l.a=90}{v2}
\fmfv{d.shape=circle,d.filled=full,d.size=3thick,l=$a_3$,l.a=90}{v3}
\fmf{half_prop}{v0,v1}
\fmf{half_prop}{v1,v2}
\fmffreeze
\fmf{half_prop,right=0.6}{v1,v3}
\end{fmfgraph*}
\end{fmffile}}

\vspace{3\baselineskip}

$\phantom{\rightarrow}$
\hspace{-10pt}
\parbox{80pt}{\begin{fmffile}{diag_tsd_const_exoff0}
\begin{fmfgraph*}(80,80)
\fmfcmd{style_def prop_pm expr p =
    draw_plain p;
    shrink(.7);
        cfill (marrow (p, .25));
        cfill (marrow (p, .75))
    endshrink;
	enddef;}
\fmfcmd{style_def prop_mm expr p =
        draw_plain p;
        shrink(.7);
            cfill (marrow (p, .75));
            cfill (marrow (reverse p, .75))
        endshrink;
        enddef;}
\fmftop{v3}\fmfbottom{v0}
\fmf{phantom}{v0,v1}
\fmfv{d.shape=square,d.filled=full,d.size=3thick}{v0}
\fmf{phantom}{v1,v2}
\fmfv{d.shape=circle,d.filled=full,d.size=3thick}{v1}
\fmf{phantom}{v2,v3}
\fmfv{d.shape=circle,d.filled=full,d.size=3thick}{v2}
\fmfv{d.shape=circle,d.filled=full,d.size=3thick}{v3}
\fmffreeze
\fmf{prop_pm,left=0.6}{v0,v2}
\fmf{prop_pm,right=0.6}{v0,v2}
\fmf{prop_pm,left=0.7}{v0,v3}
\fmf{prop_pm,right=0.7}{v0,v3}
\fmf{prop_pm,left=0.5}{v1,v2}
\fmf{prop_pm,right=0.5}{v1,v2}
\fmf{prop_mm,left=0.7}{v1,v3}
\fmf{prop_mm,right=0.7}{v1,v3}
\end{fmfgraph*}
\end{fmffile}}
\hspace{-10pt}
$\rightarrow$
\hspace{-10pt}
\parbox{80pt}{\begin{fmffile}{diag_tsd_const_exoff1}
\begin{fmfgraph*}(80,80)
\fmfcmd{style_def half_prop expr p =
    draw_plain p;
    shrink(.7);
        cfill (marrow (p, .5))
    endshrink;
	enddef;}
\fmftop{v3}\fmfbottom{v0}
\fmf{phantom}{v0,v1}
\fmfv{d.shape=square,d.filled=full,d.size=3thick}{v0}
\fmf{phantom}{v1,v2}
\fmfv{d.shape=circle,d.filled=full,d.size=3thick}{v1}
\fmf{phantom}{v2,v3}
\fmfv{d.shape=circle,d.filled=full,d.size=3thick}{v2}
\fmfv{d.shape=circle,d.filled=full,d.size=3thick}{v3}
\fmffreeze
\fmf{half_prop,left=0.6}{v0,v2}
\fmf{half_prop,right=0.6}{v0,v2}
\fmf{half_prop,left=0.7}{v0,v3}
\fmf{half_prop,right=0.7}{v0,v3}
\fmf{half_prop,left=0.6}{v1,v2}
\fmf{half_prop,right=0.6}{v1,v2}
\end{fmfgraph*}
\end{fmffile}}
\hspace{-10pt}

\vspace{2\baselineskip}

$\rightarrow$
\hspace{-10pt}
\parbox{80pt}{\begin{fmffile}{diag_tsd_const_exoff2}
\begin{fmfgraph*}(80,80)
\fmfcmd{style_def half_prop expr p =
    draw_plain p;
    shrink(.7);
        cfill (marrow (p, .5))
    endshrink;
	enddef;}
\fmftop{v3}\fmfbottom{v0}
\fmf{phantom}{v0,v1}
\fmfv{d.shape=square,d.filled=full,d.size=3thick}{v0}
\fmf{phantom}{v1,v2}
\fmfv{d.shape=circle,d.filled=full,d.size=3thick}{v1}
\fmf{phantom}{v2,v3}
\fmfv{d.shape=circle,d.filled=full,d.size=3thick}{v2}
\fmfv{d.shape=circle,d.filled=full,d.size=3thick}{v3}
\fmffreeze
\fmf{half_prop,left=0.6}{v0,v2}
\fmf{half_prop,right=0.6}{v0,v2}
\fmf{half_prop,left=0.7}{v0,v3}
\fmf{half_prop,right=0.7}{v0,v3}
\fmf{half_prop,left=0.6}{v1,v2}
\fmf{half_prop,right=0.6}{v1,v2}
\fmf{half_prop}{v0,v1}
\end{fmfgraph*}
\end{fmffile}}
\hspace{-10pt}
$\rightarrow$
\hspace{-20pt}
\parbox{80pt}{\begin{fmffile}{tsd_const_exoff}
\begin{fmfgraph*}(80,80)
\fmfcmd{style_def half_prop expr p =
    draw_plain p;
    shrink(.7);
        cfill (marrow (p, .5))
    endshrink;
	enddef;}
\fmftop{v3}\fmfbottom{v0}
\fmfv{d.shape=square,d.filled=full,d.size=3thick}{v0}
\fmfv{d.shape=circle,d.filled=full,d.size=3thick,l=$a_1$,l.a=180}{v1}
\fmf{phantom}{v2,v3}
\fmfv{d.shape=circle,d.filled=full,d.size=3thick,l=$a_2$,l.a=90}{v2}
\fmfv{d.shape=circle,d.filled=full,d.size=3thick,l=$a_3$,l.a=90}{v3}
\fmf{half_prop}{v0,v1}
\fmf{phantom}{v1,v2}
\fmffreeze
\fmf{half_prop}{v1,v2}
\fmf{half_prop,right=0.7}{v0,v3}
\end{fmfgraph*}
\end{fmffile}}
\end{center}
\caption{Production of the TSDs associated with a third-order diagonal BMBPT diagram and with an off-diagonal diagram obtained from it by turning two among the eight normal lines into anomalous ones.}
\label{f:tsd_production}
\end{figure}

The procedure is illustrated in Fig.~\ref{f:tsd_production} for a third-order diagonal BMBPT diagram and for the particular off-diagonal diagram generated from it by turning the two normal lines connecting vertex $\Omega^{40}$ to one of the two $\Omega^{04}$ vertices into anomalous lines. Cleared of other informations, the TSDs tranparently characterize the time-ordering structure underlying the diagrams. In the first one, the three $\Omega^{ij}$ vertices are at higher times than $O^{40}$ such that the two $\Omega^{04}$ vertices are at higher times than $\Omega^{40}$ without being ordered with respect to one another. From the graph theory viewpoint, the corresponding TSD is a tree, i.e., it contains no cycle, with two branches. In the second diagram, the fact that the two lines connecting $\Omega^{40}$ to (one of the two) $\Omega^{04}$ are anomalous relaxes the time-ordering between both vertices and, as a result, changes the nature of the associated TSD, i.e., $a_3$ is now directly linked to the bottom vertex

\subsection{Discussion}
\label{subsubs:discussion}

It is mandatory to generate the TSD \emph{from} its underlying BMBPT diagram. Indeed, only in this case can the rank \texttt{deg\_max} of the operators at play be employed to constrain the topology of the diagrams, eventually dictating the topology of allowed TSDs. Furthermore, going from diagonal to off-diagonal BMBPT diagrams may not only change the nature of the TSD associated to a particular diagram but also increase the list of active TSDs at a given order. 

\begin{figure}[t!]
\begin{center}
\parbox{30pt}{\begin{fmffile}{time_0_0}
\begin{fmfgraph*}(30,30)
\fmftop{t1} \fmfbottom{b1}
    \fmf{phantom}{b1,i1}
    \fmf{phantom}{i1,t1}
    \fmfv{d.shape=square,d.filled=full,d.size=3thick}{i1}
\end{fmfgraph*}
\end{fmffile}}

T0.1
\vspace{2\baselineskip}

\parbox{30pt}{\begin{fmffile}{time_1_0}
\begin{fmfgraph*}(30,30)
\fmfcmd{style_def half_prop expr p =
draw_plain p;
shrink(.7);
    cfill (marrow (p, .5))
endshrink;
enddef;}
\fmftop{v1}\fmfbottom{v0}
\fmf{phantom}{v0,v1}
\fmfv{d.shape=square,d.filled=full,d.size=3thick}{v0}
\fmfv{d.shape=circle,d.filled=full,d.size=3thick,l=$a_1$}{v1}
\fmffreeze
\fmf{half_prop}{v0,v1}
\end{fmfgraph*}
\end{fmffile}}

\vspace{\baselineskip}
T1.1
\vspace{2\baselineskip}

\parbox{60pt}{\begin{fmffile}{time_2_0}
\begin{fmfgraph*}(60,60)
\fmfcmd{style_def half_prop expr p =
draw_plain p;
shrink(.7);
    cfill (marrow (p, .5))
endshrink;
enddef;}
\fmftop{v2}\fmfbottom{v0}
\fmf{phantom}{v0,v1}
\fmfv{d.shape=square,d.filled=full,d.size=3thick}{v0}
\fmf{phantom}{v1,v2}
\fmfv{d.shape=circle,d.filled=full,d.size=3thick,l=$a_1$}{v1}
\fmfv{d.shape=circle,d.filled=full,d.size=3thick,l=$a_2$}{v2}
\fmffreeze
\fmf{half_prop}{v0,v1}
\fmf{half_prop}{v1,v2}
\end{fmfgraph*}
\end{fmffile}}
 \hspace{-20pt}
\parbox{60pt}{\begin{fmffile}{time_2_1}
\begin{fmfgraph*}(60,60)
\fmfcmd{style_def half_prop expr p =
draw_plain p;
shrink(.7);
    cfill (marrow (p, .5))
endshrink;
enddef;}
\fmftop{v2}\fmfbottom{v0}
\fmf{phantom}{v0,v1}
\fmfv{d.shape=square,d.filled=full,d.size=3thick}{v0}
\fmf{phantom}{v1,v2}
\fmfv{d.shape=circle,d.filled=full,d.size=3thick,l=$a_1$,l.a=90}{v1}
\fmfv{d.shape=circle,d.filled=full,d.size=3thick,l=$a_2$}{v2}
\fmffreeze
\fmf{half_prop}{v0,v1}
\fmf{half_prop,right=0.6}{v0,v2}
\end{fmfgraph*}
\end{fmffile}}

\vspace{\baselineskip}
T2.1 \quad \quad \quad T2.2
\vspace{2\baselineskip}

\hspace{-40pt}
\parbox{80pt}{\begin{fmffile}{time_3_0}
\begin{fmfgraph*}(80,80)
\fmfcmd{style_def half_prop expr p =
draw_plain p;
shrink(.7);
    cfill (marrow (p, .5))
endshrink;
enddef;}
\fmftop{v3}\fmfbottom{v0}
\fmf{phantom}{v0,v1}
\fmfv{d.shape=square,d.filled=full,d.size=3thick}{v0}
\fmf{phantom}{v1,v2}
\fmfv{d.shape=circle,d.filled=full,d.size=3thick,l=$a_1$,l.a=0}{v1}
\fmf{phantom}{v2,v3}
\fmfv{d.shape=circle,d.filled=full,d.size=3thick,l=$a_2$,l.a=0}{v2}
\fmfv{d.shape=circle,d.filled=full,d.size=3thick,l=$a_3$}{v3}
\fmffreeze
\fmf{half_prop}{v0,v1}
\fmf{half_prop}{v1,v2}
\fmf{half_prop}{v2,v3}
\end{fmfgraph*}
\end{fmffile}}
\hspace{-35pt}
\parbox{80pt}{\begin{fmffile}{time_3_1}
\begin{fmfgraph*}(80,80)
\fmfcmd{style_def half_prop expr p =
draw_plain p;
shrink(.7);
    cfill (marrow (p, .5))
endshrink;
enddef;}
\fmftop{v3}\fmfbottom{v0}
\fmf{phantom}{v0,v1}
\fmfv{d.shape=square,d.filled=full,d.size=3thick}{v0}
\fmf{phantom}{v1,v2}
\fmfv{d.shape=circle,d.filled=full,d.size=3thick,l=$a_1$,l.a=-60}{v1}
\fmf{phantom}{v2,v3}
\fmfv{d.shape=circle,d.filled=full,d.size=3thick,l=$a_2$,l.a=90}{v2}
\fmfv{d.shape=circle,d.filled=full,d.size=3thick,l=$a_3$}{v3}
\fmffreeze
\fmf{half_prop}{v0,v1}
\fmf{half_prop}{v1,v2}
\fmf{half_prop,right=0.6}{v1,v3}
\end{fmfgraph*}
\end{fmffile}}
\hspace{-35pt}
\parbox{80pt}{\begin{fmffile}{time_3_2}
\begin{fmfgraph*}(80,80)
\fmfcmd{style_def half_prop expr p =
draw_plain p;
shrink(.7);
    cfill (marrow (p, .5))
endshrink;
enddef;}
\fmftop{v3}\fmfbottom{v0}
\fmf{phantom}{v0,v1}
\fmfv{d.shape=square,d.filled=full,d.size=3thick}{v0}
\fmf{phantom}{v1,v2}
\fmfv{d.shape=circle,d.filled=full,d.size=3thick,l=$a_1$,l.a=90}{v1}
\fmf{phantom}{v2,v3}
\fmfv{d.shape=circle,d.filled=full,d.size=3thick,l=$a_2$,l.a=60}{v2}
\fmfv{d.shape=circle,d.filled=full,d.size=3thick,l=$a_3$}{v3}
\fmffreeze
\fmf{half_prop}{v0,v1}
\fmf{half_prop,right=0.6}{v0,v2}
\fmf{half_prop}{v2,v3}
\end{fmfgraph*}
\end{fmffile}}
\hspace{-35pt}
\parbox{80pt}{\begin{fmffile}{time_3_3}
\begin{fmfgraph*}(80,80)
\fmfcmd{style_def half_prop expr p =
draw_plain p;
shrink(.7);
    cfill (marrow (p, .5))
endshrink;
enddef;}
\fmftop{v3}\fmfbottom{v0}
\fmf{phantom}{v0,v1}
\fmfv{d.shape=square,d.filled=full,d.size=3thick}{v0}
\fmf{phantom}{v1,v2}
\fmfv{d.shape=circle,d.filled=full,d.size=3thick,l=$a_1$,l.a=90}{v1}
\fmf{phantom}{v2,v3}
\fmfv{d.shape=circle,d.filled=full,d.size=3thick,l=$a_2$}{v2}
\fmfv{d.shape=circle,d.filled=full,d.size=3thick,l=$a_3$}{v3}
\fmffreeze
\fmf{half_prop}{v0,v1}
\fmf{half_prop,right=0.6}{v0,v2}
\fmf{half_prop,right=0.6}{v0,v3}
\end{fmfgraph*}
\end{fmffile}}
\hspace{-30pt}
\parbox{80pt}{\begin{fmffile}{time_3_4}
\begin{fmfgraph*}(80,80)
\fmfcmd{style_def half_prop expr p =
draw_plain p;
shrink(.7);
    cfill (marrow (p, .5))
endshrink;
enddef;}
\fmftop{v3}\fmfbottom{v0}
\fmf{phantom}{v0,v1}
\fmfv{d.shape=square,d.filled=full,d.size=3thick}{v0}
\fmf{phantom}{v1,v2}
\fmfv{d.shape=circle,d.filled=full,d.size=3thick,l=$a_1$,l.a=0}{v1}
\fmf{phantom}{v2,v3}
\fmfv{d.shape=circle,d.filled=full,d.size=3thick,l=$a_2$,l.a=-90}{v2}
\fmfv{d.shape=circle,d.filled=full,d.size=3thick,l=$a_3$}{v3}
\fmffreeze
\fmf{half_prop}{v0,v1}
\fmf{half_prop,left=0.6}{v0,v2}
\fmf{half_prop,right=0.6}{v1,v3}
\fmf{half_prop}{v2,v3}
\end{fmfgraph*}
\end{fmffile}}
\hspace{-30pt}

\vspace{\baselineskip}
T3.1 \quad \quad\quad T3.2 \quad \quad\quad T3.3 \quad \quad\quad T3.4 \quad \quad\quad T3.5
\vspace{\baselineskip}

\end{center}
\caption{Zero-, first-, second- and third-order TSDs corresponding to off-diagonal BMBPT diagrams generated from operators containing four or six legs at most, i.e., with \texttt{deg\_max}~$=4$ or \texttt{deg\_max}~$=6$.}
\label{diagTSD0123}
\end{figure}
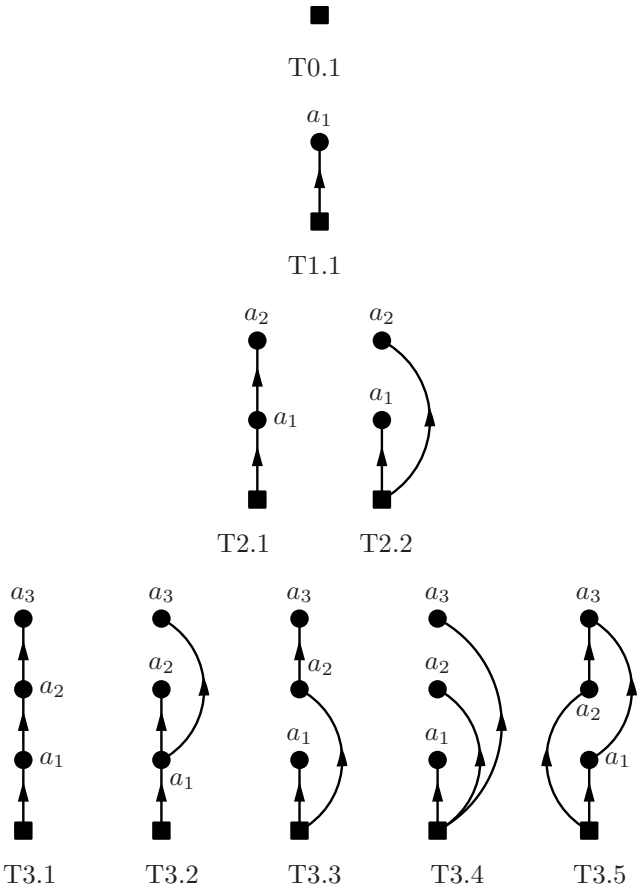

With this in mind and following the above rules, the $1/1/2/5$ TSDs of order $0/1/2/3$ associated to off-diagonal BMBPT diagrams generated from operators with \texttt{deg\_max}~$=4$ or \texttt{deg\_max}~$=6$ have been produced\footnote{The two TSDs appearing in Fig.~\ref{f:tsd_production} are denoted, respectively, as T3.3 and T3.2 (with a cyclic permutation of $(a_1, a_2, a_3)$) in Fig.~\ref{diagTSD0123}.} and systematically displayed in Fig.~\ref{diagTSD0123}. Interestingly, restricting one-self to diagonal BMBPT diagrams and \texttt{deg\_max}~$=4$, T3.4 would have to be removed from the set of active TSDs, i.e.,\@ going from diagonal to off-diagonal BMBPT or going from \texttt{deg\_max}~$=4$ to \texttt{deg\_max}~$=6$ adds one allowed third-order TSD.

\subsection{From the TSD back to the BMBPT diagram}

In the end, \emph{different} BMBPT diagrams of order $p$ can have the \emph{same} TSD, i.e., the same underlying time structure. At the same time, off-diagonal BMBPT diagrams originating from the \emph{same} diagonal diagram may have \emph{different} TSDs, i.e., different underlying time structures. Once the TSD has been extracted from the BMBPT diagram of interest, its computation follows the algorithm detailed in Ref.~\cite{arthuis18a}. In particular, the treatment of a non-tree TSD requires to turn it first into a sum of tree TSDs.

Once the expression of a tree TSD of order $p$ has been obtained, the goal is to generate the actual time-integrated expression of the BMBPT diagrams associated to it. Rather than replacing the individual factors $a_q$, $q=1,\ldots,p$, by their expressions for each BMBPT diagram, one introduces the notion of subdiagram, or subgraph, to directly obtain their combinations appearing in the denominator of the time-integrated expression of interest. In Ref.~\cite{arthuis18a}, a subdiagram of a diagonal BMBPT diagram was defined as a diagram composed by a subset of vertices plus the propagators that are exchanged between them. As each vertex label $a_q$ in a TSD eventually stands for the sum/difference of quasi-particle energies associated with the lines entering/leaving the \emph{vertex} in the associated BMBPT diagram, the combination of these labels denotes the sum/difference of quasi-particle energies associated with the lines entering/leaving the \emph{subdiagram} grouping the corresponding vertices.

In the present context of off-diagonal BMBPT diagrams, one only needs to slightly generalize the notion of subdiagrams in order to apply the algorithm stipulated in Ref.~\cite{arthuis18a} in order to determine the factors entering the denominator of the time-integrated expression of the diagram. Thus a subdiagram of an off-diagonal BMBPT diagram is now defined as a diagram composed by a subset of vertices plus the \emph{normal} propagators that are exchanged between them. This definition is obviously consistent with the one introduced in Ref.~\cite{arthuis18a} for strictly diagonal BMBPT diagrams given that the latter solely contain normal propagators.

With this definition at hand, the energy denominator resulting from a BMBPT diagram associated with a tree TSD is obtained in the following way
\begin{enumerate}
\item Consider a vertex but the bottom one in the BMBPT diagram,
\begin{enumerate}
\item determine all its descendants using the TSD,
\item form a subdiagram using the vertex and its descendants,
\item sum the quasi-particle energies corresponding to the lines entering the subdiagram,
\item add the corresponding factor to the denominator expression,
\end{enumerate}
\item Go back to 1. until all vertices have been exhausted.
\end{enumerate}
Given that anomalous lines are excluded from the definition of a subdiagram, they systematically count as entering the subdiagram whenever they connect to a vertex belonging to it.

\begin{figure}[t!]
\begin{center}
\parbox{100pt}{\begin{fmffile}{diag_ex_labelled}
\begin{fmfgraph*}(100,100)
\fmfcmd{style_def prop_pm expr p =
    draw_plain p;
    shrink(.7);
        cfill (marrow (p, .25));
        cfill (marrow (p, .75))
    endshrink;
	enddef;}
\fmftop{v3}\fmfbottom{v0}
\fmf{phantom}{v0,v1}
\fmfv{d.shape=square,d.filled=full,d.size=3thick,l=$0$}{v0}
\fmf{phantom}{v1,v2}
\fmfv{d.shape=circle,d.filled=full,d.size=3thick,l=$\tau_1$}{v1}
\fmf{phantom}{v2,v3}
\fmfv{d.shape=circle,d.filled=full,d.size=3thick,l=$\tau_2$}{v2}
\fmfv{d.shape=circle,d.filled=full,d.size=3thick,l=$\tau_3$}{v3}
\fmffreeze
\fmf{prop_pm,left=0.6,tag=1}{v0,v2}
\fmf{prop_pm,right=0.6,tag=2}{v0,v2}
\fmf{prop_pm,left=0.7,tag=3}{v0,v3}
\fmf{prop_pm,right=0.7,tag=4}{v0,v3}
\fmf{prop_pm,left=0.5,tag=5}{v1,v2}
\fmf{prop_pm,right=0.5,tag=6}{v1,v2}
\fmf{prop_pm,left=0.6,tag=7}{v1,v3}
\fmf{prop_pm,right=0.6,tag=8}{v1,v3}
\fmfposition
	\fmfipath{p[]}
	\fmfiset{p1}{vpath1(__v0,__v2)}
	\fmfiset{p2}{vpath2(__v0,__v2)}
	\fmfiset{p3}{vpath3(__v0,__v3)}
	\fmfiset{p4}{vpath4(__v0,__v3)}
	\fmfiset{p5}{vpath5(__v1,__v2)}
	\fmfiset{p6}{vpath6(__v1,__v2)}
	\fmfiset{p7}{vpath7(__v1,__v3)}
	\fmfiset{p8}{vpath8(__v1,__v3)}
	\fmfiv{label=$k_1$,l.dist=.05w,l.a=-60}{point length(p1)/2 of p1}
	\fmfiv{label=$k_2$,l.dist=.05w,l.a=-120}{point length(p2)/2 of p2}
	\fmfiv{label=$k_3$,l.dist=.03w}{point length(p3)/2 of p3}
	\fmfiv{label=$k_4$,l.dist=.03w}{point length(p4)/2 of p4}
	\fmfiv{label=$k_5$,l.dist=.03w,l.a=-30}{point length(p5)/2 of p5}
	\fmfiv{label=$k_6$,l.dist=.03w,l.a=150}{point length(p6)/2 of p6}
	\fmfiv{label=$k_7$,l.dist=.05w,l.a=60}{point length(p7)/2 of p7}
	\fmfiv{label=$k_8$,l.dist=.05w,l.a=120}{point length(p8)/2 of p8}
\end{fmfgraph*}
\end{fmffile}}
\parbox{100pt}{\begin{fmffile}{tsd_ex_labelled}
\begin{fmfgraph*}(100,100)
\fmfcmd{style_def half_prop expr p =
    draw_plain p;
    shrink(.7);
        cfill (marrow (p, .5))
    endshrink;
	enddef;}
\fmftop{v3}\fmfbottom{v0}
\fmfv{d.shape=square,d.filled=full,d.size=3thick}{v0}
\fmfv{d.shape=circle,d.filled=full,d.size=3thick,l=$a_1$,l.a=180}{v1}
\fmf{phantom}{v2,v3}
\fmfv{d.shape=circle,d.filled=full,d.size=3thick,l=$a_2$}{v2}
\fmfv{d.shape=circle,d.filled=full,d.size=3thick,l=$a_3$}{v3}
\fmf{half_prop}{v0,v1}
\fmf{half_prop}{v1,v2}
\fmffreeze
\fmf{half_prop,right=0.6}{v1,v3}
\end{fmfgraph*}
\end{fmffile}}

\vspace{3\baselineskip}

\parbox{100pt}{\begin{fmffile}{diag_exoff_labelled}
\begin{fmfgraph*}(100,100)
\fmfcmd{style_def prop_pm expr p =
    draw_plain p;
    shrink(.7);
        cfill (marrow (p, .25));
        cfill (marrow (p, .75))
    endshrink;
	enddef;}
\fmfcmd{style_def prop_mm expr p =
        draw_plain p;
        shrink(.7);
            cfill (marrow (p, .75));
            cfill (marrow (reverse p, .75))
        endshrink;
        enddef;}
\fmftop{v3}\fmfbottom{v0}
\fmf{phantom}{v0,v1}
\fmfv{d.shape=square,d.filled=full,d.size=3thick,l=$0$}{v0}
\fmf{phantom}{v1,v2}
\fmfv{d.shape=circle,d.filled=full,d.size=3thick,l=$\tau_1$}{v1}
\fmf{phantom}{v2,v3}
\fmfv{d.shape=circle,d.filled=full,d.size=3thick,l=$\tau_2$}{v2}
\fmfv{d.shape=circle,d.filled=full,d.size=3thick,l=$\tau_3$}{v3}
\fmffreeze
\fmf{prop_pm,left=0.7,tag=1}{v0,v2}
\fmf{prop_pm,right=0.7,tag=2}{v0,v2}
\fmf{prop_pm,left=0.7,tag=3}{v0,v3}
\fmf{prop_pm,right=0.7,tag=4}{v0,v3}
\fmf{prop_pm,left=0.5,tag=5}{v1,v2}
\fmf{prop_pm,right=0.5,tag=6}{v1,v2}
\fmf{prop_mm,left=0.6,tag=7}{v1,v3}
\fmf{prop_mm,right=0.6,tag=8}{v1,v3}
\fmfposition
	\fmfipath{p[]}
	\fmfiset{p1}{vpath1(__v0,__v2)}
	\fmfiset{p2}{vpath2(__v0,__v2)}
	\fmfiset{p3}{vpath3(__v0,__v3)}
	\fmfiset{p4}{vpath4(__v0,__v3)}
	\fmfiset{p5}{vpath5(__v1,__v2)}
	\fmfiset{p6}{vpath6(__v1,__v2)}
	\fmfiset{p7}{vpath7(__v1,__v3)}
	\fmfiset{p8}{vpath8(__v1,__v3)}
	\fmfiv{label=$k_1$,l.dist=.05w,l.a=-60}{point length(p1)/2 of p1}
	\fmfiv{label=$k_2$,l.dist=.05w,l.a=-120}{point length(p2)/2 of p2}
	\fmfiv{label=$k_3$,l.dist=.03w}{point length(p3)/2 of p3}
	\fmfiv{label=$k_4$,l.dist=.03w}{point length(p4)/2 of p4}
	\fmfiv{label=$k_5$,l.dist=.03w,l.a=-30}{point length(p5)/2 of p5}
	\fmfiv{label=$k_6$,l.dist=.03w,l.a=150}{point length(p6)/2 of p6}
	\fmfiv{label=$k_9$,l.dist=.03w,l.a=0}{point 3length(p7)/4 of p7}
	\fmfiv{label=$k_{10}$,l.dist=.03w,l.a=180}{point 3length(p8)/4 of p8}
	\fmfiv{label=$k_7$,l.dist=.04w,l.a=-90}{point length(p7)/5 of p7}
	\fmfiv{label=$k_8$,l.dist=.04w,l.a=-90}{point length(p8)/5 of p8}
\end{fmfgraph*}
\end{fmffile}}
\parbox{100pt}{\begin{fmffile}{tsd_exoff_labelled}
\begin{fmfgraph*}(100,100)
\fmfcmd{style_def half_prop expr p =
    draw_plain p;
    shrink(.7);
        cfill (marrow (p, .5))
    endshrink;
	enddef;}
\fmftop{v3}\fmfbottom{v0}
\fmfv{d.shape=square,d.filled=full,d.size=3thick}{v0}
\fmfv{d.shape=circle,d.filled=full,d.size=3thick,l=$a_1$,l.a=180}{v1}
\fmf{phantom}{v2,v3}
\fmfv{d.shape=circle,d.filled=full,d.size=3thick,l=$a_2$}{v2}
\fmfv{d.shape=circle,d.filled=full,d.size=3thick,l=$a_3$}{v3}
\fmf{half_prop}{v0,v1}
\fmf{half_prop}{v1,v2}
\fmffreeze
\fmf{half_prop,right=0.6}{v0,v3}
\end{fmfgraph*}
\end{fmffile}}
\end{center}
\caption{Fully-labelled third-order diagonal and off-diagonal BMBPT diagrams displayed in Fig.~\ref{f:tsd_production} and their associated TSDs. The bottom vertex corresponds to $\tilde{O}^{40}(\varphi)$.}
\label{f:tsd_labelling}
\end{figure}
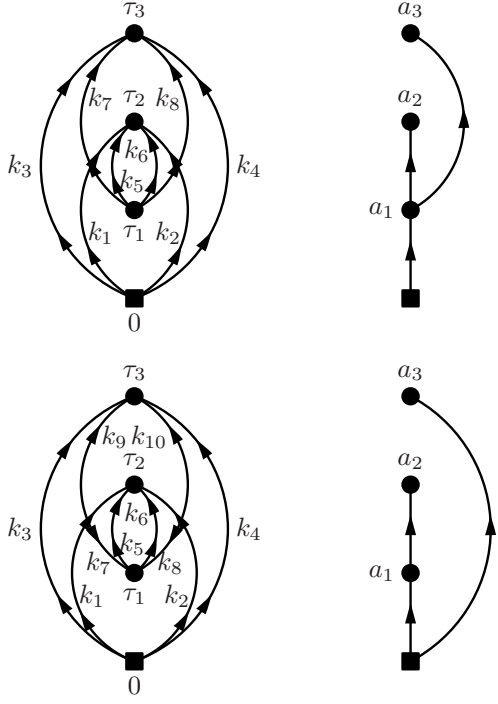

\begin{figure}
 \begin{center}
 \parbox{80pt}{\begin{fmffile}{offdiag_ex}
\begin{fmfgraph*}(80,80)
\fmfcmd{style_def prop_pm expr p =
    draw_plain p;
    shrink(.7);
        cfill (marrow (p, .25));
        cfill (marrow (p, .75))
    endshrink;
	enddef;}
\fmfcmd{style_def prop_mm expr p =
        draw_plain p;
        shrink(.7);
            cfill (marrow (p, .75));
            cfill (marrow (reverse p, .75))
        endshrink;
        enddef;}
\fmfcmd{style_def half_prop expr p =
    draw_plain p;
    shrink(.7);
        cfill (marrow (p, .5))
    endshrink;
	enddef;}
\fmfstraight
\fmftop{d1,d2,v2,d3,d4}\fmfbottom{v0}
\fmf{phantom}{v0,v1}
\fmfv{d.shape=square,d.filled=full,d.size=3thick,l=$0$}{v0}
\fmf{phantom}{v1,v2}
\fmfv{d.shape=circle,d.filled=full,d.size=3thick,l=$\tau_1$,l.a=180}{v1}
\fmfv{d.shape=circle,d.filled=full,d.size=3thick,l=$\tau_2$,l.a=180}{v2}
\fmffreeze
\fmf{prop_pm,left=0.5,tag=1,label=$k_1$}{v0,v1}
\fmf{prop_pm,right=0.5,tag=2,label=$k_2$}{v0,v1}
\fmf{prop_pm,left=0.5,tag=3,label=$k_3$}{v1,v2}
\fmf{prop_mm,right=0.5,tag=4}{v1,v2}
\fmf{half_prop,right,label=$k_7$}{d3,v2}
\fmf{half_prop,left,tag=6}{d3,v2}
\fmfposition
	\fmfipath{p[]}
	\fmfiset{p4}{vpath4(__v1,__v2)}
	\fmfiset{p6}{vpath6(__d3,__v2)}
	\fmfiv{label=$k_4$,l.dist=.03w,l.a=0}{point length(p4)/4 of p4}
	\fmfiv{label=$k_5$,l.dist=.03w,l.a=-120}{point 3length(p4)/4 of p4}
	\fmfiv{label=$k_6$,l.dist=.03w,l.a=-45}{point length(p6)/2 of p6}
\end{fmfgraph*}
\end{fmffile}}
\parbox{80pt}{\begin{fmffile}{tsd_off_diag_ex}
\begin{fmfgraph*}(80,80)
\fmfcmd{style_def half_prop expr p =
    draw_plain p;
    shrink(.7);
        cfill (marrow (p, .5))
    endshrink;
	enddef;}
\fmftop{v2}\fmfbottom{v0}
\fmfv{d.shape=square,d.filled=full,d.size=3thick}{v0}
\fmfv{d.shape=circle,d.filled=full,d.size=3thick,l=$a_1$,l.a=180}{v1}
\fmfv{d.shape=circle,d.filled=full,d.size=3thick,l=$a_2$}{v2}
\fmf{half_prop}{v0,v1}
\fmf{half_prop}{v1,v2}
\end{fmfgraph*}
\end{fmffile}}
\end{center}
\caption{Fully-labelled second-order off-diagonal BMBPT diagram and its associated TSD.}
\label{f:off_tsd_labelling}
\end{figure}
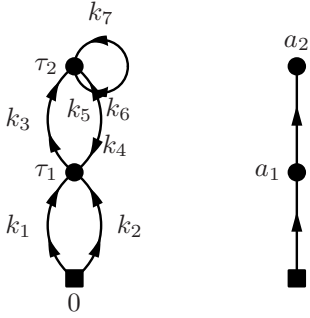

Let us illustrate the diagrammatic rule by focusing on the two third-order BMBPT diagrams\footnote{The results obtained in Eq.~\eqref{e:ex_bmbpt_goldstone} for the three second-order diagrams displayed in Fig.~\ref{f:ex_offbmbpt_diag} are straightforwardly recovered by the application of the diagrammatic rule.} displayed in Fig.~\ref{f:tsd_labelling}. 
\begin{enumerate}
\item The denominator in the time-integrated expression of the first diagram is obtained through the following steps 
\begin{enumerate}
\item The vertex at time $\tau_1$ in the BMBPT diagram corresponds to vertex $a_1$ in the TSD. Its descendants are vertices $a_2$ and $a_3$ corresponding to BMBPT vertices at times $\tau_2$ and $\tau_3$, respectively. The sum of quasi-particle energies associated to the lines entering the subgraph grouping the three vertices is $\epsilon_{k_{1}k_{2}k_{3}k_{4}}$, thus, providing the first factor entering the denominator.
\item The vertex at time $\tau_2$ in the BMBPT diagram corresponds to vertex $a_2$ in the TSD. It has no descendant such that the corresponding subgraph reduces to itself. The sum of quasi-particle energies associated to the lines entering the subgraph is $\epsilon_{k_{1}k_{2}k_{5}k_{6}}$, thus providing the second factor entering the denominator.
\item The vertex at time $\tau_3$ in the BMBPT diagram correspond to vertex $a_3$ in the TSD. It has no descendant such that the corresponding subgraph reduces to itself. The sum of quasi-particle energies associated to the lines entering the subgraph is $\epsilon_{k_{3}k_{4}k_{7}k_{8}}$, thus, providing the last factor entering the denominator. 
\item Eventually, the complete denominator reads as
\begin{equation*}
\epsilon_{k_1 k_2 k_3 k_4} \, \epsilon_{k_1 k_2 k_5 k_6} \, \epsilon_{k_3 k_4 k_7 k_8} \ ,
\end{equation*}
where each factor corresponds to a positive sum of quasi-particle energies.
\end{enumerate}
\item The denominator of the second, off-diagonal, diagram containing two anomalous lines and corresponding to a different TSD is obtained as
\begin{enumerate}
\item The vertex at time $\tau_1$ in the BMBPT diagram corresponds to vertex $a_1$ in the TSD. Contrarily to the previous case, vertex $a_3$ is not a descendant of $a_1$ anymore as is visible from the TSD such that the subgraph of interest solely groups $a_1$ and $a_2$. The sum of quasi-particle energies associated to the lines entering the subgraph in the BMBPT diagram is $\epsilon_{k_{1}k_{2}k_{7}k_{8}}$, thus, providing the first factor entering the denominator.
\item The situation of the vertex at time $\tau_2$ in the off-diagonal BMBPT diagram is strictly the same as in the previous diagonal one. Consequently, the associated factor in the denominator is $\epsilon_{k_{1}k_{2}k_{5}k_{6}}$.
\item As in the diagonal BMBPT diagram, the vertex at time $\tau_3$ has no descendant in the off-diagonal BMBPT diagram. Consequently, the subgraph corresponding to vertex $a_3$ reduces to itself. However, because the two anomalous lines carry two quasi-particle labels each, the sum of quasi-particle energies associated to the lines entering the subgraph has now become $\epsilon_{k_{3}k_{4}k_{9}k_{10}}$.
\item Eventually, the complete denominator reads as
\begin{equation*}
\epsilon_{k_{1}k_{2}k_{7}k_{8}} \, \epsilon_{k_1 k_2 k_5 k_6} \, \epsilon_{k_{3}k_{4}k_{9}k_{10}} \ ,
\end{equation*}
where each factor corresponds to a positive sum of quasi-particle energies.
\end{enumerate}
\end{enumerate}
For completeness, let us work out another example highlighting additional features of interest, i.e., the second-order off-diagonal diagram displayed in Fig.~\ref{f:off_tsd_labelling} together with its associated TSD\footnote{It is worth noting that the TSD of the off-diagonal diagram of interest is unchanged compared to the diagonal diagram it is generated from. However, the companion diagram with one more anomalous line joining the first and second $\Omega$ vertices relates to a different TSD.}. Applying the diagrammatic rule, one obtains
\begin{enumerate}
\item The vertex at time $\tau_1$ in the BMBPT diagram corresponds to vertex $a_1$ in the TSD. Because it remains one normal line connecting to the vertex at time $\tau_2$, $a_2$ is indeed its descendant. The subgraph of interest thus groups $a_1$ and $a_2$. Due to the more general definition of subgraphs at play in the context of off-diagonal BMBPT, the anomalous line connecting the two vertices is excluded from it, together with the self contraction on the upper vertex. Consequently, the sum of quasi-particle energies associated to the lines entering the subgraph is $\epsilon_{k_{1}k_{2}k_{4}k_{5}k_{6}k_{7}}$, thus, providing the first factor entering the denominator.
\item The vertex at time $\tau_2$ in the BMBPT diagram corresponds to vertex $a_2$ in the TSD. It has no descendant such that the corresponding subgraph reduces to itself, excluding the self contraction that the vertex exchanges with itself. The sum of quasi-particle energies associated to the lines entering the subgraph is $\epsilon_{k_{3}k_{5}k_{6}k_{7}}$, thus providing the second factor entering the denominator.
\item Eventually, the complete denominator reads as
\begin{equation*}
\epsilon_{k_{1}k_{2}k_{4}k_{5}k_{6}k_{7}} \, \epsilon_{k_{3}k_{5}k_{6}k_{7}} \ .
\end{equation*}
\end{enumerate}

\section{Output of the \texttt{ADG} program}

The typical output of the code associated with an off-diagonal BMBPT diagram is:

\paragraph{Diagram 3.3:}
\begin{align*}
\text{PO}2.3.3
&= \lim\limits_{\tau \to \infty}\frac{(-1)^2 }{2(3!)}\sum_{k_i}\tilde{O}^{40}_{k_{1}k_{2}k_{3}k_{4}} (\varphi) \Omega^{04}_{k_{1}k_{2}k_{3}k_{5}} \Omega^{04}_{k_{6}k_{4}k_{7}k_{8}} \\
& \times R^{--}_{k_{6}k_{5}}(\varphi) R^{--}_{k_{8}k_{7}}(\varphi) \\
& \times \int_{0}^{\tau}\!\!\mathrm{d}\tau_1\mathrm{d}\tau_2\theta(\tau_2-\tau_1) e^{-\tau_1 \epsilon^{}_{k_{1}k_{2}k_{3}k_{5}}}e^{-\tau_2 \epsilon^{}_{k_{4}k_{6}k_{7}k_{8}}} \\
&= \frac{(-1)^2 }{2(3!)}\sum_{k_i}\frac{\tilde{O}^{40}_{k_{1}k_{2}k_{3}k_{4}} (\varphi) \Omega^{04}_{k_{1}k_{2}k_{3}k_{5}} \Omega^{04}_{k_{6}k_{4}k_{7}k_{8}} }{ \epsilon^{}_{k_{1}k_{2}k_{3}k_{5}}\ \epsilon^{}_{k_{4}k_{6}k_{8}k_{7}}\ } \\
&\times R^{--}_{k_{6}k_{5}}(\varphi) R^{--}_{k_{8}k_{7}}(\varphi)
\end{align*}

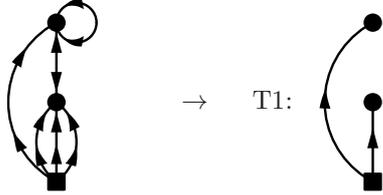
\begin{center}
\parbox{60pt}{\begin{fmffile}{diag_2_2}
\begin{fmfgraph*}(60,60)
\fmfcmd{style_def prop_pm expr p =
draw_plain p;
shrink(.7);
	cfill (marrow (p, .25));
	cfill (marrow (p, .75))
endshrink;
enddef;}
\fmfcmd{style_def prop_mm expr p =
draw_plain p;
shrink(.7);
	cfill (marrow (p, .75));
	cfill (marrow (reverse p, .75))
endshrink;
enddef;}
\fmftop{v2}\fmfbottom{v0}
\fmf{phantom}{v0,v1}
\fmfv{d.shape=square,d.filled=full,d.size=3thick}{v0}
\fmf{phantom}{v1,v2}
\fmfv{d.shape=circle,d.filled=full,d.size=3thick}{v1}
\fmfv{d.shape=circle,d.filled=full,d.size=3thick}{v2}
\fmffreeze
\fmfcmd{style_def half_prop expr p =
draw_plain p;
shrink(.7);
	cfill (marrow (p, .5))
endshrink;
enddef;}
\fmfv{}{a20}
\fmffixed{(15pt,0)}{v2,a20}
\fmf{half_prop,right}{a20,v2}
\fmf{half_prop,left}{a20,v2}
\fmffreeze
\fmf{prop_pm,left=0.5}{v0,v1}
\fmf{prop_pm}{v0,v1}
\fmf{prop_pm,right=0.5}{v0,v1}
\fmf{prop_pm,left=0.6}{v0,v2}
\fmf{prop_mm}{v1,v2}
\end{fmfgraph*}
\end{fmffile}}
\hspace{10pt} $\rightarrow$ \hspace{10pt} T1:\parbox{60pt}{\begin{fmffile}{time_0}
\begin{fmfgraph*}(60,60)
\fmfcmd{style_def half_prop expr p =
draw_plain p;
shrink(.7);
	cfill (marrow (p, .5))
endshrink;
enddef;}
\fmftop{v2}\fmfbottom{v0}
\fmf{phantom}{v0,v1}
\fmfv{d.shape=square,d.filled=full,d.size=3thick}{v0}
\fmf{phantom}{v1,v2}
\fmfv{d.shape=circle,d.filled=full,d.size=3thick}{v1}
\fmfv{d.shape=circle,d.filled=full,d.size=3thick}{v2}
\fmffreeze
\fmf{half_prop}{v0,v1}
\fmf{half_prop,left=0.6}{v0,v2}
\end{fmfgraph*}
\end{fmffile}}
\end{center}

\begin{equation*}
\text{T}1 = \frac{1}{a_1a_2}
\end{equation*}
\begin{align*}
a_1 &= \epsilon^{}_{k_{1}k_{2}k_{3}k_{5}}\\
a_2 &= \epsilon^{}_{k_{4}k_{6}k_{7}k_{8}}
\end{align*}

\section{Connection to time-ordered diagrammatics}

In Ref.~\cite{arthuis18a}, time-unordered and time-ordered diagrammatics emerging, respectively, from the time-dependent and the time-independent formulations of straight, i.e.,\@ diagonal, BMBPT were compared at length. The main outcome of the analysis related to the capacity of the time-unordered diagrammatics to resum at once large classes of time-ordered diagrams. Correspondingly, it was shown that the new diagrammatic rule allowing for the direct obtention of the time-integrated results on the basis of time-unordered diagrams generalizes the resolvent rule at play for time-ordered diagrams.

As in Ref.~\cite{arthuis18a}, the formal and numerical developments presented in this paper rely on the time-dependent formulation of PNP-BMBPT~\cite{Duguet:2015yle}. While it is traditionally more customary to design many-body perturbation theories on the basis of a time-independent formalism~\cite{Shavitt_Bartlett_2009}, this task has so far not been attempted for PNP-BMBPT. While the end result has to be the same, the partitioning\footnote{A valid partitioning relates to splitting the complete order $p$ in a sum of terms that are individually proportional to a fraction of the form $1/(\epsilon_{k_i\ldots k_j}\ldots\epsilon_{k_u\ldots k_v})$ with $p$ energy factors in the denominator. Any other form does not constitute a valid partitioning in the present context.} of the complete order-$p$ contribution to the observable $\text{O}^{\text{A}}_0$ will differ in both approaches. In the absence of the time-ordered diagrammatics associated to PNP-BMBPT, we cannot proceed to the same analysis as done in Ref.~\cite{arthuis18a} for straight BMBPT. Leaving this analysis for the future, it can however be anticipated that time-unordered off-diagonal BMBPT diagrams will feature the same capacity to resum large classes of time-ordered diagrams at play in the, yet-to-be-formulated\footnote{The fact that anomalous propagators/contractions are not diagonal in their quasi-particle indices should lead to a rather unconventional time-ordered diagrammatics that shall itself lead to an interesting variant of the resolvent rule.}, time-independent version of PNP-BMBPT.

\section{Use of the \texttt{ADG} program}
\label{programuse}

\texttt{ADG} has been designed to work on any computer with a Python3 distribution, and successfully tested on recent GNU/Linux distributions and on MacOS. Additionally to Python, \emph{setuptools} and \emph{distutils} packages must already be installed, which is the case on most standard recent distributions. Having \emph{pip} installed eases the process but is not technically required. The \emph{NumPy}, \emph{NetworkX} and \emph{SciPy} libraries are automatically downloaded during the install process. Additionally, one needs a \LaTeX\ distribution installed with the PDF\LaTeX\ compiler for \texttt{ADG} to produce the pdf file associated to the output if desired.

\subsection{Installation}

\subsubsection{From the Python Package Index}

The easiest way to install\footnote{As the previous version of ADG was developped with Python2, the installation will be made next to the previous version, though it should not cause any problem using it. Users who would want to uninstall the previous version first might do it by entering \texttt{pip2 uninstall adg}.} \texttt{ADG} is to obtain it from the Python Package Index\footnote{\url{https://pypi.org/project/adg/}} by entering the following command
\begin{verbatim}
pip3 install adg
\end{verbatim}
Provided \emph{setuptools} is already installed, \emph{pip} takes care of downloading and installing \texttt{ADG} as well as \emph{NumPy}, \emph{SciPy} and \emph{NetworkX}. Once a new version of \texttt{ADG} is released, one can install it by entering the command
\begin{verbatim}
pip3 install --upgrade adg
\end{verbatim}

\subsubsection{From the source files}

Once the \texttt{ADG} source files are downloaded from the CPC library or the GitHub repository\footnote{\url{https://github.com/adgproject/adg}}, one must enter the project folder and either run
\begin{verbatim}
pip3 install .
\end{verbatim}
or
\begin{verbatim}
python3 setup.py install
\end{verbatim}
With this method, \emph{pip}\footnote{Depending on the system, it might be necessary either to use the "--user" flag to install it only for a specific user or to run the previous command with "sudo -H" to install it system-wide.} also takes care of downloading and installing \emph{NumPy}, \emph{NetworkX} and \emph{SciPy}.

\subsection{Run the program}

\subsubsection{Batch mode}

The most convenient way to use \texttt{ADG} is to run it in batch mode with the appropriate flags. For example, to run the program and generate off-diagonal BMBPT diagrams at order 3 for example, one can use
\begin{verbatim}
adg -o 3 -t PBMBPT -d -c
\end{verbatim}
where the \texttt{-o} flag is for the order, \texttt{-t} for the type of theory, \texttt{-d} indicates that the diagrams must be drawn and \texttt{-c} that \texttt{ADG} must compile the \LaTeX\ output. A complete list of the program's options can be obtained via the program's documentation (see Sec.~\ref{documentation}) or through
\begin{verbatim}
adg -h
\end{verbatim}

Currently, \texttt{ADG} can be run either in relation to HF-MBPT by using \texttt{-t MBPT}, to straight BMBPT by using \texttt{-t BMBPT} or to off-diagonal BMBPT by using \texttt{-t PBMBPT}. Though the algorithms described in the previous sections can be used regardless of the diagrams' orders, \texttt{ADG} has been arbitrarily restricted to order 10 or lower to avoid major overloads of the system. Future users are nevertheless advised to first launch calculations at low orders (2, 3 or 4 typically) as the time and memory needed for computations rise rapidly with the perturbative order.

\subsubsection{Interactive mode}

As an alternative to the batch mode, \texttt{ADG} can be run on a terminal by entering the command
\begin{verbatim}
adg -i
\end{verbatim}
A set of questions must be answered using the keyboard to configure and launch the calculation. The interactive mode then proceeds identically to the batch mode.

\subsection{Steps of a program run}

Let us briefly recapitulate the different steps of a typical \texttt{ADG} run
\begin{itemize}
\item Run options are set either by using the command-line flags entered by the user or during the interactive session via keyboard input.
\item \texttt{ADG} creates a list of adjacency matrices for the appropriate theory and perturbative order using \emph{NumPy}, and feeds them to \emph{NetworkX} that creates \emph{MultiDiGraph} objects.
\item Checks are performed on the list of graphs to remove topologically equivalent or ill-defined graphs.
\item The list of topologically unique graphs is used to produce \emph{Diagram} objects that store the graph as well as some of its associated properties depending on the theory (HF status, excitation level, etc.). For off-diagonal BMBPT, this is done in two-steps, first genereating the straight BMBPT ones. The expression associated to the graphs are eventually extracted.
\item The program prints on the terminal the number of diagrams per category and writes the \LaTeX\ output file, the details of which depend on the options selected by the user, as well as a list of adjacency matrices associated to the diagrams. Other output files may be produced, depending on the theory and the user's input.
\item If asked by the user, the program performs the PDF\LaTeX\ compilation.
\item Unnecessary temporary files are removed and the programs exits.
\end{itemize}

\subsection{Documentation}
\label{documentation}

\subsubsection{Local documentation}

Once the source files have been downloaded, a quick start guide is available in the \texttt{README.md} file. Once \texttt{ADG} is installed, it is possible to read its manpages through
\begin{verbatim}
man adg
\end{verbatim}
or a brief description of the program and its options through
\begin{verbatim}
adg -h
\end{verbatim}
A more detailed HTML documentation can be generated directly from the source files by going into the \texttt{docs} directory and running
\begin{verbatim}
make html
\end{verbatim}
The documentation is then stored in \texttt{docs/build/html}, with the main file being \texttt{index.html}. A list of other possible types of documentation format is available by running
\begin{verbatim}
make help
\end{verbatim}

\subsubsection{Online documentation}

The full HTML documentation is available online under \url{https://adg.readthedocs.io/}, and help with eventual bugs of the program can be obtained by opening issues on the GitHub repository at \url{https://github.com/adgproject/adg}.

\section{Conclusions}
\label{conclusions}

The motivations underlining our work were explained at length in Ref.~\cite{arthuis18a} where the first version of the code \texttt{ADG} was described. The long-term rationale relates to the possibility to automatically (i) generate and (ii) algebraically evaluate diagrams in various quantum many-body methods of interest. In Ref.~\cite{arthuis18a}, the focus was put on Bogoliubov many-body perturbation theory (BMBPT) that has been recently formulated~\cite{Duguet:2015yle} and first implemented at low orders~\cite{Tichai:2018mll} to tackle (near) degenerate Fermi systems, e.g., open-shell nuclei displaying a superfluid character. Given the need to tackle three-nucleon interactions, i.e., six-leg vertices, in nuclear physics and the implementation of high-order contributions authorized by the rapid progress of computational power, the first version of the code \texttt{ADG} makes possible to generate all valid BMBPT diagrams and to evaluate their algebraic expression to be implemented in a numerical application. This is realized at an arbitrary order $p$ for a Hamiltonian containing both two-body (four-legs) and three-body (six-legs) interactions (vertices). The formal advances and the numerical methods necessary to achieve this goal can be found in Ref.~\cite{arthuis18a}.

Bogoliubov MBPT perturbatively expands the exact solution of the Schrödinger equation around a so-called Bogoliubov reference state, i.e., a general product state breaking $U(1)$ global-gauge symmetry associated with the conservation of good particle number in the system. This results into a (hopefully small) symmetry contamination as soon as the expansion is truncated in actual calculations. Given that the breaking of a symmetry cannot actually be realized in a finite quantum system, $U(1)$ symmetry must eventually be restored at any truncation order, which is made possible thanks to the recent formulation of the particle-number projected Bogoliubov many-body perturbation theory (PNP-BMBPT)~\cite{Duguet:2015yle} that extends straight BMBPT on the basis of a more general diagrammatic expansion. 

Consequently, the present paper details the systematic generation and evaluations of diagrams at play in PNP-BMBPT operated by the second version (v2.0.0) of the code \texttt{ADG}. While the automated evaluation of the diagrams only requires a mere extension of the diagrammatic rule unrevealed in Ref.~\cite{arthuis18a}, the method used to first generate all allowed diagrams is different from the one employed in Ref.~\cite{arthuis18a}. Taking advantage of the capacity of the code \texttt{ADG} to already produce all valid BMBPT diagrams of order $p$, the set of rules to generate all those at play in PNP-BMBPT \emph{from} those appearing in BMBPT was identified and implemented. 

Eventually, the second version of the code \texttt{ADG} is kept flexible enough to be expanded throughout the years to tackle the diagrammatics at play in yet other many-body formalisms that either already exist or are yet to be formulated.

\section*{Acknowledgments}

The authors thank M.\@ Drissi for reporting a bug in the previous version of the program. 
This publication is based on work supported by the UK Science and Technology Facilities Council (STFC) through grants ST/P005314/1 and ST/L005516/1, the Max Planck Society, the Deutsche Forschungsgemeinschaft (DFG, German Research Foundation) -- Project-ID 279384907 -- SFB 1245, and the framework of the Espace de Structure et de r\'eactions Nucl\'eaires Th\'eorique (ESNT) at CEA.


\appendix

\section{Reference states}
\label{ingredients}

\subsection{Bogoliubov vacuum}

We consider the Bogoliubov reference state defined as
\begin{equation}
\label{e:bogvac}
| \Phi \rangle \equiv \mathcal{C} \prod_{k} \beta_{k} | 0 \rangle \, ,
\end{equation}
where quasi-particle operators are related to particle ones through a Bogoliubov transformation
\begin{equation}
\begin{pmatrix}
\beta \\
\beta^{\dagger}
\end{pmatrix}
 = W^{\dagger}
\begin{pmatrix}
c \\
c^{\dagger}
\end{pmatrix}
=
\begin{pmatrix}
U^{\dagger} & V^{\dagger} \\
V^{T} &  U^{T}
\end{pmatrix}
\begin{pmatrix}
c \\
c^{\dagger}
\end{pmatrix}
 \, . \label{transfobogo}
\end{equation}
The product state $| \Phi \rangle$ is a vacuum for the set of quasi-particle operators, i.e.,\@ $\beta_k | \Phi \rangle = 0$ for all $k$.

\subsection{Rotated Bogoliubov vacuum}

One introduces the gauge-rotated Bogoliubov vacuum
\begin{equation}
| \Phi (\varphi) \rangle \equiv R(\varphi) |  \Phi \rangle \equiv \mathcal{C} \displaystyle \prod_{k} \bar{\beta}_{k} | 0 \rangle \, , \label{rotationvacuum}
\end{equation}
where rotated quasi-particle operators are defined through
\begin{align}
\begin{pmatrix}
\bar{\beta} \\
\bar{\beta}^{\dagger}
\end{pmatrix}
 (\varphi)
&\equiv 
R(\varphi)
\begin{pmatrix}
\beta \\
\beta^{\dagger}
\end{pmatrix}
R^{-1}(\varphi) \\
&\equiv W^{\varphi \, \dagger}
\begin{pmatrix}
c \\
c^{\dagger}
\end{pmatrix}
\, , \nonumber
\end{align}
with the associated Bogoliubov transformation reading as
\begin{align}
W^{\varphi \, \dagger} &\equiv 
\begin{pmatrix}
U^{\varphi \, \dagger} & V^{\varphi \, \dagger} \\
V^{\varphi \, T} &  U^{\varphi \, T}
\end{pmatrix} \\
&= 
\begin{pmatrix}
U^{\dagger} e^{-i\varphi} & V^{\dagger} e^{+i\varphi} \\
V^{T}  e^{-i\varphi} &  U^{T} e^{+i\varphi}
\end{pmatrix} \, . \nonumber
\end{align}

\subsection{Thouless transformation between vacua}
\label{SecThouless}

As stipulated by Eq.~\eqref{rotationvacuum}, $| \Phi (\varphi) \rangle $ is obtained from $| \Phi \rangle $ via the unitary transformation $R(\varphi)$ whose generator is $A$. One can rather express $| \Phi(\varphi) \rangle$ via a non-unitary Thouless transformation of $| \Phi \rangle$ according to~\cite{thouless60}
\begin{equation}
\label{thoulessbetweenbothvacua}
| \Phi(\varphi) \rangle \equiv \langle \Phi | \Phi(\varphi) \rangle  e^{Z(\varphi)} | \Phi \rangle \, ,
\end{equation}
where the one-body Thouless operator 
\begin{equation}
\label{thoulessoprot}
Z(\varphi) \equiv \frac{1}{2} \sum_{k_1k_2} Z^{20}_{k_1k_2} (\varphi) \beta^{\dagger}_{k_1} \beta^{\dagger}_{k_2}
\end{equation}
only contains a pure excitation part over $| \Phi \rangle $. The corresponding Thouless matrix
\begin{equation}
\label{thoulessoprot2}
Z^{20} (\varphi) \equiv  N^{ \ast} (\varphi) M^{\ast -1} (\varphi)
\end{equation}
is expressed in terms of the Bogoliubov transformation connecting quasi-particle operators of $| \Phi(\varphi) \rangle$ to those of $| \Phi \rangle$
\begin{align}
\begin{pmatrix}
\bar{\beta} \\
\bar{\beta}^{\dagger}
\end{pmatrix} (\varphi)
&\equiv 
\mathcal{W}^{\dagger}(\varphi)
\begin{pmatrix}
\beta \\
\beta^{\dagger}
\end{pmatrix} \, ,
\end{align}
where
\begin{align}
\mathcal{W}^{\dagger}(\varphi) &= W^{\varphi \, \dagger}W \nonumber \\
&\equiv 
\begin{pmatrix}
M^{\dagger}(\varphi) & N^{\dagger}(\varphi) \\
N^{T}(\varphi) &  M^{T}(\varphi)
\end{pmatrix} \label{transfobogotransition} \\
&=
\begin{pmatrix}
U^{\varphi \, \dagger}U + V^{\varphi \, \dagger}V & V^{\varphi \, \dagger}U^{\ast} + U^{\varphi \, \dagger}V^{\ast} \\
V^{\varphi \, T}U + U^{\varphi \, T}V &  U^{\varphi \, T}U^{\ast} + V^{\varphi \, T}V^{\ast}
\end{pmatrix} \, . \nonumber
\end{align}

\subsection{Elementary contractions}
\label{sectioncontractions}

The elementary contractions of quasi-particle operators that are in use when employing the off-diagonal Wick theorem~\cite{balian69a} are given by
\begin{align}
\textbf{R}(\varphi) &\equiv
\begin{pmatrix}
\frac{\langle \Phi | \beta^{\dagger}\beta^{\phantom{\dagger}} | \Phi(\varphi) \rangle}{\langle \Phi | \Phi(\varphi) \rangle} & \frac{\langle \Phi | \beta^{\phantom{\dagger}}\beta^{\phantom{\dagger}} | \Phi(\varphi) \rangle}{\langle \Phi | \Phi(\varphi) \rangle} \\
\frac{\langle \Phi | \beta^{\dagger}\beta^{\dagger} | \Phi(\varphi) \rangle}{\langle \Phi | \Phi(\varphi) \rangle} &  \frac{\langle \Phi | \beta^{\phantom{\dagger}}\beta^{\dagger} | \Phi(\varphi) \rangle}{\langle \Phi | \Phi(\varphi) \rangle}
\end{pmatrix} \nonumber \\
&\equiv
\begin{pmatrix}
R^{+-}(\varphi) & R^{--}(\varphi) \\
R^{++}(\varphi) &  R^{-+}(\varphi)
\end{pmatrix} \nonumber \\
&=
\begin{pmatrix}
0 & -Z^{20} (\varphi) \\
0 &  1
\end{pmatrix}  \, . \label{offdiagonalcontract}
\end{align}
Most of the above contractions are easily obtained by using the fact that $| \Phi \rangle$ is the vacuum of the quasi-particle operators, i.e.,\@ $\beta_k | \Phi \rangle =\langle \Phi | \beta^{\dagger}_k =0$ for all $k$. The single non-trivial (anomalous) contraction is obtained on the basis of standard Wick's theorem as
\begin{align}
R^{--}_{k_1k_2}(\varphi) &= \frac{\langle \Phi | \beta_{k_1}\beta_{k_2} | \Phi(\varphi) \rangle}{\langle \Phi | \Phi(\varphi) \rangle} \nonumber \\
&= \langle \Phi |  \beta_{k_1}\beta_{k_2} e^{Z(\varphi)} | \Phi \rangle  \nonumber \\
&= \frac{1}{2} \sum_{kk'} Z^{20}_{kk'} (\varphi) \langle \Phi |  \beta_{k_1}\beta_{k_2} \beta^{\dagger}_{k} \beta^{\dagger}_{k'} | \Phi \rangle  \nonumber \\
&= \frac{1}{2}  \left(Z^{20}_{k_2k_1}(\varphi) - Z^{20}_{k_1k_2}(\varphi)\right)  \nonumber\\
&= - Z^{20}_{k_1k_2}(\varphi) \, , \label{nontrivialcontract}
\end{align}
and is zero in the diagonal case, i.e.,\@ $R^{--}_{k_1k_2}(0)=0$.

\section{Transformed operator $\tilde{O}(\varphi)$}
\label{transformME}

The gauge-dependent similarity transformed operator\footnote{The Hermitian character of an operator $O$ is lost by the application of similarity transformation.} of $O$ is defined through
\begin{align}
  \tilde{O}(\varphi) &\equiv e^{-Z(\varphi)} O e^{Z(\varphi)} \, . \label{transformedopApp}
\end{align}
Taking as an example one term in the normal-ordered expression of $O$, e.g.,
\begin{equation}
  O^{ij} \equiv \frac{1}{i!}\frac{1}{j!} \!\!\sum_{k_1\ldots k_{i+j}}\!\! O^{ij}_{k_1\ldots k_{i+j}} \beta^\dagger_{k_1} \ldots \beta^\dagger_{k_i}\beta_{k_{i+j}} \ldots \beta_{k_{i+1}}\, , \nonumber
\end{equation}
its transformed partner reads as
\footnote{The notation $\tilde{O}^{(ij)}(\varphi)$ denotes the transformed operator of $O^{ij}$ such that the upper label $(ij)$ is a sole reminder of the normal-ordered nature of the original operator but does \emph{not} characterize the normal-ordered nature of the transformed operator.
Contrarily, $\tilde{O}^{mn}(\varphi)$ does denote the normal-ordered part of the transformed operator $\tilde{O}(\varphi)$ of $O$ containing $m$ ($n$) quasi-particle creation (annihilation) operators.}
\begin{align}
  \tilde{O}^{(ij)}(\varphi)
  &\equiv e^{-Z(\varphi)} O^{ij} e^{Z(\varphi)} \label{examplerotatedop1} \\
  &= \frac{1}{i!}\frac{1}{j!} \!\!\sum_{k_1\ldots k_{i+j}}\!\! O^{ij}_{k_1\ldots k_{i+j}} \tilde{\beta}^\dagger_{k_1} \ldots \tilde{\beta}^\dagger_{k_i}\tilde{\beta}_{k_{i+j}} \ldots \tilde{\beta}_{k_{i+1}} \, , \nonumber
\end{align}
where the transformed quasi-particle operators are
\begin{subequations}
\label{transforqptilde}
\begin{align}
\tilde{\beta}_{k}(\varphi) &\equiv e^{-Z(\varphi)}\beta_{k}e^{Z(\varphi)} \nonumber \\
&= \beta_{k} - [Z(\varphi),\beta_{k}] + \frac{1}{2!} [Z(\varphi),[Z(\varphi),\beta_{k}]] + \ldots  \nonumber \\
&= \beta_{k} + \sum_{k'} Z^{20}_{kk'}(\varphi) \beta^{\dagger}_{k'} \, , \label{transforqptilde1} \\
\tilde{\beta}^{\dagger}_{k}(\varphi) &\equiv e^{-Z(\varphi)}\beta^{\dagger}_{k}e^{Z(\varphi)} \nonumber  \\
&= \beta^{\dagger}_{k} - [Z(\varphi),\beta^{\dagger}_{k}] + \frac{1}{2!} [Z(\varphi),[Z(\varphi),\beta^{\dagger}_{k}]] + \ldots \nonumber \\
&= \beta^{\dagger}_{k} \, , \label{transforqptilde2}
\end{align}
\end{subequations}
were use was made of the elementary commutators
\begin{subequations}
\begin{align}
 \Big[ \beta^\dagger_{k} \beta^\dagger_{k'} , \beta_{k_1} \Big]   &= \beta^\dagger_{k} \delta_{k' k_1} - \beta^\dagger_{k'} \delta_{k k_1} \, , \\
 \Big[ \beta^\dagger_{k} \beta^\dagger_{k'} , \beta^\dagger_{k_1} \Big] &= 0 \, . 
\end{align}
\end{subequations}

Exploiting Eq.~\eqref{transforqptilde} and normal ordering the resulting terms with respect to $| \Phi \rangle$, the transformed operator in Eq.~\eqref{examplerotatedop1} is eventually written as
\begin{align}
  \tilde{O}^{(ij)}(\varphi)
  &\equiv \sum_{m=i}^{i+j} \sum_{\substack{n=0\\m+n\leq i+j}}^{j} \frac{1}{m!}\frac{1}{n!} \nonumber \\
  & \times \sum_{k_1\ldots k_{m+n}}\!\! \tilde{O}^{mn(ij)}_{k_1\ldots k_{m+n}}(\varphi) \beta^\dagger_{k_1} \ldots \beta^\dagger_{k_m}\beta_{k_{m+n}} \ldots \beta_{k_{m+1}} \, , \label{rotatedop}
\end{align}
thus defining a sum of normal-ordered terms. Each term has at least as many quasi-particle creation operators ($i$) as the original operator $O^{ij}$ and possibly up to the total number of original quasi-particle operators ($i+j$). The number of annihilation operators ranges from 0 to the original number ($j$) such that the overall number of quasi-particle operators is bound to remain between $i$ and $i+j$ in each term. One notices that the only structural difference between the original and the transformed normal-ordered operators relates to the fact that matrix elements of the latter depend on the gauge angle.
Of course, the original operator is recovered in the unrotated limit, i.e.,\@ $\tilde{O}(0)=O$.

Applying the above procedure to the complete operator $O$ provides the normal-ordered form of the transformed operator
\begin{align} \label{e:h3qpastransformed}
\tilde{O}(\varphi) &\equiv \tilde{O}^{[0]}(\varphi) + \tilde{O}^{[2]}(\varphi) + \tilde{O}^{[4]}(\varphi) + \tilde{O}^{[6]}(\varphi)  \, ,
\end{align}
in which the term $\tilde{O}^{nm}(\varphi)$ collects various contributions $\tilde{O}^{nm(ij)}(\varphi)$. Each term $\tilde{O}^{nm}(\varphi)$ possesses the same operator structure as the corresponding term in Eq.~\eqref{e:oqpas}, except that the original matrix elements are replaced by gauge-dependent ones, e.g.,\@ $O^{31}_{k_1 k_2 k_3 k_4}$ is formally replaced by $\tilde{O}^{31}_{k_1 k_2 k_3 k_4}(\varphi)$. The expressions of the matrix elements of each normal-ordered contribution $\tilde{O}^{nm}(\varphi)$ in terms of the matrix elements of the original normal-ordered contributions to an operator $O$ with $\texttt{deg\_max}=4$ can be found in Ref.~\cite{Duguet:2015yle}.

\section{Changelog}

Since the previous main version of \emph{ADG}, the main modifications to the software have focused on:
\begin{itemize}
\item Adding the particle-number projected BMBPT formalism.
\item Fixing an error in a symmetry factor in BMBPT diagrams.
\item Porting the code to Python3 while maintaining compatibility with Python2.7.
\item Changing the matrices data structures to \emph{NumPy} arrays.
\item Removing deprecated calls to \emph{NetworkX}.
\item Various optimisations to the BMBPT diagram generation process.
\item Reducing the memory requirements of diagrams.
\item Fixing several errors in the documentation.
\item Fixing the installation process for additional dependencies.
\end{itemize}

\section*{References}
\bibliographystyle{apsrev4-1}
\bibliography{Bibliographie}







\end{document}